\documentclass[reprint,preprintnumbers,amsmath,amssymb,aps,nofootinbib,showkeys, superscriptaddress]{revtex4-2}
\usepackage[utf8]{inputenc}  % UTF-8 enabled
\usepackage{graphicx}  % Include figure files
\usepackage{xcolor}  % Allow for a color text
\usepackage{amsmath}  % math fonts
\usepackage{amsfonts}  % math fonts
\usepackage{latexsym}  % math fonts
\usepackage{amssymb,bbm}  % math fonts
\usepackage{bm}  % bold math fonts
\usepackage[colorlinks=true,allcolors=blue]{hyperref}  % add hypertext capabilities
\usepackage{lipsum}  % dummy text
\usepackage{newunicodechar}
\newunicodechar{′}{\ensuremath{'} }
\usepackage{setspace}
\usepackage{braket}
\usepackage{siunitx}
\usepackage{tabularx}
\usepackage{enumitem}

\usepackage[table]{xcolor}

\usepackage{booktabs}

\definecolor{stronggreen}{RGB}{158,214,127}   % dark green
\definecolor{softgreen}{RGB}{222,242,211}     % light green
\definecolor{softyellow}{RGB}{237,235,118}    % yellow

\newcolumntype{C}[1]{>{\centering\arraybackslash}p{#1}}
\newcolumntype{L}[1]{>{\raggedright\arraybackslash}p{#1}}
\definecolor{softgreen}{RGB}{222,242,211}   % typical
\definecolor{stronggreen}{RGB}{158,214,127} % best / demonstrated
\definecolor{softyellow}{RGB}{245,241,175}  % projected / future

\newcommand{\xd}{\delta}

\newcommand{\tm}{{\text -}}

\newcommand{\tacq}{t_{\R{acq}}}

\newcommand{\xt}{\vartheta}

\newcommand{\xr}{\rho}

\newcommand{\xph}{\phi}

\newcommand{\app}{\approx}

\newcommand{\fsens}{f_{\R{sens}}}

\newcommand{\Cs}{{}^{13}\R{C}}

\newcommand{\Bsens}{B_{\R{target}}}

\newcommand{\degree}{^{\circ}}
\newcommand{\xD}{\Delta}

\newcommand{\fr}[2]{\frac{#1}{#2}}

\newcommand{\sq}[1]{\sqrt{#1}}

\newcommand{\dg}{\dagger}

\newcommand{\beq}{\begin{equation}}
\newcommand{\eeq}{\end{equation}}
                  
\newcommand{\benum}{\begin{enumerate}}
\newcommand{\eenum}{\end{enumerate}}
                    
\newcommand{\bit}{\begin{itemize}}
\newcommand{\eit}{\end{itemize}}
\newcommand{\xhat}{\hat{\T{x}}}
\newcommand{\yhat}{\hat{\T{y}}}
\newcommand{\zhat}{\hat{\T{z}}}

\newcommand{\bea}{\begin{eqnarray}}
\newcommand{\eea}{\end{eqnarray}}

\newcommand{\bev}{\begin{verbatim}}
\newcommand{\eev}{\end{verbatim}}

\newcommand{\noi}{\noindent}

\newcommand{\pll}{\parallel}

\newcommand{\T}[1]{\textbf{#1}}
\newcommand{\I}[1]{\textit{#1}}
\newcommand{\R}[1]{\textrm{#1}}
\newcommand{\zfr}[1]{\figurename\,\ref{fig:#1}}

\newcommand{\ba}{\left\{ \begin{array}{lr}}
\newcommand{\ea}{\end{array}\right.}

\newcommand{\blist}[1]{
 \begin{list}{#1}%$\ast\circ\bullet\Right
 \begin{align}
	 arrow
 \end{align}
 $\checkmark\star
  { \setlength{\itemsep}{3pt}
     \setlength{\parsep}{2pt}
     \setlength{\topsep}{3pt}
     \setlength{\partopsep}{0pt}
     \setlength{\leftmargin}{1em}
     \setlength{\labelwidth}{1em}
     \setlength{\labelsep}{0.5em} } }
\newcommand{\elist}{
  \end{list}  }

\DeclareMathSymbol{\vartheta}{\mathalpha}{letters}{"12}
\DeclareMathSymbol{\theta}{\mathalpha}{letters}{"23}
\DeclareMathSymbol{\phi}{\mathalpha}{letters}{"27}
\DeclareMathSymbol{\varphi}{\mathalpha}{letters}{"1E}

\newcommand{\bef}
{
\begin{figure}[htbp]
\centering
}

\newcommand{\eef}{\end{figure}}

\renewcommand{\figurename}{Fig.}

\counterwithin{paragraph}{section} % 1. Reset paragraph counter for each new section
\makeatletter
\renewcommand{\p@paragraph}{\thesection} % Tells LaTeX that the prefix for a \ref{paragraph} is JUST the section number
\makeatother
\begin{document}

% \begin{bibunit}
%%%%%%%%%%%%%%%%%%%%%%%%%%%%%%%%%%%%%%%%%%%%%%%%%%%%%%%%%%%%%%%%%%%%%%%%%%%
%                        Main Document
%%%%%%%%%%%%%%%%%%%%%%%%%%%%%%%%%%%%%%%%%%%%%%%%%%%%%%%%%%%%%%%%%%%%%%%%%%%
%%%%%%%%%%%%%%%%%%%%%%%%%%%%%%%%%%%%%%%%%%%%%%%%%%%%%%%%%%%%%%%%%%%%%%%%%%%

\title{Robust Quantum Sensing via Prethermal Spin Orbits}

\author{Enrico Daniel Richter}
\affiliation{Department of Chemistry, University of California, Berkeley, Berkeley, CA 94720, USA}
\affiliation{Accelerator Technology and Applied Physics Division, Lawrence Berkeley National Laboratory, Berkeley, CA 94720, USA}
\affiliation{Faculty of Physics and Earth System Sciences, Felix Bloch Institute for Solid State Physics, Applied Quantum Systems, Leipzig University, Linnéstraße 5, 04103 Leipzig, Germany}

\author{Ryan J. Smith}
\affiliation{Department of Chemistry, University of California, Berkeley, Berkeley, CA 94720, USA}
\affiliation{Chemical Sciences Division, Lawrence Berkeley National Laboratory, Berkeley, CA 94720, USA}

\author{Brayden Glockzin}
\affiliation{Department of Chemistry, University of California, Berkeley, Berkeley, CA 94720, USA}

\author{Emanuel Druga}
\affiliation{Department of Chemistry, University of California, Berkeley, Berkeley, CA 94720, USA}

\author{Thomas Schenkel}
\affiliation{Accelerator Technology and Applied Physics Division, Lawrence Berkeley National Laboratory, Berkeley, CA 94720, USA}

\author{Ashok Ajoy}
\email{ashokaj@berkeley.edu}
\affiliation{Department of Chemistry, University of California, Berkeley, Berkeley, CA 94720, USA}
\affiliation{Chemical Sciences Division, Lawrence Berkeley National Laboratory, Berkeley, CA 94720, USA}

\begin{abstract}
Practical performance of quantum sensors is often curtailed by uncontrolled environmental drift (bias-field instability, temperature fluctuations, mechanical vibration), background fields, and imperfect control pulses. This motivates developing physical mechanisms that intrinsically compensate for such perturbations while retaining high sensitivity to target fields.
We introduce an interaction-protected magnetometry scheme where periodic driving steers the collective magnetization onto two long-lived, prethermal Floquet “orbit” axes well-separated on the Bloch sphere. Rapid toggling between these axes encodes target fields as a differential signal, whereas background fields appear as common-mode motion that is strongly rejected, achieving ${>}{10^3}$-fold suppression while canceling prethermal transients. This enables accurate reconstruction of rapidly varying audio-band magnetic signals without predictive filtering or spectral tuning. We provide an experimental proof-of-principle using a dense ensemble of coupled nuclear spins, operated here as a broadband (0–\SI{1}{\kilo\hertz}) magnetometer. The protocol is remarkably tolerant to imperfections, operating robustly across millions of pulses under pulse-angle (${\sim}{10^\circ}$) and pulse frequency (${>}\SI{1}{\kilo\hertz}$) errors, large bias-field drifts (${>}\SI{50}{\micro\tesla}$), temperature variations over $\SI{150}{\kelvin}$, and harsh mechanical vibrations. 
These results establish Floquet prethermalization as a resource for robust quantum sensors that combines broadband magnetic-field sensitivity with intrinsic immunity to diverse environmental and control perturbations, opening a path toward stable quantum metrology beyond controlled laboratory conditions.

\end{abstract}

\maketitle

\begin{figure*}[t]
    \centering  \includegraphics[width=0.95\textwidth]{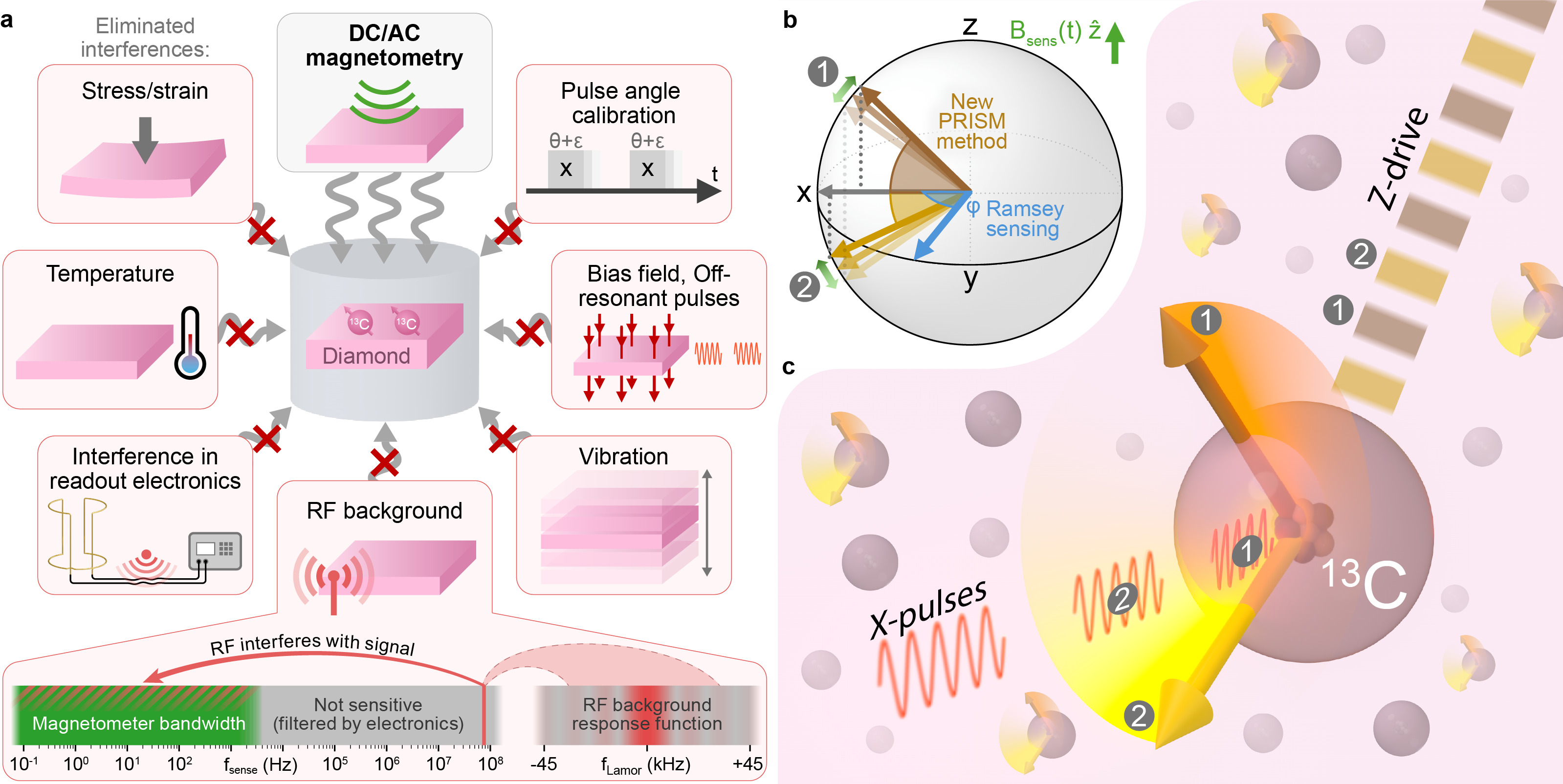}
    \caption{\T{Robust magnetometry with prethermal orbits.}
(a) \T{\I{System and objectives.}} 
Interacting ensemble of diamond $\Cs$ nuclear spins functions as a quantum sensor probed via an RF resonator. Tailored control fields render the ensemble sensitive to DC/AC magnetic fields (top) while rejecting various noise sources (crossed arrows).
\T{\I{Inset: influence of RF background}}. Resonator response shown in red (App.~\ref{secSI:RFsusceptibility}) covers ${\approx\pm}$\SI{13}{\kilo\hertz} about the Larmor frequency (here \SI{75}{\mega\hertz}). Background RF fields within this range can mimic spin responses in magnetometer bandwidth (green) (App.~\ref{secSI:samplingrate}). Other out-of-band signals (gray) are filtered out.
(b) \T{\I{PRISM Operating principle.}} Conventional Ramsey sensing (blue) accumulates a single phase $\xph$. Instead in PRISM periodic driving toggles the magnetization between two long-lived, well-separated prethermal orientations (\textcircled{1} and \textcircled{2}; gold/brown). The target field $B_{target}$ imparts common elevation on both orientations (light to dark arrows), manifesting in inverted response in the projected $\xhat$‑components (dotted lines), enabling differential readout and cancellation of the common-mode noise sources in (a).
(c) \T{\I{Schematic of PRISM prethermal orbit formation}} under simultaneous transverse ($\xhat$) and longitudinal ($\zhat$) drives of the interacting $\Cs$ nuclei. \textcircled{1} 
After first $\xhat$ pulse, a positive $\zhat$ drive is applied, aligning the magnetization along upper axis. \textcircled{2} Subsequent $\xhat$ pulse with inversion of the $\zhat$ drive aligns it along lower axis.
    }
    \label{fig:fig1}
\end{figure*}

Quantum sensors derive their excellent sensitivity from the fragility of quantum states to external perturbations~\cite{Degen17},  yet the same fragility also renders them vulnerable to uncontrolled noise. Several parasitic influences can readily degrade performance. While ideal quantum coherence can be maintained under controllable laboratory conditions, e.g. through vibration isolation~\cite{kono2024mechanically}, vacuum chambers or shielding~\cite{harrington2025synchronous}, or laser or microwave stabilization~\cite{jing2024practical,berzins2024impact}, maintaining comparable performance in noisy or mobile environments remains challenging. For practical quantum sensing therefore robustness must accompany sensitivity.

In the context of magnetometry with solid-state spins, four dominant noise channels define limits of precision (\zfr{fig1}a): (1) environmental perturbations such as temperature, strain, and vibration; (2) control imperfections, including pulse infidelity, spatial inhomogeneity of control fields, and residual offsets from bias field drifts; (3) material disorder that promotes decoherence; and (4) background fields and spurious signals that contaminate readout and are indistinguishable from genuine spin responses. 
Existing approaches address subsets of these limitations—for example, dynamical decoupling and robust pulse engineering suppress certain noise channels but often yield narrowband response; dual-transition or double-quantum schemes mitigate temperature/strain cross-sensitivity; modulation/lock-in and gradiometric readout reject specific backgrounds (App.~\ref{secSI:existingtechniques}). An ideal solution would combine broadband reconstruction with intrinsic rejection of multiple drifts and backgrounds simultaneously.

Here we introduce a new methodology for intrinsic spin stabilization that achieves this goal in a disordered network of \I{interacting} nuclear spins subject to periodic Floquet driving. The resulting effective Hamiltonian hosts long‑lived prethermal many‑body phases~\cite{maricq1987spin, d2014long,abanin2015exponentially,kuwahara2016floquet, weidinger2017floquet}; these metastable states are steered into multi‑axial Floquet orbits whose relative positions encode desired magnetic fields. Crucially, the evolution of these orbits exhibits simultaneous immunity to pulse errors, vibration, and background field fluctuations (\zfr{fig1}a), while yielding a transient-free response to rapidly changing fields over a wide bandwidth. Collectively, this establishes a general framework for interaction‑protected quantum metrology, which we term \emph{Prethermal Robust Internally Modulated Spin Magnetometry} (PRISM).

\vspace{0.5em}
\noi{\normalsize\bfseries Sensing Principle\par}
\noi 
As a concrete experimental example, we utilize an ensemble of $\Cs$ nuclear spins randomly distributed in a single-crystal diamond at natural isotopic abundance (Methods), serving as a prototypical solid-state nuclear-spin ensemble. The nuclei form a dense (${\sim}\SI[parse-numbers=false]{10^{21}}{\per\cubic\centi\meter}$) coupled network governed by long-range magnetic dipolar interactions~\cite{duer2008solid},
$
H_{\mathrm{nn}}=\sum_{i<j} d_{ij}\big(3I_i^z I_j^z-\mathbf{I}_i\cdot\mathbf{I}_j\big), 
$ 
where $I$ are spin-$\fr{1}{2}$ Pauli matrices, and $d_{ij}{\propto\:} r_{ij}^3$ depends on interspin separation. Macroscopically large nuclear polarization is generated optically via lattice Nitrogen Vacancy (NV) centers at room temperature~\cite{ajoy2019hyperpolarized, doherty2013nitrogen} (Methods). The collective $\Cs$ magnetization is detected inductively through a radio-frequency (RF) resonator tuned to the nuclear Larmor frequency (${\nu_L}{\approx}\SI{75}{\mega\hertz}$ at bias field $B_0=\SI{7}{\tesla}$)~\cite{ajoy2019wide,fukushima2018experimental}.

In conventional Ramsey sensing~\cite{ramsey1990experiments,taylor2008high} (\zfr{fig1}b, blue), a phase $\xph$ accumulates under a target field, but sensitivity is limited by finite dephasing times $T_2^*$ ($\app$\SI{1.3}{\milli\second} here), dominated by bias-field fluctuations and interspin interactions~\cite{Beatrez21_90s}. Dynamical-decoupling approaches~\cite{schmitt2017submillihertz, glenn2018high, barry2024sensitive, devience2015nanoscale} can suppress such bias drifts but are susceptible to pulse errors and yield narrow resonant sensing bands.

Our approach instead exploits Floquet prethermalization~\cite{ho2023quantum, Beatrez21_90s,peng2021floquet} as a key design principle for robust sensing (Methods). By interleaving rapid ($\SI{\approx100}{\micro\second}$) $\xhat$- and sinusoidal $\zhat$-axis drives (\zfr{fig1}c), the nuclei alternate between two long-lived prethermal orientations—golden and brown arrows, labeled \textcircled{1} and \textcircled{2} in \zfr{fig1}b—that sample points along a circular Bloch-sphere trajectory (shaded traces, \zfr{fig1}c). These ``orbital axes'' can differ by as much as $\SI{\approx100}{\degree}$ (App.~\ref{SIMethods:pulseangle}), and exhibit a significantly extended lifetime $T_2'\approx \SI{98}{\second}$ (App.~\ref{secSI:lifetime}), exceeding $T_2^{\ast}$ by almost $10^5$-fold.

Upon exposure to an AC magnetic field, the two axes acquire common elevation in the $\xhat\tm\zhat$ plane (light‑to‑dark arrows, \zfr{fig1}b). Although both axes undergo identical polar motion (green double arrows), their positional offset causes a difference in their $\xhat$‑components when projected into the $\xhat\tm\yhat$‑plane, and this encodes the target field (App.~\ref{secSI:theorytrajectories}). Effectively a \I{time‑domain gradiometer} is realized: rapid alternation between the prethermal axes results in background signals (such as readout-chain interferences), transient signals and pulse angle errors manifesting as common‑mode signals that can be suppressed. 
Measurement of the orbit geometry itself enables additional internal calibration and cancellation, of other interferences in \zfr{fig1}a such as pulse offset or bias drift.

\begin{figure*}[t]
   \centering  \includegraphics[width=0.99\textwidth]{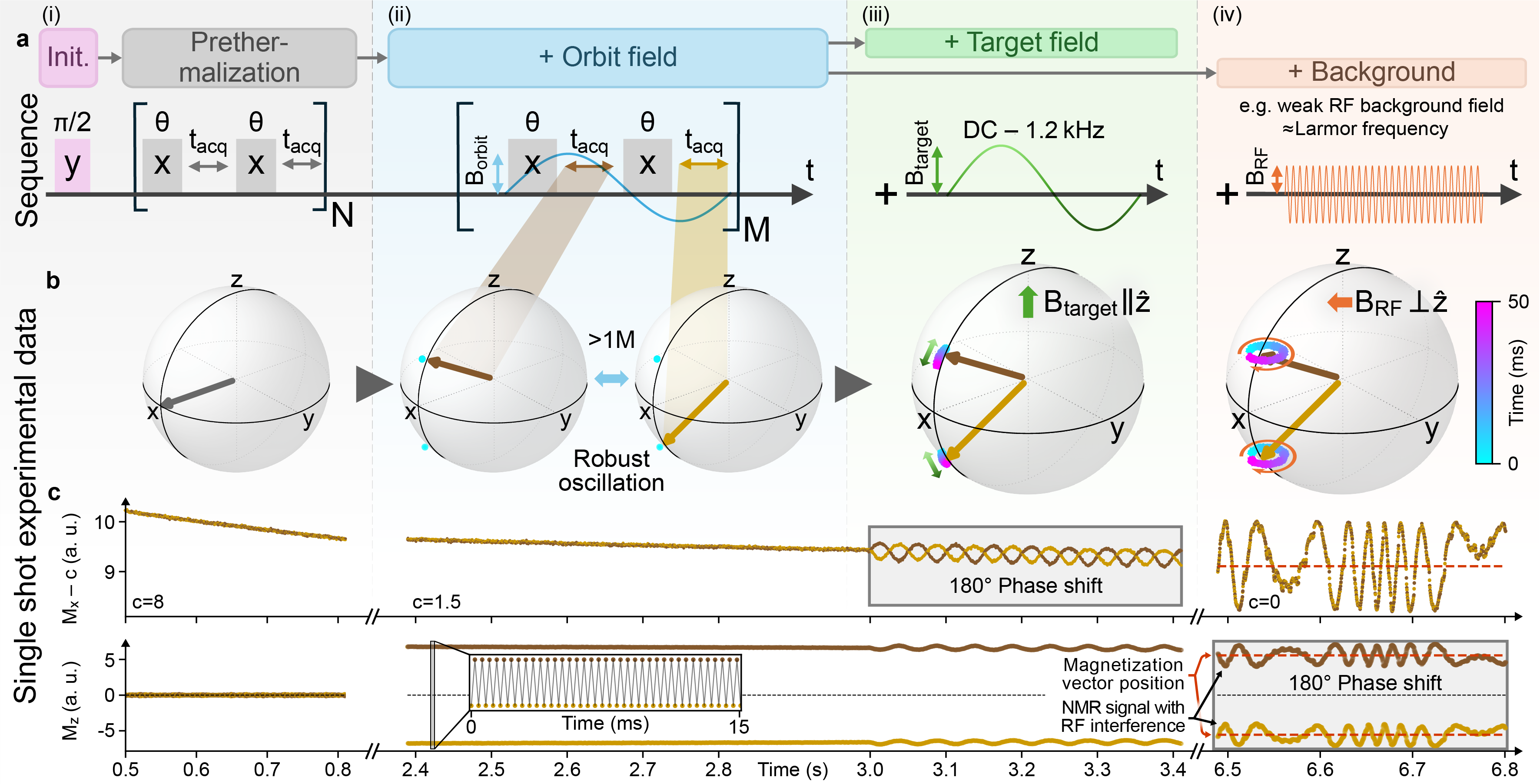}
    \caption{\T{Operating principle of PRISM sensing and background suppression.} Sequential experiments demonstrate key operating principles.
(a) \T{\I{Top panel:}} Pulse sequences applied in stepwise blocks.  (b) \T{\I{Middle panel:}} Measured spin trajectories on the Bloch sphere (App.~\ref{secSI:3Dextraction}), color-coded by time. 
(c) \T{\I{Lower panel:}} Ensemble $\Cs$ magnetization components $M_x$ and $M_z$ (bottom), with alternate orbit axes in gold and brown.
(i) \T{\I{Initialization and prethermalization}} (pink and gray blocks). After an initial $(\pi/2)_y$ pulse, $\xt_x$ pulses (here ${\approx}\SI{166}{\degree}$) generate a long-lived prethermal plateau in $M_x$ ($M_z = 0$). Read out occurs during each acquisition window $\tacq$ (here \SI{100}{\micro\second}).
(ii) \T{\I{Orbit excitation}} (blue block). At $t > \SI{2.0}{\second}$, a resonant $\zhat$-axis drive (a, light blue) matching the pulse period establishes two distinct, prethermal orbits (brown/gold). (b) Tracked motion shows that both share identical $\xhat$-projections, yielding a single $M_x$ plateau (c), while $M_z$ bifurcates into two stable plateaus that alternate rapidly (see inset zoomed in over \SI{15}{\milli\second}).
(iii) \T{\I{AC-field response}} (green block). At $t {=} \SI{3.0}{\second}$, a target low-frequency AC magnetic field to be sensed is applied (a). The two orbits trace synchronous arcs across opposite hemispheres (b), with $M_x$ oscillations carrying an imprint of the field and being exactly $\SI{180}{\degree}$ out of phase, encoding the differential field signal.
(iv) \T{\I{Background response}} (orange block). At $t > \SI{6.5}{\second}$, exposure to a sinusoidal RF background close to the Larmor frequency (a), approximately five times stronger than the sensing signal, induces an apparently complex elliptical motion but displays identical evolution in both hemispheres (b). Resulting $M_x$ responses remain in phase, even for broadband RF backgrounds (c), identifying the background as a common-mode signal that can be suppressed by differential readout (\zfr{fig3}).
}
    \label{fig:fig2}
\end{figure*}

\begin{figure}[t]
    \centering
    \includegraphics[width=0.5\textwidth]{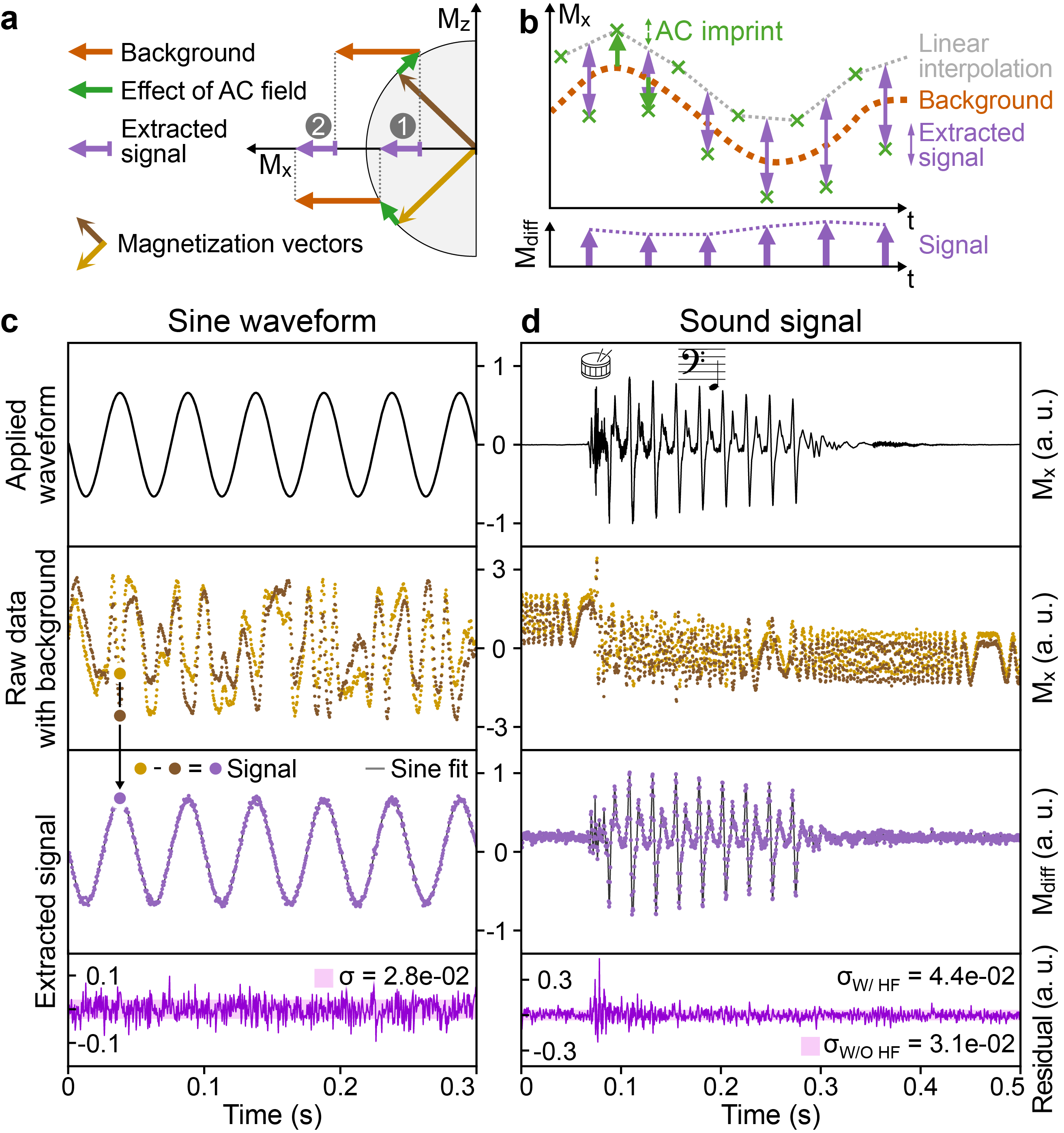}
    \caption{\T{AC magnetometry with concurrent background suppression.} 
(a) \T{\I{Principle.}} Brown and gold arrows denote the two prethermal orbit axes; green arrows indicate spin motion under an applied magnetic field. It drives trajectories in the same direction across both hemispheres (cf. Fig. \ref{fig:fig1}b), producing distinct $\xhat$-axis projections (dotted lines), whereas background fields (orange) generate identical out-of-plane motion in both. Differential correction (see b) thus cancels the common-mode background.
(b) \T{\I{Signal extraction.}} Green crosses mark alternating readout points carrying the signal imprint. Linear interpolation (gray) aligns sampling intervals, enabling subtraction between successive points to remove common-mode components (orange), yielding signal in lower panel (App.~\ref{secSI:signalextraction}).
(c-d)  \T{\I{Demonstration for representative signals}} with ${\sim}2\times$ larger background. (c) \T{\I{Response to a sinusoidal field}} ($f = \SI{20}{\hertz}$, $B_{\R{target}} = \SI{16.5}{\micro\tesla}$). Top: applied field; second: raw magnetometer traces under additional broadband RF background, with alternating points shown in brown/gold; third: reconstruction following (b) recovering the AC waveform; bottom panel: residuals (standard deviation) are $< 3$\% over the \SI{0.3}{\second} window. (d) \T{\I{Analogous measurement for a sound-wave}} signal ($B_{\R{target}}=\SI{33}{\micro\tesla}$) under chirped background, showing near-perfect recovery. Residuals (bottom panel) remain low, except where drum spectral components exceed the sensor bandwidth. For "W/O HF" drum was excluded from standard deviation.
    }
    \label{fig:fig3}
\end{figure}

\begin{figure*}[t]
    \centering
    \includegraphics[width=0.99\textwidth]{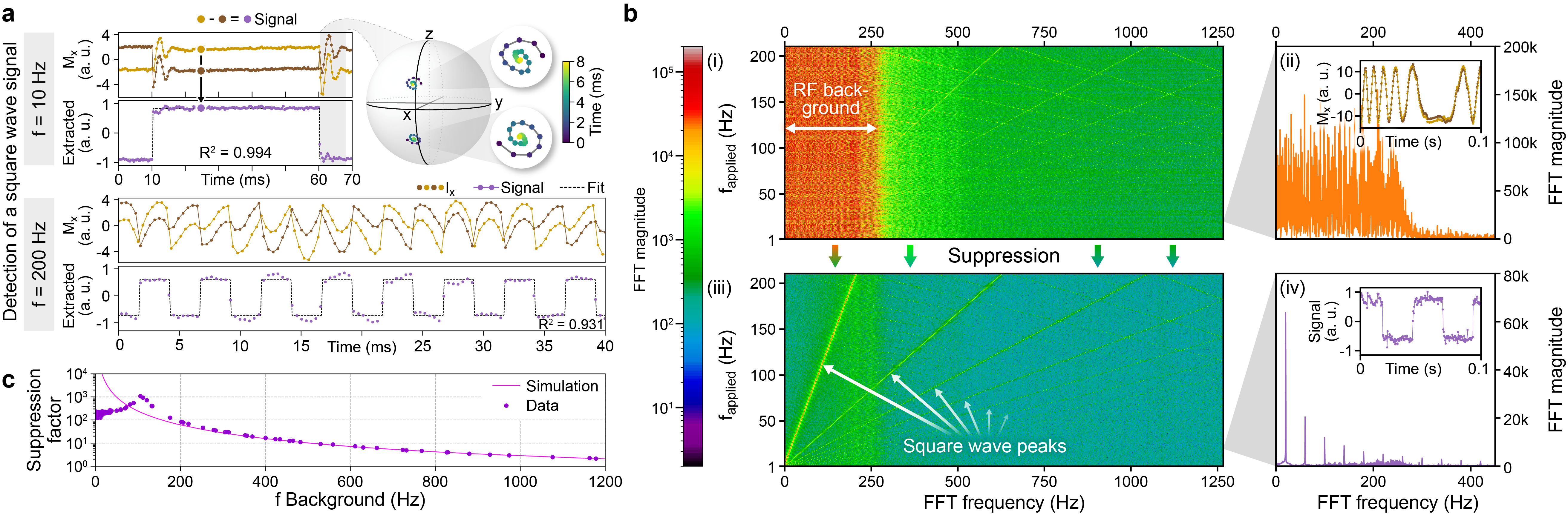}
    \caption{\T{Transient-free, background-suppressed magnetometry.}
(a) \T{\I{Transient suppression}}. \I{Upper panel:} measured $M_x$ magnetization under a step-change AC field (square wave, \SI{10}{\hertz}). Gold and brown points mark the two prethermal orbit axes. \I{Insets:} Transient response manifests as spiral trajectories that evolve with same handedness in both hemispheres. Protocol of Fig. \ref{fig:fig3}b yields cancellation resulting in a sharp, transient-free signal (second panel). \I{Lower panels:} identical measurements at \SI{200}{\hertz}; subtraction recovers the square-wave response with minimal residuals
(b) \T{\I{Broadband background suppression}}. Spectral map (Fourier transforms) of spin response to square-wave AC signal (frequency $f_{\R{applied}}=1–\SI{210}{\hertz}$, step size \SI{1}{\hertz}) in presence of a strong broadband background (0-\SI{250}{\kilo\hertz}). (i) Single-orbit data reveals contamination, dominated by low-frequency interference. (ii) Spectral response map at $f_{\R{applied}}=\SI{20}{\hertz}$ and corresponding time-domain signal (ii inset). (iii) Differential operation strongly suppresses this background, isolating AC signal harmonics up to 11th order at $f_{\R{applied}}=\SI{20}{\hertz}$ (iv) and cleanly reconstructing the time-domain waveform (iv inset). Noise is reduced even beyond the background band (iii, lower right region).
(c) \T{\I{Suppression factor}} measured versus frequency $f$, exceeding $10^3$ (\SI{-60}{dB}) and saturating at low $f$. Solid line: simulated suppression factor limited by interpolation, shows good agreement with data at high $f$ (App.~\ref{secSI:increasesamplingrate}).
    }
\label{fig:fig4}
\end{figure*}

\vspace{0.5em}
\noi{\normalsize\bfseries Prethermal Orbit Quantum Sensing (PRISM)\par}
\noi \zfr{fig2} shows the PRISM protocol in a step-wise fashion over $\SI{\sim 7}{\second}$. Panels in \zfr{fig2}a-c depict the pulse sequence, the resulting spin trajectories on the Bloch sphere, and ensemble $\Cs$ magnetization components along $\xhat,\zhat$ ($M_x$, $M_z$) respectively. Here $M_{x}{=}\R{Tr}\{\xr^{\dg} I_{x}\}$, $\xr$ being the state density matrix. 
Quasi-continuous readout~\cite{Beatrez21_90s, moon2025high} captures the spin evolution in real-time without reinitialization, enabling time-resolved reconstruction of the full three-dimensional magnetization vector $\T{M}$ during the protocol. The algorithm for 3D spin-tracking, detailed in App.~\ref{secSI:3Dextraction}, represents a substantial refinement over previous implementations~\cite{sahin22_trajectory}. 

The experimental procedure proceeds as follows (\zfr{fig2}a). An initial $\pi/2$-pulse tips the spins along $\xhat$, followed by $N$ cycles of spin-locking pulses~\cite{ostroff1966multiple, rhim1976multiple} (rotation angle $\xt$) with inductive detection during $\tacq$ windows between them.
Under this drive, the internuclear Hamiltonian is transformed to $H_{\mathrm{eff}}=-\frac{1}{2}\sum_{i<j} d_{ij}\big(3I_i^x I_j^x-\mathbf{I}_i\cdot\mathbf{I}_j\big)$ at zeroth-order in the Magnus expansion~\cite{magnus1954exponential,haeberlen2012high}, yielding rapid stabilization ($\app$\SI{10}{\milli\second}) to a Floquet prethermal plateau~\cite{Beatrez21_90s}, decay from which follows a universal form $\propto e^{-\sq{R_pt}}e^{-R_dt}$~\cite{selco2025emergent}. The resulting $1/e$ transverse lifetime is long, $T_2′ \approx \SI{56}{\second}$, and is aided by disordered arrangement of the spins in the lattice~\cite{selco2025emergent}. 

Superimposing a $\zhat$ sinusoidal drive $B_{\R{orbit}}$ (light blue, \zfr{fig2}a), matched in period to two pulses (App.~\ref{secSI:timingzdrive}), induces orbital motion that alternates the polarization between two hemispheric states (gold and brown arrows, \zfr{fig2}b). This yields rapidly switching prethermal plateaus (\zfr{fig2}c), with the stable toggling every \SI{200}{\micro\second} evident in the lower inset. These stable oscillations bear resemblance to discrete-time-crystal (DTC) behavior~\cite{sacha2017time,khemani2019brief, ho2017critical}. In App.~\ref{SIMethods:pulseangle}, we map the alternating $M_z$ dynamics versus $\xt$, showing a pronounced stability dome at $\xt\app\pi$, akin to a DTC phase diagram~\cite{DTC_Beatrez2022}. Unlike genuine DTCs, no period-doubling or symmetry breaking emerges, and a single sharp instability appears at $\xt=\pi$.

Moreover, while the  intrinsic $T_2'$ lifetime is already long, remarkably, it is further extended ($T_2′ \approx \SI{98}{\second}$) by applying the orbit field (App.~\ref{secSI:lifetime}). The extension arises from the finite energy density imposed by the orbit drive~\cite{moon2024discrete}.

The prethermal orbit axes function as higher-dimensional analogues of the initial Ramsey state in \zfr{fig1}b. When subjected to a weak target field ${B_\mathrm{target}(t) \pll \zhat}$, both the upper-hemisphere and lower-hemisphere trajectories shift in a common direction. The resulting spin motions, tracked in \zfr{fig2}b(iii), trace stable arcs on each hemisphere, directly encoding the sensed AC field following the schematic of \zfr{fig1}b and as illustrated in \zfr{fig3}a (green arrows). The corresponding magnetizations $M_x$ and $M_z$ exhibit oscillations carrying an imprint of the field, with $M_x$ displaying a \SI{180}{\degree} phase inversion between the two axes (\zfr{fig2}c, gray highlighted region).

We contrast this with the sensor response to a parasitic RF background (lower panel, \zfr{fig1}a). “Background” here denotes any interference in readout electronics or any response within the resonator detection bandwidth that does not originate from the $\Cs$ nuclei yet are spectrally indistinguishable from them. The \I{apparent} spin trajectories in this case (\zfr{fig2}b) deform into paths that oscillate along the $\xhat$ and $\yhat$ directions at the Bloch sphere  (\zfr{fig3}a, orange arrows)—distinct from the orbital arcs produced by true DC and AC signals.
Correspondingly, the relative phase between $M_x$ and $M_z$ inverts (\zfr{fig2}c), with $M_x$ components now in phase across the two orbits. Taking the differential $M_x$ response between the two orbit manifolds therefore cancels these common-mode background contributions while retaining the true response.

\vspace{0.5em}
\noi{\normalsize\bfseries Background-Suppressed Broadband Sensing\par}
\noi 
RF background suppression forms a convenient and experimentally tunable means to evaluate the performance of PRISM, and serves as test case for the broader noise correction in \zfr{fig1}a. 
Most parasitic background fields are outside the resonator bandwidth and thus rejected (lower panel of \zfr{fig1}a, gray region; see Methods)~\cite{moon2025high,fukushima2018experimental}; however, spurious components within $\pm 13$ kHz of the Larmor frequency can pass through, generating responses that mimic genuine spin signals within the magnetometer bandwidth (green region). The red-shaded trace in \zfr{fig1}a depicts the measured resonator transfer function, quantifying this residual background response (App.~\ref{secSI:RFsusceptibility}).

\zfr{fig3} examines magnetometry when the spins experience desired AC and undesired background fields simultaneously.
To isolate the AC signal even when the background is substantially stronger, we analyze the $M_x$ component, noting that the two orbits are sampled alternately in time. Interpolation of the quasi-continuous readout allows fairly accurate estimation of instantaneous positions and accurate two-point subtraction (see App.~\ref{secSI:signalextraction} for detailed algorithm). \zfr{fig3}b depicts this: 
The interpolated traces are subtracted, which recovers the AC-induced signal, as shown in the lower panel in \zfr{fig3}b.

\zfr{fig3}c demonstrates this concept experimentally. A sinusoidal $\Bsens$ field is applied atop a background twice as large (extendable to $>$100-fold larger; see \zfr{fig4}). The raw $M_x$ signal (second panel) appears strongly distorted. Applying the differential subtraction procedure of \zfr{fig3}b yields a cleanly reconstructed AC response (purple trace) corresponding to $\Bsens$. Residuals relative to a pure sinusoidal fit remain below 3\% over the full acquisition window, demonstrating high-fidelity background rejection, with the remainder set mostly by Johnson noise.

The method generalizes to non-stationary or rapidly varying signals~\cite{herb2025quantum}. \zfr{fig3}d presents an analogous experiment to \zfr{fig3}c, in which a drum hit marks the onset of the open $E_1$ ($\SI{{\approx}41}{\hertz}$) note of a bass guitar, used as the $\Bsens$ input. The top panel of \zfr{fig3}d displays the target waveform; the middle shows the signal measured with a superimposed broadband “swish” background. 
The reconstructed $M_{\text{diff}}(2t) := \left( \frac{M_x(t-1) + M_x(t+1)}{2} - M_x(t) \right)/2$ trace (third panel) faithfully reproduces the input, with the residuals (bottom panel) remaining low. The brief deviations at \SI{\approx 0.08}{\second} are caused by drum spectral components exceeding the sensor bandwidth (Methods).
App.~\ref{secSI:soundsignal} shows a more complex example of a musical riff, several seconds long, that is accurately recovered even under strong broadband interference.

Operation in \zfr{fig3}d and App.~\ref{secSI:soundsignal} implicitly relies on the sensor’s flat response over a wide bandwidth, experimentally confirmed in App.~\ref{secSI:sensorresponse}, where the sensor response is found to be almost frequency-independent from DC to \SI{1}{\kilo\hertz} (extendable with shorter pulses). This broadband behavior contrasts with dynamical-decoupling schemes requiring resonance tuning~\cite{taylor2008high} and is not susceptible to spurious harmonic artifacts like those methods~\cite{loretz2015spurious}.

\begin{figure*}[t]
    \centering
    \includegraphics[width=0.99\textwidth]{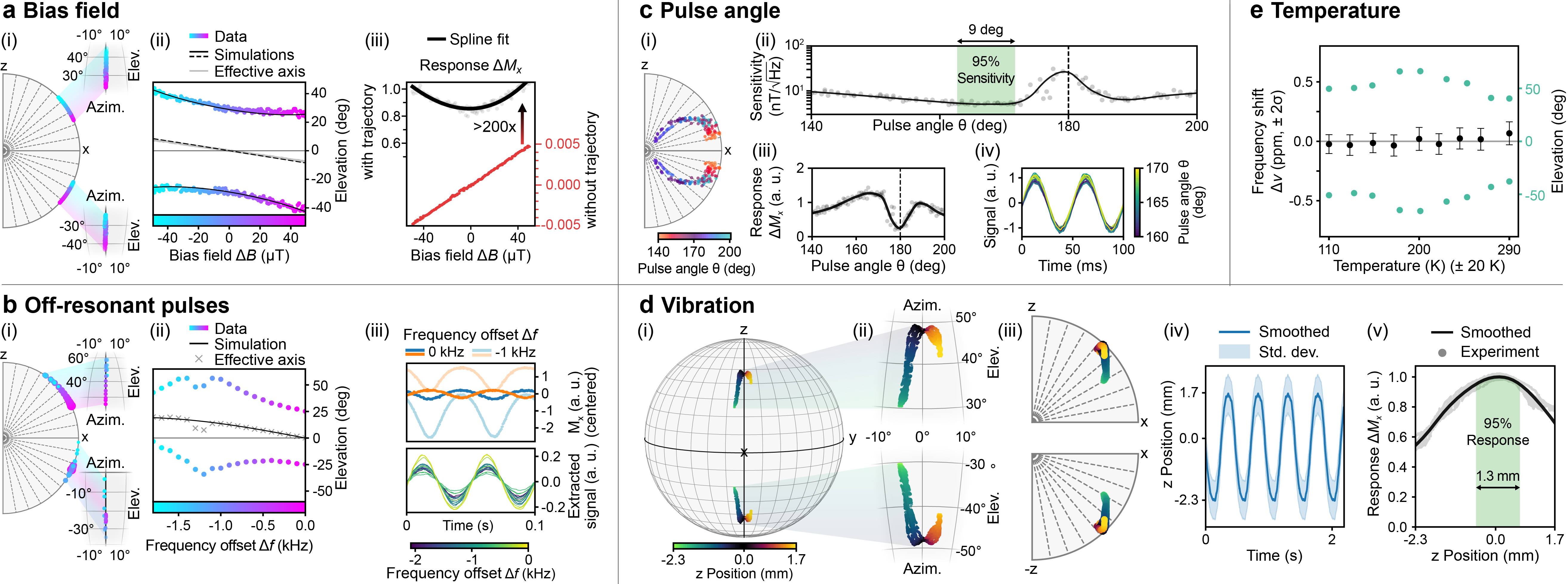}
    \caption{\T{Comprehensive robustness of PRISM against various environmental and control errors.} Test signals are \SI{20}{\hertz}, \SI{1.8}{\micro\tesla}.
    (a) \T{\I{Bias-field drift.}} (i) Tracked Bloch-sphere trajectories of both orbit axes under DC bias shift $\xD B$ (colorbar). (ii) Extracted elevation angles for pulsed spin-locking (gray) and for orbit axes (colored). Simulations (black lines, App.~\ref{secSI:SimulationRotationMatricesBRAYDEN}) match experiment. (iii) Magnetometer response $\xD M_x$ to test signal. Parabolic differential response (black spine guide) yields a ${>}200$-fold enhanced AC response over response of spin-locking, which varies linearly with $|\xD B|$.
    (b) \T{\I{Off-resonant pulses.}} (i-ii) Tracked trajectories and elevation angles under RF-frequency detuning reveal robust generation of the two prethermal axes across the entire range, with only a slight elevation drift, correctable a posteriori analogous to bias field corrections.
    Narrow dips occur near \SI{1.2}{\kilo\hertz}.
    Crosses: effective axis. Solid line: simulation of conventional spin lock axis. (iii) \I{Upper panel:} Distortion of representative magnetometry traces of a test signal (blue/orange) for on-resonant (dark) and \SI{1}{\kilo\hertz}-detuned (light) pulses (zero-centered). \I{Lower panel:} Subtraction recovers a near-ideal sinusoid, largely insensitive to detuning (colorbar).
    (c) \T{\I{Pulse-angle errors.}} (i) Trajectories under $\xt$ variations (colorbar) for sequence in Fig.~\ref{fig:fig2}a remain symmetric over a wide $\xt$ range. (ii) AC sensitivity versus $\xt$ under a test field is flat over 163$^{\circ}$–172$^{\circ}$, and becomes worse near 180$^{\circ}$ where Floquet stabilization breaks down. (iii) Corresponding $\xD M_x$ response exhibits same plateau. (iv) Representative AC-field responses across 160$^{\circ}$–170$^{\circ}$, showing nearly identical waveforms.
    (d) \T{\I{Vibration tolerance}} (i–iii) 3D tracked trajectories under ${\pm}$2 mm mechanical motion of the diamond within the RF resonator at \SI{2.1}{\hertz}. (iv) Applied vibration profile. (v) Extracted $\xD M_x$ remains flat (${<}$5\% variation) over a \SI{1.3}{\milli\meter} displacement interval demonstrating vibration immunity.
    (e) \T{\I{Temperature robustness.}} Elevation angles recorded from \SI{110}{\kelvin} to RT show only weak temperature dependence, which can be corrected similar to (b).
    }
\label{fig:fig5}
\end{figure*}

\vspace{0.5em}
\noi{\normalsize\bfseries Transient-Free Sensing of Rapidly Varying Fields\par}
\noi The framework in \zfr{fig1}b extends to the elimination of dynamical transients. These transients originate because, under nonperiodic driving, each change in the applied field $\Bsens$ induces a perturbation $\xd H(t)$ in $H_{\R{eff}}$. For rapidly varying, nonrepetitive fields, the system must continuously re-establish a new prethermal equilibrium, producing a transient as it re-enters a quasi-conserved subspace~\cite{Beatrez21_90s,sahin22_trajectory}. They are typically millisecond-scale responses (${\sim} 5T_2^{\ast}$) and become prominent for large step-changes in $\Bsens$, and at first glance, appear to be a strong limitation to prethermal sensing~\cite{Sahin22,sahin22_trajectory}.  \zfr{fig4}a illustrates these transients for a square-wave magnetic modulation with amplitude step $\Delta B{\approx}{\SI{33}{\micro\tesla}}$ every \SI{50}{\milli\second}. The two $M_x$ orbit axes (brown and gold) display oscillatory distortions as the prethermal plateaus re-establish. 

Interestingly, however, when the spin motion corresponding to these transients are tracked on the Bloch sphere (insets \zfr{fig4}a), we observe that the two manifolds trace spiral trajectories in opposite hemispheres, converging toward distinct prethermal axes, but evolve with identical handedness. The time-encoded color scale highlights this symmetry: $\hat{\mathbf{y}}$ and $\hat{\mathbf{z}}$ components of the two transient manifolds map onto one another through a ${\approx}180^\circ$ rotation about the pulse axis, ensuring their $M_x$ projections remain in phase, while preserving the imprint of $B_{\mathrm{target}}$ (App.~\ref{secSI:transients}). Numerical simulations in App.~\ref{secSI:transients} reproduce the observations in good agreement. 

The second panel of \zfr{fig4}a demonstrates the transient cancellation experimentally, following the protocol of \zfr{fig3}b. The recovered signal (purple) closely reproduces the ideal square-wave response with sharply defined steps. With transients suppressed, the effective sensing bandwidth becomes limited only by the Floquet drive period, not by the intrinsic prethermalization rate (App.~\ref{secSI:sensorresponse}). The lower panels of \zfr{fig4}a show results for a faster modulated $\Bsens$ field ($f = \SI{200}{\hertz}$). In the raw data, transient oscillations dominate the response; after cancellation, these distortions vanish, yielding an accurate reconstruction of the square-wave waveform (purple). The small, time-dependent residuals arise from interpolation errors in estimating instantaneous spin phases, and could, in principle, be further minimized through higher-order prediction schemes.

\vspace{0.5em}
\noi{\normalsize\bfseries Characterizing Suppression Factors \par}
\noi 
To quantify suppression, we use background-field rejection as a test case, since these fields can be applied in a well-controlled fashion. In \zfr{fig4}b, a square-wave magnetic signal $\Bsens$ with variable frequency ($f_{\R{applied}}$) is applied alongside RF sweeping from 0 to \SI{250}{\hertz} offset from the Larmor frequency to emulate broadband background.
The background amplitude exceeds the signal by over an order of magnitude. For each frequency slice ($f_{\R{applied}}$), the full $M_x$ response is recorded. The resulting Fourier transforms, logarithmically plotted (colorbar, \zfr{fig4}b), form a spectral response map of the sensor with and without differential correction.

\zfr{fig4}b(i) shows the raw data from one orbit axis, where the background (orange region) overwhelms the response. \zfr{fig4}b(ii) displays representative frequency- and time-domain slices, highlighting the two orbit manifolds (brown and gold), both dominated by the applied background field.

After subtracting the two axes (\zfr{fig3}b), data in \zfr{fig4}b(iii) shows that the background is nearly eliminated, with the orange band effectively extinguished. The fundamental harmonic at $\fsens$ emerges as a bright feature, while the diagonal ridges correspond to higher-order harmonics of the square-wave field, confirming faithful, transient-free recovery of the signal. \zfr{fig4}b(iv) displays this reconstructed signal, showing clear harmonic content up to the 11th order.

Further comparison of the panels in \zfr{fig4}b reveals the method also mitigates broadband noise contributions, such as from the amplifier. For instance, in the lower right of \zfr{fig4}b(iii), the noise power is reduced by about an order of magnitude---well beyond the $\sq{2}$ improvement expected from simple dual-manifold averaging. 

Finally, \zfr{fig4}c quantifies the suppression factor as a function of frequency on a logarithmic scale. The maximum suppression exceeds a factor of $10^3$ (\SI{-60}{dB}, App.~\ref{secSI:evalsuppfactor}) and plateaus below $\sim$\SI{100}{\hertz}. The solid line shows a theoretical scaling (App.~\ref{secSI:sim_suppressionfactor}) of the suppression, demonstrating good agreement with the data. Performance at higher frequencies is limited by the finite sampling interval between successive orbit axes, and can be reduced by shortening the pulses or using higher-order prediction schemes beyond simple linear interpolation. 

Collectively, \zfr{fig4} shows transient-free operation and broadband noise suppression without predictive filtering.

\vspace{0.5em}
\noi{\normalsize\bfseries Robustness to Bias Field Drift \par}
\noi Variations due to unstable magnetic field environments~\cite{halde2025let} or motion through inhomogeneous fields~\cite{lang2019nonvanishing}
pose important challenges encountered in unshielded field-deployed sensing; however, PRISM enhances robustness against these issues. \zfr{fig5}a(i) shows tracked Bloch-sphere trajectories under varying bias fields, and \zfr{fig5}a(ii) presents the corresponding elevation angles.

Consider first an on-resonance pulse train without orbit field, i.e. spins aligned along $\xhat$. Small bias field changes perturb the spins' elevation angle $\phi=\cos^{-1}(M_x/|\T{M}|)$ linearly (\zfr{fig5}a(ii), gray line) due to the field's impact during the pulse periods. For DC sensing, or AC sensing in the presence of bias drifts, this yields errors because only $M_x$ is directly measured, and a precise estimation is subject to the assumed value of $|\T{M}|$. The latter is confounded by variables affecting signal amplitude, such as hyperpolarization variation; slow variations of the field on timescales $\gtrsim T_2$ are similarly obscured by any variations in the relaxation dynamics of the sensor, for example due to drift in pulse sequence parameters.

Application of a trajectory field lifts these degeneracies, such that static or quasi-static fields can be estimated accurately. As shown in the insets of Fig~$\ref{fig:fig5}$a(i), bias field variation causes both manifolds of the trajectory to simultaneously lift; and any tilt of the central axis of the trajectories manifests as an imbalance of the two manifolds. This relative measure enables extraction of the static field (demonstrated in Fig~\ref{fig:fig5}a(ii)) in a manner immune to the absolute signal amplitude. Consequently sensing can be carried out at the true DC limit, concomitant with detection of AC fields.

Indeed, quasi-continuous estimation of the bias field within the measurement window serves as internal calibration of the bias-dependent response shown in Fig~$\ref{fig:fig5}$a(iii), enforcing faithful reconstruction of AC field amplitudes. Importantly, cancellation of the bias dependence of the two manifolds yields a first-order-insensitive AC response $\xD M_x$ (\zfr{fig5}a(iii), black line; App.~\ref{secSI:RobustnessDCOffRes}).
Notably, the AC response is amplified by over $200$ times compared to simple spin-locking (cf. Fig~$\ref{fig:fig5}$a(iii)). 

\vspace{0.5em}
\noi{\normalsize\bfseries Robustness to Pulse Offset Errors \par}
\noi \zfr{fig5}b addresses persistent detuning errors in the pulse resonance frequency, mimicking static bias offsets. We tested detuning up to \SI{2}{\kilo\hertz}, which exceeds that studied in \zfr{fig5}a and is larger than the $\Cs$ nuclear linewidth (${\approx}\SI{0.25}{\kilo\hertz}$)~\cite{Beatrez21_90s}.
In high detuning regimes, the spin dynamics become complex, with first-order effects on the pulse axis and second-order perturbations affecting manifold separation. When the offset approaches the effective Rabi field strength (here ${\approx}\SI{400}{\micro\tesla}$), these higher-order effects produce sharp, nonperturbative features in the elevation angles near ${\app}\SI{1.2}{\kilo\hertz}$ as the inter-vector angle between the two prethermal axes approaches its maximum for the given pulse length (App.~\ref{secSI:RobustnessDCOffRes}).
Nonetheless, applying an off-resonant orbit field (as described in App.~\ref{secSI:3Dextraction}) allows full constraints of individual elevation angles, enabling calibration despite stronger detunings without relying on fixed absolute signal amplitude $|\T{M}|$.

The top panel of \zfr{fig5}b(iii) shows representative AC magnetometry traces under a sinusoidal test field. On-resonance pulses (dark points) produce the expected pair of inverted sinusoids. At large detuning, the individual responses, determined by the gradient of the elevation angle curves in \zfr{fig5}b(ii), become distorted; at 1~kHz detuning (\zfr{fig5}b(iii), light points) each manifold is nonsinusoidal and the inversion is lost, yet these distortions cancel in the subtracted signal (bottom), suppressing the impact of detuning error across the full 0–\SI{2}{\kilo\hertz} range.

\vspace{0.5em}
\noi{\normalsize\bfseries Robustness to Pulse-Angle Errors\par}
\noi \zfr{fig5}c examines robustness to pulse-angle errors, a dominant instability in ensemble magnetometers, where spatial RF inhomogeneity, drift, and calibration inaccuracies are typically challenging to contend with~\cite{zhang2016microwave,bayat2014efficient}. We probe errors over a fairly large flip-angle range ($\xt{=}$140$^{\circ}$–200$^{\circ}$) in the sequence of \zfr{fig2}a. 

We find the orbital dynamics remain largely insensitive to any particular flip-angle condition, as long as the $\zhat$-drive frequency is resonant with two $\xhat$-pulses, which is a condition that can be easily ensured in practice. Indeed, with the orbit field applied, \zfr{fig5}c(i) shows that the dynamics in the $\xhat\tm\zhat$-plane remain symmetric across the two hemispheres over the full range of $\xt$ (colorbar). Simultaneous changes in the elevation angle and norm have a compensating effect, resulting in a very broad plateau in the magnetometer sensitivity shown in \zfr{fig5}c(ii), that remains remarkably flat, varying by less than 5\% across a $\xt{=}\SI{163}{\degree}–\SI{172}{\degree}$ window (shaded green).
We observe that sensitivity (\zfr{fig5}c(ii)) and AC response (\zfr{fig5}c(iii)) collapse at $\xt{=}\SI{180}{\degree}$, due to significant delocalization of the two axes (App. \ref{SIMethods:pulseangle}). Slightly shorter rotations, however, yield excellent stability: \zfr{fig5}c(iv) shows nearly identical AC responses across this wide $\xt$ band. The robustness displayed here exceeds that of dynamical-decoupling schemes~\cite{souza2011robust} when considering the number of pulses (${>}10^5$) employed.

\begin{table}[t]
    \centering
    \caption{\T{Summary of robust performance} demonstrated here. See App.~\ref{secSI:performancemetrics} for a more detailed description and projected improvement.}
    \includegraphics[width=0.5\textwidth]{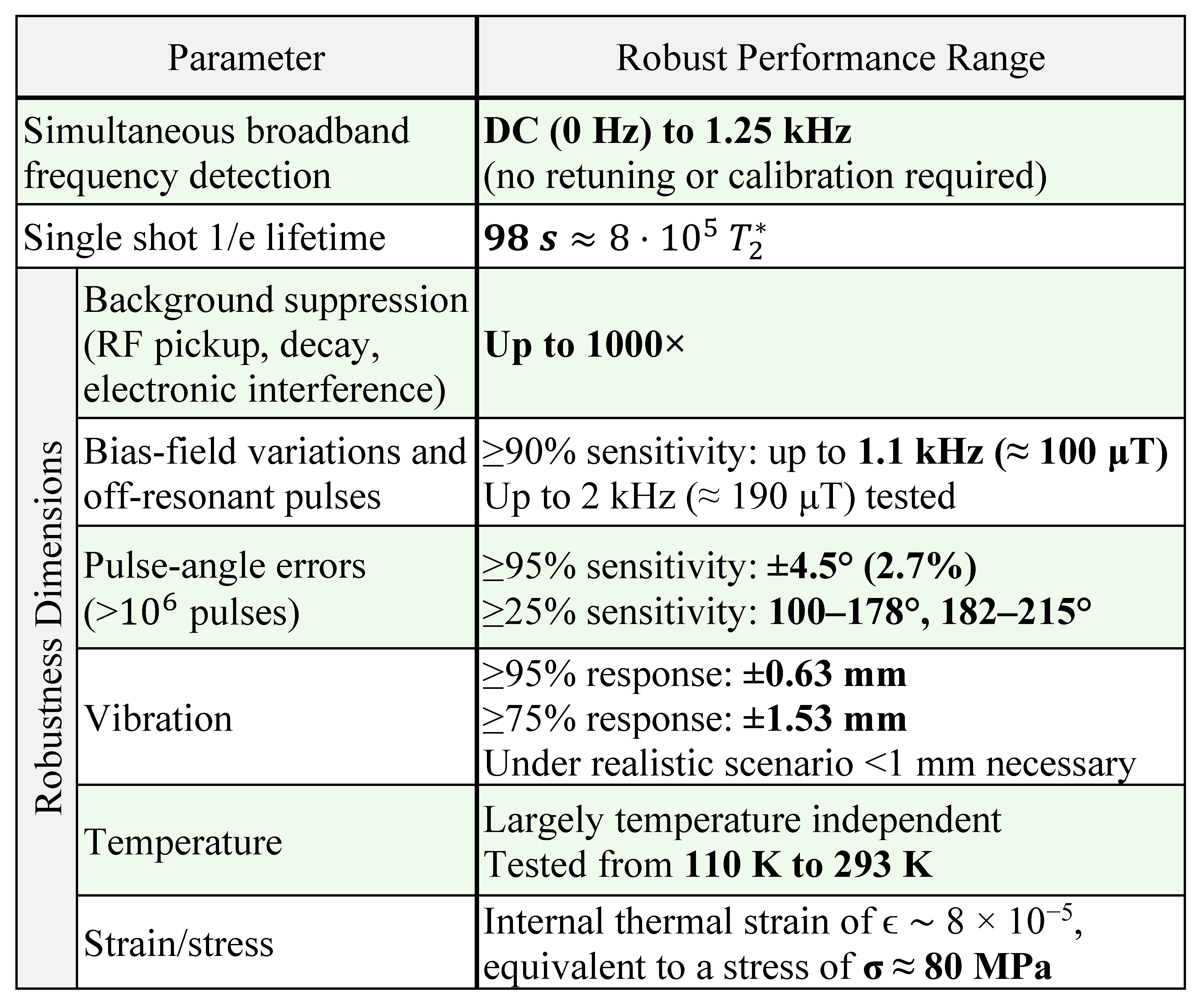}
    \label{tab:fig6}
    \vspace{-1em}
\end{table}

\vspace{0.5em}
\noi{\normalsize\bfseries Robustness to Vibration\par}
\noi Stability under sample vibration is a critical requirement for field deployment~\cite{bongs2019taking,halde2025let}. We impose internal motion by translating the sample relative to the RF resonator by ${\app}2\times$ the sample size while continuously sensing a test signal.
\zfr{fig5}d(iv) displays the tracked sample motion (App.~\ref{SI:MeasurementOfVibration}).
As the sample moves through the resonator’s fringe fields, the spins experience simultaneous variations in bias field, flip angle and possibly strain; the error channels of \zfr{fig5}a and \zfr{fig5}c now act jointly in a position-dependent manner. At the largest displacements the sample approaches the resonator’s physical boundaries, encountering strongly inhomogeneous fields. 

Spin dynamics are now three-dimensional and highly complex, and in either manifold the AC magnetometry signal is dominated by the large vibration induced variations (cf. App.~\ref{secSI:vibrationNote}). Despite this, \zfr{fig5}d(i-iii) demonstrates that motion in the two manifolds remains nearly identical, allowing to correct for the vibrational effects and recover the desired AC signal. \zfr{fig5}d(v) shows that the extracted response $\xD M_x$ remains flat and first-order insensitive to vibration, with ${<}5$\% variation over \SI{{\approx}1.3}{\milli\meter} displacement (App.~\ref{secSI:vibration3Dextraction} and \ref{SI:MeasurementOfVibration}).

\vspace{0.5em}
\noi{\normalsize\bfseries Robustness to Temperature and Strain\par}
\noi
Nuclear spins couple only weakly to lattice phonons, as reflected in their long $T_1$ times~\cite{ajoy2019hyperpolarized}. Any significant thermal dependence could be anticipated to arise indirectly via the electronic spin bath (NV and P1 centers), whose relaxation times are temperature-dependent~\cite{jarmola2012temperature,ernst2023temperature}.

\zfr{fig5}e reports the orbit-axis excitation from \SI{110}{\kelvin} to room temperature, recorded with no retuning or recalibration between the temperature steps~\cite{d2025cryogenic}. The elevation angles vary only weakly over this entire range. We attribute this minor variation not to a fundamental spin-bath interaction, but to small, hardware-related pulse angle deviations arising from the temperature-dependent properties of the RF resonator~\cite{d2025cryogenic}. Resilience to these temperature-driven drifts stems from its insensitivity across a broad plateau of flip angles, as established in \zfr{fig5}c. For applications requiring even greater stability, this residual dependence could be further suppressed by recalibrating the prethermal axes in-situ during the single-shot measurement.

Temperature can also impact sensing by shifting the energy levels of the spin system, as seen in NV centers, where zero-field splitting acquires temperature dependence due to lattice phonon coupling~\cite{jarmola2012temperature}. In this case, the energy shift from temperature can cause an identical signature to an external magnetic field, complicating magnetometry in thermally unstable environments. \zfr{fig5}e indicates the observed Larmor frequency over the experimentally probed temperature range. As expected, the data shows no significant temperature dependence in the $^{13}$C Zeeman splitting, rendering the spin system inherently resilient to thermal fluctuation. 

Finally, considering the impact of strain fluctuations, we note that they couple through the same electronic channels~\cite{udvarhelyi2018spin} and are orders of magnitude smaller than the macroscopic vibrations tested in \zfr{fig5}d. PRISM's demonstrated insensitivity to both thermal and mechanical perturbations therefore suggests a comparable strong resilience to strain (App.~\ref{secSI:tempstrain}).

\vspace{0.5em}
\noi{\normalsize\bfseries Conclusions and Outlook \par}
\noi Our results develop Floquet prethermal orbits as a resource to achieve intrinsically compensated quantum sensing. This PRISM method (\zfr{fig1}b) renders the sensor stable in the presence of background fields, fluctuations in bias or pulse detuning, and transient spin dynamics, and retains our prethermal nuclear spin platform's robustness to pulse angle miscalibration, vibration, and temperature (summarized in Table~\ref{tab:fig6}). App.~\ref{secSI:existingtechniques} presents a comparison with existing robust sensing methods that can provide subsets of these characteristics; no single alternative method achieves all facets simultaneously.

The approach extends quantum sensing beyond the conventional paradigm of isolated (non-interacting) sensors~\cite{barry2020sensitivity} to one where interactions are harnessed.  Indeed, the collectively stabilized long-lived spin orbits allow sensing protocols that go beyond conventional Ramsey or dynamical decoupling techniques~\cite{Degen17}. While this work uses the simplest two-point orbits, more intricate trajectories could support higher-dimensional correction schemes, exploiting orbital symmetries and field-induced deformations to extract targeted information about the sensed fields. We anticipate that these principles will be broadly applicable across interacting spin or qubit ensembles, including NV centers~\cite{salhov2024protecting, eisenach2021cavity, wolf2015subpicotesla}, SiC~\cite{jiang2023quantum}, 2D quantum sensors~\cite{gottscholl2021spin}, cold atoms~\cite{geiger2020high}, and trapped ions~\cite{reiter2017dissipative}.

Complementarily, in a classical analogy, the spins serve as collective oscillators switching between bistable orientations, resembling features of proposed levitating compass-needle magnetometers~\cite{jackson2016precessing}. The effective $Q$, given by the ratio of Larmor frequency to measurable lifetime approaches $10^{10}$ at room temperature, attractive even with respect to classical oscillators such as YIG.

The approach offers several practical advantages. Unlike conventional quantum-sensing schemes~\cite{Degen17}, it requires no spectral tuning or resonance matching, enabling continuous tracking of nonstationary signals across a wide bandwidth. Hyperpolarization is largely insensitive to laser-amplitude noise~\cite{Sarkar22}, permitting the use of inexpensive diode lasers~\cite{Ajoy19}. Likewise, the field stability shown in \zfr{fig5}a supports permanent-magnet biasing~\cite{soltner2010dipolar} with only modest, compensable sensitivity loss (App.~\ref{secSI:optimization}). Robust pulse-error (\zfr{fig5}b-c)  compensation enables inexpensive, compact amplifiers, while resonator-based detection eliminates bulky optics. Combined with strong background rejection, vibration and temperature stability, these features could potentially enable compact, portable sensors deployable even on noisy, dynamic platforms such as drones.

The measured sensitivity here is \SI{4.77}{\nano\tesla\per\sqrt{\hertz}} (App.~\ref{secSI:sensitivity}), but it was not optimized and is not central to this study. Moreover, the wide-band, flat frequency response (App.~\ref{secSI:sensorresponse}) inherently trades bandwidth for sensitivity. Even within the present system, sensitivity below \SI{10}{\pico\tesla\per\sqrt{\hertz}}  is feasible by reintroducing frequency selectivity; this will be the subject of a future paper. Additional routes for sensitivity improvement (detailed in App.~\ref{secSI:optimization}), include increasing the resonator filling factor, using larger sample volumes, and exploiting molecular crystals~\cite{singh2025room} that naturally offer increased hyperpolarizability, spin density, lifetime, and gyromagnetic ratio. 

Future opportunities include electron–nuclear comagnetometry using $\Cs$ nuclear spins as stable references for co-located, potentially higher-sensitivity NV electron sensors within the same volume~\cite{jaskula2019cross,kornack2005nuclear}. We anticipate that the ideas in the paper will also contribute new stabilization principles for nuclear gyroscopes~\cite{kornack2005nuclear, ajoy2012stable, ledbetter2012gyroscopes, jarmola2021demonstration}, nuclear clocks~\cite{beeks2021thorium, zhang2024frequency}, magnetic-resonance-based dark-matter searches~\cite{budker2014proposal, garcon2017cosmic}, and nanoscale NMR sensors~\cite{budakian2024roadmap}.

\clearpage

%%%%%%%%%%%%%%%%%%%%%%%%%%%%%%%%%%%%%%%%%%%%%%%%%%%%%%%%%%%%%%%%%%%%%%%%%%%
%                        Methods
%%%%%%%%%%%%%%%%%%%%%%%%%%%%%%%%%%%%%%%%%%%%%%%%%%%%%%%%%%%%%%%%%%%%%%%%%%%
%%%%%%%%%%%%%%%%%%%%%%%%%%%%%%%%%%%%%%%%%%%%%%%%%%%%%%%%%%%%%%%%%%%%%%%%%%%
%\beginmethods
\section*{Methods}

    \vspace{0.5em}
    \noi{\normalsize\bfseries Sample and nuclear hyperpolarization\par}
    \noi
        Experiments were performed on a disordered network of dipole-coupled $^{13}$C nuclear spins in a single-crystal chemical-vapor-deposition (CVD) diamond at natural $^{13}$C abundance (1.1\%). The crystal measured ${3.4\times 3.2\times \SI{2.1}{\milli\meter\cubed}}$ and contained $[\mathrm{NV}^-] \SI{\approx 1}{ppm}$ and $[N_\mathrm{S}^0] \SI{\approx 20}{ppm}$. Macroscopic $^{13}$C polarization was generated at low magnetic field ($B_{\mathrm{pol}} \SI{\approx 38}{\milli\tesla}$) via optical dynamic nuclear polarization mediated by nitrogen--vacancy (NV$^-$) centers. The NV centers were optically pumped with $15$--\SI{20}{\watt} of continuous-wave green light at \SI{520}{\nano\meter}, and polarization was transferred to the surrounding $^{13}$C nuclei using frequency-chirped microwave irradiation applied during illumination. The total hyperpolarization time was $60$--\SI{120}{\second}. Further details of the polarization-transfer protocol are provided in Ref.~\cite{ajoy2018orientation, elanchezhian2021galton}. Unless stated otherwise, all measurements were carried out at room temperature.

    \vspace{0.5em}
    \noi{\normalsize\bfseries Sample transfer and high-field detection\par}
    \noi
        Following hyperpolarization, the diamond was transferred using a mechanical shuttler into a $7~\mathrm{T}$ superconducting magnet. Inside the magnet, a radio-frequency (RF) resonator tuned and matched to the $^{13}$C Larmor frequency ($\omega_0/2\pi \approx \SI{75}{\mega\hertz}$ at $B_0 = \SI{7}{\tesla}$) enabled both the delivery of RF pulses and the inductive detection of the collective $^{13}$C magnetization.

    \vspace{0.5em}
    \noi{\normalsize\bfseries Control and acquisition electronics\par}
    \noi
        RF pulses were synthesized and delivered to the resonator using a Tabor Proteus arbitrary waveform transceiver; during each acquisition window $t_{\mathrm{acq}}$, the same instrument digitized the down-converted RF signal. The orbit field (``$\zhat$-drive''), as well as calibrated target fields and artificial backgrounds, were generated with arbitrary-function generators (Tektronix AFG31000 series), amplified by an AE Techron~7224 power amplifier, and applied using an auxiliary coil mounted around the RF resonator. The auxiliary coil was nominally aligned with the $B_0$ axis and intentionally tilted by ${\sim} 5^{\circ}$ to introduce a small transverse component, enabling injection of RF fields with $\xhat\tm\yhat$ polarization. Additional details of the instrument configuration are provided in Ref.~\cite{moon2025high}.

    \vspace{0.5em}
    \noi{\normalsize\bfseries Calibration procedures\par}
        \textbf{Resonator tuning and matching.} Prior to extended measurements like parameter scans, and at regular intervals, the RF resonator was tuned and matched by adjusting its capacitors and by using a vector network analyzer.
        
            \textbf{Larmor frequency.} A coarse estimate of the $^{13}$C Larmor frequency was obtained from a single-shot free-induction decay (FID), accurate to $\pm \SI{40}{\hertz}$. When required, a fine adjustment was performed using the PRISM sequence by adjusting the pulse frequency in small steps and identifying the setting at which the sequence reported zero apparent DC bias field. Alternatively, any residual DC bias inferred from the sequence was subtracted during analysis.
        
        \textbf{Pulse-angle calibration.} Accurate flip-angle calibration was performed using a Rabi-style procedure. An initial $(\pi/2)_y$ pulse rotated the net magnetization to $+\xhat$. Several hundred spin-lock pulses $(\pi/2)_x$ were then applied along $\xhat$ to prethermalize the spins exactly along $\xhat$. Subsequently, repeated $(\pi/2)_y$ pulses rotated the magnetization around the Bloch sphere. Deviations of the measured signal amplitudes from their expected values (near zero magnetization along $\pm z$, maximal along $\pm x$) yielded the pulse-angle error, which was used to calibrate the nominal flip angle.

    \vspace{0.5em}
    \noi{\normalsize\bfseries Pulse sequence\par}
    \noi
        The pulse sequence was preprogrammed and temporally synchronized with the AFG31000 instruments output. A hardware trigger from the Tabor Proteus initiated the AFGs. During measurements, selected parameters (e.g., AFG waveform properties) were updated in real-time by a Python control script.
        
        Unless stated otherwise, a pulse angle of approximately $\vartheta=166^{\circ}$ was used. The RF power was held fixed, and the flip angle was set by adjusting the pulse duration, resulting in pulse lengths of about \SI{100}{\micro\second}. The interpulse spacing was \SI{100}{\micro\second}. At the start of each spacing, a \SI{12}{\micro\second} receiver-delay interval suppressed resonator ring-down from the preceding pulse. The nuclear magnetization was then sampled for $t_{\mathrm{acq}} = 48$--\SI{76}{\micro\second} (depending on the experiment), followed by a second \SI{12}{\micro\second} guard interval to avoid overlap with the subsequent pulse.
        
        During each acquisition window, in-phase ($I$) and quadrature ($Q$) components were obtained by coherent demodulation in the Tabor Proteus against a phase-stable reference at a sampling rate $f_s=\SI{168.75}{\mega\hertz}$. 
        Averaging the $I$ and $Q$ samples over $t_{\mathrm{acq}}$ yielded $M_x$ and $M_y$, and consequently, the amplitude and phase of the transverse magnetization vector, effectively filtering out parasitic fields that fall outside the detectable frequency range set by the acquisition window length.

    \vspace{0.5em}
    \noi{\normalsize\bfseries Magnetization calibration and vector reconstruction\par}
    \noi
        At the start of each measurement, we calibrated the net nuclear magnetization to allow three-dimensional reconstruction of the trajectory states for illustrative purposes; this reconstruction is not necessary for the actual sensing.  Specifically, we detuned the orbit-field frequency by a few hertz (\SIrange{0.5}{2}{\hertz}) to induce a slow rotation of the two trajectory states about the $\xhat$-axis. After a few (two to five) revolutions, the rotation was halted by returning the orbit-field frequency to its nominal value at the moment when the two states crossed the $\xhat\tm\zhat$ plane.
        
            The three-dimensional magnetization vector was then reconstructed using the net nuclear magnetization and baseline extrapolation. Full details of the calibration and reconstruction procedures are provided in the App.~\ref{secSI:3Dextraction}.

        For measurements involving imposed vibrations, three-dimensional reconstruction used calibration-derived mappings between response and flip angle, and between flip angle and elevation angle. Algorithmic details and validation are provided in App.~\ref{secSI:vibration3Dextraction}.

    \vspace{0.5em}
    \noi{\normalsize\bfseries Signal extraction and background suppression\par}
    \noi
        Data was analyzed using custom Python code. To extract the measurement signal while suppressing common-mode backgrounds, the time-domain $\xhat$-axis magnetization $M_x$ was partitioned into even- and odd-indexed samples. The odd-indexed series was linearly interpolated to compensate residual timing offsets, and the difference between the even and (interpolated) odd series yielded the differential signal with suppressed common-mode noise. A detailed explanation is provided in App.~\ref{secSI:signalextraction}.

    \vspace{0.5em}
    \noi{\normalsize\bfseries Robustness tests\par}
    \noi
        Robustness was assessed by: (i) injecting artificial RF backgrounds through the auxiliary coil, (ii) applying a static (DC) bias field with the auxiliary coil, (iii) varying the RF pulse duration (and thus the pulse angle), (iv) detuning the RF carrier frequency emitted by the transceiver, and (v) intentionally oscillating the shuttler along the $B_0$ axis during acquisition (App.~\ref{SI:MeasurementOfVibration}). A comprehensive description of these tests is provided in the Supplementary Information (App.~\ref{secSI:vibrationNote}-\ref{secSI:signalextraction}).

    \vspace{0.5em}
    \noi{\normalsize\bfseries Temperature-dependent measurements\par}
    \noi
        To assess robustness against ambient-temperature variations, we used an equivalent NMR setup in which the diamond, RF resonator, auxiliary (AC) coil, and the hyperpolarization coil and laser were housed inside a cryostat, as described in Ref.~\cite{d2025cryogenic}. The sample and associated hardware were initially cooled to \SI{110}{\kelvin} and subsequently warmed in discrete steps to room temperature using a resistive heater, without any recalibration or retuning of the experiment.
        
        At each temperature, three measurements were performed: (i) an FID without any applied DC field, (ii) an FID with a fixed-voltage DC magnetic field to determine the field strength generated by the auxiliary coil, and (iii) a measurement using the PRISM protocol to extract the elevation angle. For (iii), the two trajectory states were continuously rotated about the $\xhat$-axis at a few hertz to enable estimation of the elevation angle.

    \vspace{0.5em}
    \noi{\normalsize\bfseries Simulation and theoretical framework\par}
    \noi
        A concise theoretical overview of the trajectories and their response to small fields is provided in the App.~\ref{secSI:theorytrajectories}, along with matrix-based simulations (App.~\ref{secSI:SimulationRotationMatricesBRAYDEN}). For completeness, the essential theoretical elements are also summarized here.
        
        When the orbit field is period-matched to the two spin-lock pulses (Fig. \ref{fig:fig2}a(ii)), the stroboscopic dynamics are governed by an effective Hamiltonian composed of a spin--spin interaction term ($H_\mathrm{int}$) and a non-commuting emergent field ($\boldsymbol{w}_k(\varphi)$) ~\cite{sahin22_trajectory}:
        \begin{equation}
        H_{\mathrm{eff},k} = H_{\mathrm{int}} + \boldsymbol{w}_k(\varphi)\cdot\boldsymbol{I},
        \label{eq:floquet_hamiltonian_MAIN}
        \end{equation}
        where $\varphi$ is the phase of the orbit field and $\boldsymbol{I}$ is the vector of spin-$1/2$ operators. Because the magnetization is measured after each pulse, every acquisition corresponds to a distinct stroboscopic frame defined relative to either the first or the second spin-lock pulse. Accordingly, Eq.~\ref{eq:floquet_hamiltonian_MAIN} defines two effective Hamiltonians ($k = 1,2$), each conditioned on the pulse with respect to which the period is defined. As a result, the magnetization direction alternates between $\mathbf{n}_1$ and $\mathbf{n}_2$ ($\mathbf{n}_k = \boldsymbol{w}_k/|\boldsymbol{w}_k|$), giving rise to two distinct prethermal expectation values of the magnetization.
        
        For on-resonance pulses, the orbit is centered along $\mathbf{\xhat}$. However, the sensed AC field introduces a time-dependent perturbation that tilts the effective pulse axis in the $\pm \mathbf{\zhat}$ directions. This perturbation modulates the transverse-plane projections of the two magnetization vectors, producing phase-shifted signatures of the sensed AC field.
        
    \vspace{0.5em}
    \noi{\normalsize\bfseries Sound signal measurement\par} \label{secmethods:soundmeasurement}
    \noi
        A sound waveform (sampling rate \SI{22.05}{\kilo\hertz}) was exported as \texttt{.csv}, played using a Tektronix AFG31000 arbitrary-function generator, summed with the orbit-field drive, and amplified with an AE Techron~7224 power amplifier. The combined output was delivered via coaxial cabling to the auxiliary coil, producing a weak AC magnetic field (target field \SI{\approx 33}{\micro\tesla}) that was continuously detected via the $^{13}$C nuclear spins.
        
        Experimental parameters were: pulse length $\tau_{\mathrm{p}}=\SI{94.72}{\micro\second}$ and interpulse delay $\tau_s = \SI{100.03}{\micro\second}$, corresponding to an effective sampling rate of \SI{2.567}{\kilo\hertz} (Nyquist limit \SI{1.284}{\kilo\hertz}). Each acquisition window (\SI{64}{\micro\second}) began \SI{12}{\micro\second} after the RF pulse. An artificial RF background was applied within $\approx\pm \SI{400}{\hertz}$ of the Larmor frequency, with its frequency modulated at \SI{2.5}{\hertz} using a triangular waveform.

        The measurement was repeated with an extended acquisition time, as detailed in App.~\ref{secSI:soundsignal}.

\vspace{0.5em}
\noi\paragraph*{Acknowledgements}
\noi We gratefully acknowledge discussions with M. Bukov and P. Schindler. We thank C. Selco and C. Shah for support with the low-temperature measurements, and T. Splettstößer for his assistance with the illustrations.
This work was supported in part by the U.S. Department of Energy National Nuclear Security Administration through the NNSA Office of Defense Nuclear Nonproliferation R\&D through the LB24-NV center $^{13}$C quantum sensor-PD3Ta project and the Nonproliferation Stewardship Program (NSP). We additionally acknowledge funding from ONR (N00014-20-1-2806), DOE SBIR, and instrumentation support from AFOSR DURIP (FA9550-
22-1-0156) and NSF MRI (2320520). 

\vspace{0.5em}
\noi\paragraph*{Contributions.} 
\noi EDR and RJS performed experiments and analyzed data. EDR developed use of two-point trajectories for sensing, established robustness towards control errors, transient and vibration suppression, and method for 3D reconstruction. EDR and RJS developed background suppression. EDR and BG conducted simulations with support from RJS. BG developed theory with assistance from RJS. AA supervised the experimental work. All authors contributed to the manuscript.

\vspace{0.5em}
\noi\paragraph*{Competing Interest.}
\noi EDR, RJS, and AA are inventors on a provisional patent application by LBNL related to this work. The other authors declare no competing interests.

%\paragraph*{Data availability.}

%\paragraph*{Code availability.}

%%%%%%%%%%%%%%%%%%%%%%%%%%%%%%%%%%%%%%%%%%%%%%%%%%%%%%%%%%%%%%%%%%%%%%%%%%%
%                    Bibliography
%%%%%%%%%%%%%%%%%%%%%%%%%%%%%%%%%%%%%%%%%%%%%%%%%%%%%%%%%%%%%%%%%%%%%%%%%%%
%%%%%%%%%%%%%%%%%%%%%%%%%%%%%%%%%%%%%%%%%%%%%%%%%%%%%%%%%%%%%%%%%%%%%%%%%%%
% \putbib
% \end{bibunit}
%\clearpage
\bibliography{bilbio}

@article{Sarkar22,
  title = {Rapidly Enhanced Spin-Polarization Injection in an Optically Pumped Spin Ratchet},
  author = {Sarkar, Adrisha and Blankenship, Brian and Druga, Emanuel and Pillai, Arjun and Nirodi, Ruhee and Singh, Siddharth and Oddo, Alexander and Reshetikhin, Paul and Ajoy, Ashok},
  journal = {Phys. Rev. Appl.},
  volume = {18},
  issue = {3},
  pages = {034079},
  numpages = {12},
  year = {2022},
  month = {Sep},
  publisher = {American Physical Society},
  doi = {10.1103/PhysRevApplied.18.034079},
  url = {https://link.aps.org/doi/10.1103/PhysRevApplied.18.034079}
}

@book{duer2008solid,
  title={Solid state NMR spectroscopy: principles and applications},
  author={Duer, Melinda J},
  year={2008},
  publisher={John Wiley \& Sons},
  isbn={978-0-632-05351-3}
}

@article{doherty2013nitrogen,
  title={The nitrogen-vacancy colour centre in diamond},
  author={Doherty, Marcus W and Manson, Neil B and Delaney, Paul and Jelezko, Fedor and Wrachtrup, J{\"o}rg and Hollenberg, Lloyd CL},
  journal={Physics Reports},
  volume={528},
  number={1},
  pages={1--45},
  year={2013},
  publisher={Elsevier},
  doi={10.1016/j.physrep.2013.02.001}
}

@article{ajoy2019wide,
  title={Wide dynamic range magnetic field cycler: Harnessing quantum control at low and high fields},
  author={Ajoy, A and Lv, X and Druga, E and Liu, K and Safvati, B and Morabe, A and Fenton, M and Nazaryan, R and Patel, S and Sjolander, TF and others},
  journal={Review of Scientific Instruments},
  volume={90},
  pages={013112},
  number={1},
  year={2019},
  publisher={AIP Publishing},
  doi={10.1063/1.5064685}
}

@book{fukushima2018experimental,
  title={Experimental pulse NMR: a nuts and bolts approach},
  author={Fukushima, Eiichi},
  year={2018},
  publisher={CRC press},
  doi={10.1201/9780429493867}
}

@article{ramsey1990experiments,
  title={Experiments with separated oscillatory fields and hydrogen masers},
  author={Ramsey, Norman F},
  journal={Reviews of modern physics},
  volume={62},
  number={3},
  pages={541},
  year={1990},
  publisher={APS},
  doi={10.1103/RevModPhys.62.541}
}

@article{schmitt2017submillihertz,
  title={Submillihertz magnetic spectroscopy performed with a nanoscale quantum sensor},
  author={Schmitt, Simon and Gefen, Tuvia and St{\"u}rner, Felix M and Unden, Thomas and Wolff, Gerhard and M{\"u}ller, Christoph and Scheuer, Jochen and Naydenov, Boris and Markham, Matthew and Pezzagna, Sebastien and others},
  journal={Science},
  volume={356},
  number={6340},
  pages={832--837},
  year={2017},
  publisher={American Association for the Advancement of Science},
  doi={10.1126/science.aam5532}
}

@article{devience2015nanoscale,
  title={Nanoscale {NMR} spectroscopy and imaging of multiple nuclear species},
  author={DeVience, Stephen J and Pham, Linh M and Lovchinsky, Igor and Sushkov, Alexander O and Bar-Gill, Nir and Belthangady, Chinmay and Casola, Francesco and Corbett, Madeleine and Zhang, Huiliang and Lukin, Mikhail and others},
  journal={Nature nanotechnology},
  volume={10},
  number={2},
  pages={129--134},
  year={2015},
  publisher={Nature Publishing Group UK London},
  doi={10.1038/nnano.2014.313}
}

@unpublished{khemani2019brief,
  title={A Brief History of Time Crystals}, 
  author={Vedika Khemani and Roderich Moessner and S. L. Sondhi},
  year={2019},
  eprint={1910.10745},
  archivePrefix={arXiv},
  primaryClass={cond-mat.str-el},
  url={https://arxiv.org/abs/1910.10745}, 
}

@article{unden2016quantum,
  title={Quantum metrology enhanced by repetitive quantum error correction},
  author={Unden, Thomas and Balasubramanian, Priya and Louzon, Daniel and Vinkler, Yuval and Plenio, Martin B and Markham, Matthew and Twitchen, Daniel and Stacey, Alastair and Lovchinsky, Igor and Sushkov, Alexander O and others},
  journal={Physical review letters},
  volume={116},
  number={23},
  pages={230502},
  year={2016},
  publisher={APS},
  doi={10.1103/PhysRevLett.116.230502}
}

@article{louzonCPDD25,
  title = {Robust Noise Suppression and Quantum Sensing by Continuous Phased Dynamical Decoupling},
  author = {Louzon, Daniel and Genov, Genko T. and Staudenmaier, Nicolas and Frank, Florian and Lang, Johannes and Markham, Matthew L. and Retzker, Alex and Jelezko, Fedor},
  journal = {Phys. Rev. Lett.},
  volume = {134},
  issue = {12},
  pages = {120802},
  numpages = {7},
  year = {2025},
  month = {Mar},
  publisher = {American Physical Society},
  doi = {10.1103/PhysRevLett.134.120802},
  url = {https://link.aps.org/doi/10.1103/PhysRevLett.134.120802}
}

@article{zeng2024wide,
  title={Wide-band unambiguous quantum sensing via geodesic evolution},
  author={Zeng, Ke and Yu, Xiaohui and Plenio, Martin B and Wang, Zhen-Yu},
  journal={Physical Review Letters},
  volume={132},
  number={25},
  pages={250801},
  year={2024},
  publisher={APS},
  doi={10.1103/PhysRevLett.132.250801}
}

@article{aiello2013composite,
  title={Composite-pulse magnetometry with a solid-state quantum sensor},
  author={Aiello, Clarice D and Hirose, Masashi and Cappellaro, Paola},
  journal={Nature communications},
  volume={4},
  number={1},
  pages={1419},
  year={2013},
  publisher={Nature Publishing Group UK London},
  doi={10.1038/ncomms2375}
}

@article{hiroseCDD12,
  title = {Continuous dynamical decoupling magnetometry},
  author = {Hirose, Masashi and Aiello, Clarice D. and Cappellaro, Paola},
  journal = {Phys. Rev. A},
  volume = {86},
  issue = {6},
  pages = {062320},
  numpages = {5},
  year = {2012},
  month = {Dec},
  publisher = {American Physical Society},
  doi = {10.1103/PhysRevA.86.062320},
  url = {https://link.aps.org/doi/10.1103/PhysRevA.86.062320}
}

@unpublished{moon2024discrete,
  title={Discrete time crystal sensing},
  author={Leo Joon Il Moon and Paul M. Schindler and Ryan J. Smith and Emanuel Druga and Zhuo-Rui Zhang and Marin Bukov and Ashok Ajoy},
  year={2024},
  eprint={2410.05625},
  archivePrefix={arXiv},
  primaryClass={quant-ph},
  url={https://arxiv.org/abs/2410.05625}, 
}

@article{ho2017critical,
  title={Critical time crystals in dipolar systems},
  author={Ho, Wen Wei and Choi, Soonwon and Lukin, Mikhail D and Abanin, Dmitry A},
  journal={Physical review letters},
  volume={119},
  number={1},
  pages={010602},
  year={2017},
  publisher={APS},
  doi={10.1103/PhysRevLett.119.010602}
}

@article{sacha2017time,
  title={Time crystals: a review},
  author={Sacha, Krzysztof and Zakrzewski, Jakub},
  journal={Reports on Progress in Physics},
  volume={81},
  number={1},
  pages={016401},
  year={2017},
  publisher={IOP Publishing},
  doi={10.1088/1361-6633/aa8b38}
}

@article{budker2014proposal,
  title={Proposal for a cosmic axion spin precession experiment ({CASPEr})},
  author={Budker, Dmitry and Graham, Peter W and Ledbetter, Micah and Rajendran, Surjeet and Sushkov, Alexander O},
  journal={Physical Review X},
  volume={4},
  number={2},
  pages={021030},
  year={2014},
  publisher={APS},
  doi={10.1103/PhysRevX.4.021030}
}

@article{souza2011robust,
  title={Robust dynamical decoupling for quantum computing and quantum memory},
  author={Souza, Alexandre M and Alvarez, Gonzalo A and Suter, Dieter},
  journal={Physical review letters},
  volume={106},
  number={24},
  pages={240501},
  year={2011},
  publisher={APS},
  doi={10.1103/PhysRevLett.106.240501}

}

@article{harrington2025synchronous,
  title={Synchronous detection of cosmic rays and correlated errors in superconducting qubit arrays},
  author={Harrington, Patrick M and Li, Mingyu and Hays, Max and Van De Pontseele, Wouter and Mayer, Daniel and Pinckney, H Douglas and Contipelli, Felipe and Gingras, Michael and Niedzielski, Bethany M and Stickler, Hannah and others},
  journal={Nature Communications},
  volume={16},
  number={1},
  pages={6428},
  year={2025},
  publisher={Nature Publishing Group UK London},
  doi = {10.1038/s41467-025-61385-x}
}

@article{berzins2024impact,
  title={Impact of microwave phase noise on diamond quantum sensing},
  author={Berzins, Andris and Saleh Ziabari, Maziar and Silani, Yaser and Fescenko, Ilja and Damron, Joshua T and Barry, John F and Jarmola, Andrey and Kehayias, Pauli and Richards, Bryan A and Smits, Janis and others},
  journal={Physical Review Research},
  volume={6},
  number={4},
  pages={043148},
  year={2024},
  publisher={APS},
  doi={10.1103/PhysRevResearch.6.043148}
}

@article{kono2024mechanically,
  title={Mechanically induced correlated errors on superconducting qubits with relaxation times exceeding 0.4 ms},
  author={Kono, Shingo and Pan, Jiahe and Chegnizadeh, Mahdi and Wang, Xuxin and Youssefi, Amir and Scigliuzzo, Marco and Kippenberg, Tobias J},
  journal={Nature Communications},
  volume={15},
  number={1},
  pages={3950},
  year={2024},
  publisher={Nature Publishing Group UK London},
  doi={10.1038/s41467-024-48230-3}
}

@article{budakian2024roadmap,
  title={Roadmap on nanoscale magnetic resonance imaging},
  author={Budakian, Raffi and Finkler, Amit and Eichler, Alexander and Poggio, Martino and Degen, Christian L and Tabatabaei, Sahand and Lee, Inhee and Hammel, P Chris and Eugene, S Polzik and Taminiau, Tim H and others},
  journal={Nanotechnology},
  volume={35},
  number={41},
  pages={412001},
  year={2024},
  publisher={IOP Publishing},
  doi={10.1088/1361-6528/ad4b23}
}

@article{reiter2017dissipative,
  title={Dissipative quantum error correction and application to quantum sensing with trapped ions},
  author={Reiter, Florentin and S{\o}rensen, Anders S{\o}ndberg and Zoller, Peter and Muschik, C A},
  journal={Nature communications},
  volume={8},
  number={1},
  pages={1822},
  year={2017},
  publisher={Nature Publishing Group UK London},
  doi={10.1038/s41467-017-01895-5}
}

@article{soltner2010dipolar,
  title={Dipolar Halbach magnet stacks made from identically shaped permanent magnets for magnetic resonance},
  author={Soltner, Helmut and Bl{\"u}mler, Peter},
  journal={Concepts in magnetic resonance part a},
  volume={36},
  number={4},
  pages={211--222},
  year={2010},
  publisher={Wiley Online Library},
  doi={10.1002/cmr.a.20165}
}

@article{jacobson2019thermal,
  title={Thermal expansion coefficient of diamond in a wide temperature range},
  author={Jacobson, P and Stoupin, S},
  journal={Diamond and Related Materials},
  volume={97},
  pages={107469},
  year={2019},
  publisher={Elsevier},
  doi={10.1016/j.diamond.2019.107469}
}

@article{jing2024practical,
  title={Practical ultra-low frequency noise laser system for quantum sensors},
  author={Jing, Mingyong and Xue, Shiyu and Zhang, Hao and Zhang, Linjie and Xiao, Liantuan and Jia, Suotang},
  journal={EPJ Quantum Technology},
  volume={11},
  number={1},
  pages={84},
  year={2024},
  publisher={Springer Berlin Heidelberg},
  doi={10.1140/epjqt/s40507-024-00297-z}
}

@article{salhov2024protecting,
  title={Protecting quantum information via destructive interference of correlated noise},
  author={Salhov, Alon and Cao, Qingyun and Cai, Jianming and Retzker, Alex and Jelezko, Fedor and Genov, Genko},
  journal={Physical Review Letters},
  volume={132},
  number={22},
  pages={223601},
  year={2024},
  publisher={APS},
  doi={10.1103/PhysRevLett.132.223601}
}

@article{ajoy2018orientation,
  title={Orientation-independent room temperature optical {${}^{13}C$} hyperpolarization in powdered diamond},
  author={Ajoy, Ashok and Liu, Kristina and Nazaryan, Raffi and Lv, Xudong and Zangara, Pablo R and Safvati, Benjamin and Wang, Guoqing and Arnold, Daniel and Li, Grace and Lin, Arthur and others},
  journal={Science advances},
  volume={4},
  number={5},
  pages={eaar5492},
  year={2018},
  publisher={American Association for the Advancement of Science},
  doi={10.1126/sciadv.aar5492}
}

@article{elanchezhian2021galton,
  title={Electron-to-nuclear spectral mapping via dynamic nuclear polarization},
  author={Pillai, Arjun and Elanchezhian, Moniish and Virtanen, Teemu and Conti, Sophie and Ajoy, Ashok},
  journal={The Journal of Chemical Physics},
  volume={159},
  number={15},
  year={2023},
  pages={154201},
  publisher={AIP Publishing},
  doi={10.1063/5.0157954}
}

@unpublished{halde2025let,
  title={Who Let the Diamonds Out?}, 
  author={Vincent Halde and Olivier Bernard and Mathieu Brochu and Laurier Dufresne and Nicolas Fleury and Kayla Johnson and Benjamin Moffett and David Roy-Guay},
  year={2025},
  eprint={2509.19179},
  archivePrefix={arXiv},
  primaryClass={quant-ph},
  url={https://arxiv.org/abs/2509.19179}, 
}

@article{bongs2019taking,
  title={Taking atom interferometric quantum sensors from the laboratory to real-world applications},
  author={Bongs, Kai and Holynski, Michael and Vovrosh, Jamie and Bouyer, Philippe and Condon, Gabriel and Rasel, Ernst and Schubert, Christian and Schleich, Wolfgang P and Roura, Albert},
  journal={Nature Reviews Physics},
  volume={1},
  number={12},
  pages={731--739},
  year={2019},
  publisher={Nature Publishing Group UK London},
  doi={10.1038/s42254-019-0117-4}
}

@article{barry2020sensitivity,
  title={Sensitivity optimization for NV-diamond magnetometry},
  author={Barry, John F and Schloss, Jennifer M and Bauch, Erik and Turner, Matthew J and Hart, Connor A and Pham, Linh M and Walsworth, Ronald L},
  journal={Reviews of Modern Physics},
  volume={92},
  number={1},
  pages={015004},
  year={2020},
  publisher={APS},
  doi={10.1103/RevModPhys.92.015004}
}

@article{jiang2023quantum,
  title={Quantum sensing of radio-frequency signal with {NV} centers in {SiC}},
  author={Jiang, Zhengzhi and Cai, Hongbing and Cernansky, Robert and Liu, Xiaogang and Gao, Weibo},
  journal={Science Advances},
  volume={9},
  number={20},
  pages={eadg2080},
  year={2023},
  publisher={American Association for the Advancement of Science},
  doi={10.1126/sciadv.adg2080}
}

@article{loretz2015spurious,
  title={Spurious harmonic response of multipulse quantum sensing sequences},
  author={Loretz, Michael and Boss, JM and Rosskopf, Tobias and Mamin, HJ and Rugar, D and Degen, Christian L},
  journal={Physical Review X},
  volume={5},
  number={2},
  pages={021009},
  year={2015},
  publisher={APS},
  doi={10.1103/PhysRevX.5.021009}
}

@article{ernst2023temperature,
  title={Temperature dependence of photoluminescence intensity and spin contrast in nitrogen-vacancy centers},
  author={Ernst, Stefan and Scheidegger, Patrick J and Diesch, Simon and Lorenzelli, Luca and Degen, Christian L},
  journal={Physical review letters},
  volume={131},
  number={8},
  pages={086903},
  year={2023},
  publisher={APS},
  doi={10.1103/PhysRevLett.131.086903}
}

@article{herb2025quantum,
  title={Quantum magnetometry of transient signals with a time resolution of 1.1 nanoseconds},
  author={Herb, Konstantin and V{\"o}lker, Laura A and Abendroth, John M and Meinhardt, Nicholas and van Schie, Laura and Gambardella, Pietro and Degen, Christian L},
  journal={Nature Communications},
  volume={16},
  number={1},
  pages={822},
  year={2025},
  publisher={Nature Publishing Group UK London},
  doi={10.1038/s41467-025-55956-1}
}

@article{bayat2014efficient,
  title={Efficient, uniform, and large area microwave magnetic coupling to {NV} centers in diamond using double split-ring resonators},
  author={Bayat, Khadijeh and Choy, Jennifer and Farrokh Baroughi, Mahdi and Meesala, Srujan and Loncar, Marko},
  journal={Nano letters},
  volume={14},
  number={3},
  pages={1208--1213},
  year={2014},
  publisher={ACS Publications},
  doi={10.1021/nl404072s}
}

@article{lang2019nonvanishing,
  title={Nonvanishing effect of detuning errors in dynamical-decoupling-based quantum sensing experiments},
  author={Lang, J E and Madhavan, T and Tetienne, J-P and Broadway, D A and Hall, L T and Teraji, Tokuyuki and Monteiro, TS and Stacey, Alastair and Hollenberg, L C L},
  journal={Physical Review A},
  volume={99},
  number={1},
  pages={012110},
  year={2019},
  publisher={APS},
  doi={10.1103/PhysRevA.99.012110}
}

@article{zhang2016microwave,
  title={Microwave magnetic field coupling with nitrogen-vacancy center ensembles in diamond with high homogeneity},
  author={Zhang, Ning and Zhang, Chen and Xu, Lixia and Ding, Ming and Quan, Wei and Tang, Zheng and Yuan, Heng},
  journal={Applied Magnetic Resonance},
  volume={47},
  number={6},
  pages={589--599},
  year={2016},
  publisher={Springer},
  doi={10.1007/s00723-016-0777-5}
}

@article{plotzki2025defect,
  title={Defect interplay for nuclear hyperpolarization in type {Ib} diamond},
  author={Plotzki, David and Engel, Johannes and P{\"o}ppl, Andreas and Knolle, Wolfgang and Wunderlich, Ralf},
  journal={Diamond and Related Materials},
  pages={112554},
  year={2025},
  publisher={Elsevier},
  doi={10.1016/j.diamond.2025.112554}
}

@article{d2025cryogenic,
  title={Cryogenic field-cycling instrument for optical {NMR} hyperpolarization studies},
  author={D'Souza, Noella and Harkins, Kieren A and Selco, Cooper and Basumallick, Ushoshi and Breuer, Samantha and Zhang, Zhuorui and Reshetikhin, Paul and Ho, Marcus and Nayak, Aniruddha and McAllister, Maxwell and others},
  journal={Journal of Magnetic Resonance},
  volume={375},
  pages={107874},
  year={2025},
  publisher={Elsevier},
  doi={10.1016/j.jmr.2025.107874}
}

@article{jarmola2012temperature,
  title={Temperature- and magnetic-field-dependent longitudinal spin relaxation in nitrogen-vacancy ensembles in diamond},
  author={Jarmola, A and Acosta, V M and Jensen, K and Chemerisov, S and Budker, D},
  journal={Physical review letters},
  volume={108},
  number={19},
  pages={197601},
  year={2012},
  publisher={APS},
  doi={10.1103/PhysRevLett.108.197601}
}

@article{ajoy2019hyperpolarized,
  title={Hyperpolarized relaxometry based nuclear ${T}_1$ noise spectroscopy in diamond},
  author={Ajoy, Ashok and Safvati, Ben and Nazaryan, Raffi and Oon, JT and Han, Ben and Raghavan, Priyanka and Nirodi, Ruhee and Aguilar, Alessandra and Liu, Kristina and Cai, Xiao and others},
  journal={Nature communications},
  volume={10},
  number={1},
  pages={5160},
  year={2019},
  publisher={Nature Publishing Group UK London},
  doi={10.1038/s41467-019-13042-3}
}

@article{udvarhelyi2018spin,
  title={Spin-strain interaction in nitrogen-vacancy centers in diamond},
  author={Udvarhelyi, P{\'e}ter and Shkolnikov, Vladyslav O and Gali, Adam and Burkard, Guido and P{\'a}lyi, Andr{\'a}s},
  journal={Physical Review B},
  volume={98},
  number={7},
  pages={075201},
  year={2018},
  publisher={APS},
  doi={10.1103/PhysRevB.98.075201}
}

@article{garcon2017cosmic,
  title={The cosmic axion spin precession experiment ({CASPEr}): a dark-matter search with nuclear magnetic resonance},
  author={Garcon, Antoine and Aybas, Deniz and Blanchard, John W and Centers, Gary and Figueroa, Nataniel L and Graham, Peter W and Kimball, Derek F Jackson and Rajendran, Surjeet and Sendra, Marina Gil and Sushkov, Alexander O and others},
  journal={Quantum Science and Technology},
  volume={3},
  number={1},
  pages={014008},
  year={2017},
  publisher={IOP Publishing},
  doi={10.1088/2058-9565/aa9861}
}

@article{beeks2021thorium,
  title={The thorium-229 low-energy isomer and the nuclear clock},
  author={Beeks, Kjeld and Sikorsky, Tomas and Schumm, Thorsten and Thielking, Johannes and Okhapkin, Maxim V and Peik, Ekkehard},
  journal={Nature Reviews Physics},
  volume={3},
  number={4},
  pages={238--248},
  year={2021},
  publisher={Nature Publishing Group UK London},
  doi={10.1038/s42254-021-00286-6}
}

@article{zhang2024frequency,
  title={Frequency ratio of the {${}^{229m}Th$} nuclear isomeric transition and the {${}^{87}Sr$} atomic clock},
  author={Zhang, Chuankun and Ooi, Tian and Higgins, Jacob S and Doyle, Jack F and von der Wense, Lars and Beeks, Kjeld and Leitner, Adrian and Kazakov, Georgy A and Li, Peng and Thirolf, Peter G and others},
  journal={Nature},
  volume={633},
  number={8028},
  pages={63--70},
  year={2024},
  publisher={Nature Publishing Group UK London},
  doi={10.1038/s41586-024-07839-6}
}

@article{ledbetter2012gyroscopes,
  title={Gyroscopes based on nitrogen-vacancy centers in diamond},
  author={Ledbetter, MP and Jensen, Kasper and Fischer, Ran and Jarmola, Andrey and Budker, Dmitry},
  journal={Physical Review A},
  volume={86},
  number={5},
  pages={052116},
  year={2012},
  publisher={APS},
  doi={10.1103/PhysRevA.86.052116}
}

@article{jarmola2021demonstration,
  title={Demonstration of diamond nuclear spin gyroscope},
  author={Jarmola, Andrey and Lourette, Sean and Acosta, Victor M and Birdwell, A Glen and Bl{\"u}mler, Peter and Budker, Dmitry and Ivanov, Tony and Malinovsky, Vladimir S},
  journal={Science advances},
  volume={7},
  number={43},
  pages={eabl3840},
  year={2021},
  publisher={American Association for the Advancement of Science},
  doi={10.1126/sciadv.abl3840}
}

@article{ajoy2012stable,
  title={Stable three-axis nuclear-spin gyroscope in diamond},
  author={Ajoy, Ashok and Cappellaro, Paola},
  journal={Physical Review A},
  volume={86},
  number={6},
  pages={062104},
  year={2012},
  publisher={APS},
  doi = {10.1103/PhysRevA.86.062104}
}

@article{kornack2005nuclear,
  title={Nuclear spin gyroscope based on an atomic comagnetometer},
  author={Kornack, Thomas W and Ghosh, Rajat K and Romalis, Michael V},
  journal={Physical review letters},
  volume={95},
  number={23},
  pages={230801},
  year={2005},
  publisher={APS},
  doi={10.1103/PhysRevLett.95.230801}
}

@article{jaskula2019cross,
  title={Cross-sensor feedback stabilization of an emulated quantum spin gyroscope},
  author={Jaskula, J-C and Saha, Kasturi and Ajoy, Ashok and Twitchen, Daniel J and Markham, Matthew and Cappellaro, Paola},
  journal={Physical review applied},
  volume={11},
  number={5},
  pages={054010},
  year={2019},
  publisher={APS},
  doi={10.1103/PhysRevApplied.11.054010}
}

@article{singh2025room,
  title={Room-temperature quantum sensing with photoexcited triplet electrons in organic crystals},
  author={Singh, Harpreet and D'Souza, Noella and Zhong, Keyuan and Druga, Emanuel and Oshiro, Julianne and Blankenship, Brian and Montis, Riccardo and Reimer, Jeffrey A and Breeze, Jonathan D and Ajoy, Ashok},
  journal={Physical Review Research},
  volume={7},
  number={1},
  pages={013192},
  year={2025},
  publisher={APS},
  doi={10.1103/PhysRevResearch.7.013192}
}

@article{jackson2016precessing,
  title={Precessing ferromagnetic needle magnetometer},
  author={Jackson Kimball, Derek F and Sushkov, Alexander O and Budker, Dmitry},
  journal={Physical review letters},
  volume={116},
  number={19},
  pages={190801},
  year={2016},
  publisher={APS},
  doi={10.1103/PhysRevLett.116.190801}
}

@article{geiger2020high,
  title={High-accuracy inertial measurements with cold-atom sensors},
  author={Geiger, Remi and Landragin, Arnaud and Merlet, S{\'e}bastien and Pereira Dos Santos, Franck},
  journal={AVS Quantum Science},
  volume={2},
  pages={024702},
  number={2},
  year={2020},
  publisher={AIP Publishing},
  doi={10.1116/5.0009093}
}

@article{gottscholl2021spin,
  title={Spin defects in {hBN} as promising temperature, pressure and magnetic field quantum sensors},
  author={Gottscholl, Andreas and Diez, Matthias and Soltamov, Victor and Kasper, Christian and Krau{\ss}e, Dominik and Sperlich, Andreas and Kianinia, Mehran and Bradac, Carlo and Aharonovich, Igor and Dyakonov, Vladimir},
  journal={Nature communications},
  volume={12},
  number={1},
  pages={4480},
  year={2021},
  publisher={Nature Publishing Group UK London},
  doi={10.1038/s41467-021-24725-1}
}

@article{wolf2015subpicotesla,
  title={Subpicotesla diamond magnetometry},
  author={Wolf, Thomas and Neumann, Philipp and Nakamura, Kazuo and Sumiya, Hitoshi and Ohshima, Takeshi and Isoya, Junichi and Wrachtrup, J{\"o}rg},
  journal={Physical Review X},
  volume={5},
  number={4},
  pages={041001},
  year={2015},
  publisher={APS},
  doi={10.1103/PhysRevX.5.041001}
}

@article{eisenach2021cavity,
  title={Cavity-enhanced microwave readout of a solid-state spin sensor},
  author={Eisenach, Erik R and Barry, John F and O’Keeffe, Michael F and Schloss, Jennifer M and Steinecker, Matthew H and Englund, Dirk R and Braje, Danielle A},
  journal={Nature communications},
  volume={12},
  number={1},
  pages={1357},
  year={2021},
  publisher={Nature Publishing Group UK London},
  doi={10.1038/s41467-021-21256-7}
}

@article{hart2021n,
  title={{N-V}--diamond magnetic microscopy using a double quantum 4-{Ramsey} protocol},
  author={Hart, Connor A and Schloss, Jennifer M and Turner, Matthew J and Scheidegger, Patrick J and Bauch, Erik and Walsworth, Ronald L},
  journal={Physical Review Applied},
  volume={15},
  number={4},
  pages={044020},
  year={2021},
  publisher={APS},
  doi={10.1103/PhysRevApplied.15.044020}
}

@article{maricq1987spin,
  title={Spin thermodynamics of periodically time-dependent systems: The quasistationary state and its decay},
  author={Maricq, M Matti},
  journal={Physical Review B},
  volume={36},
  number={1},
  pages={516},
  year={1987},
  publisher={APS},
  doi={10.1103/PhysRevB.36.516}
}

@article{kuwahara2016floquet,
  title={Floquet--Magnus theory and generic transient dynamics in periodically driven many-body quantum systems},
  author={Kuwahara, Tomotaka and Mori, Takashi and Saito, Keiji},
  journal={Annals of Physics},
  volume={367},
  pages={96--124},
  year={2016},
  publisher={Elsevier},
  doi={10.1016/j.aop.2016.01.012}
}

@article{weidinger2017floquet,
  title={Floquet prethermalization and regimes of heating in a periodically driven, interacting quantum system},
  author={Weidinger, Simon A and Knap, Michael},
  journal={Scientific reports},
  volume={7},
  number={1},
  pages={45382},
  year={2017},
  publisher={Nature Publishing Group UK London},
  doi={10.1038/srep45382}
}

@article{d2014long,
  title={Long-time behavior of isolated periodically driven interacting lattice systems},
  author={D’Alessio, Luca and Rigol, Marcos},
  journal={Physical Review X},
  volume={4},
  number={4},
  pages={041048},
  year={2014},
  publisher={APS},
  doi={10.1103/PhysRevX.4.041048}
}

@article{abanin2015exponentially,
  title={Exponentially slow heating in periodically driven many-body systems},
  author={Abanin, Dmitry A and De Roeck, Wojciech and Huveneers, Fran{\c{c}}ois},
  journal={Physical review letters},
  volume={115},
  number={25},
  pages={256803},
  year={2015},
  publisher={APS},
  doi={10.1103/PhysRevLett.115.256803}
}

@article{peng2021floquet,
  title={Floquet prethermalization in dipolar spin chains},
  author={Peng, Pai and Yin, Chao and Huang, Xiaoyang and Ramanathan, Chandrasekhar and Cappellaro, Paola},
  journal={Nature Physics},
  volume={17},
  number={4},
  pages={444--447},
  year={2021},
  publisher={Nature Publishing Group UK London},
  doi={10.1038/s41567-020-01120-z}
}

@article{haithamNVMWLockIn17,
author = {Haitham A. R. El-Ella and Sepehr Ahmadi and Adam M. Wojciechowski and Alexander Huck and Ulrik L. Andersen},
journal = {Opt. Express},
number = {13},
pages = {14809--14821},
publisher = {Optica Publishing Group},
title = {Optimised frequency modulation for continuous-wave optical magnetic resonance sensing using nitrogen-vacancy ensembles},
volume = {25},
month = {Jun},
year = {2017},
url = {https://opg.optica.org/oe/abstract.cfm?URI=oe-25-13-14809},
doi = {10.1364/OE.25.014809},
}

@unpublished{selco2025emergent,
  title={Emergent Decoherence Dynamics in Doubly Disordered Spin Networks}, 
  author={Cooper M. Selco and Christian Bengs and Chaitali Shah and Zhuorui Zhang and Ashok Ajoy},
  year={2025},
  eprint={2511.07785},
  archivePrefix={arXiv},
  primaryClass={quant-ph},
  url={https://arxiv.org/abs/2511.07785}
}

@article{magnus1954exponential,
  title={On the exponential solution of differential equations for a linear operator},
  author={Magnus, Wilhelm},
  journal={Communications on pure and applied mathematics},
  volume={7},
  number={4},
  pages={649--673},
  year={1954},
  publisher={Wiley Online Library},
  doi={10.1002/cpa.3160070404}
}

@book{haeberlen2012high,
  author    = {Haeberlen, Ulrich},
  title     = {High Resolution {NMR} in Solids: Selective Averaging},
  series    = {Advances in Magnetic Resonance, Supplement 1},
  publisher = {Academic Press},
  year      = {1976},
  address   = {New York},
  doi={10.1016/B978-0-12-025561-0.X5001-1}
}

@article{ostroff1966multiple,
  title={Multiple spin echoes and spin locking in solids},
  author={Ostroff, ED and Waugh, JS},
  journal={Physical Review Letters},
  volume={16},
  number={24},
  pages={1097},
  year={1966},
  publisher={APS},
 doi={10.1103/PhysRevLett.16.1097}
}

@article{rhim1976multiple,
  title={Multiple-pulse spin locking in dipolar solids},
  author={Rhim, W-K and Burum, D P and Elleman, D D},
  journal={Physical Review Letters},
  volume={37},
  number={26},
  pages={1764},
  year={1976},
  publisher={APS},
  doi={10.1103/PhysRevLett.37.1764}
}

@article{moon2025high,
  title = {High-speed, high-memory {NMR} spectrometer and hyperpolarizer},
  journal = {Journal of Magnetic Resonance},
  volume = {380},
  pages = {107952},
  year = {2025},
  issn = {1090-7807},
  doi = {10.1016/j.jmr.2025.107952},
  author = {Leo Joon Il Moon and William Beatrez and Jason Ball and Joan Mercade and Mark Elo and Angad Singh and Emanuel Druga and Ashok Ajoy}
}

@article{ho2023quantum,
  title={Quantum and classical Floquet prethermalization},
  author={Ho, Wen Wei and Mori, Takashi and Abanin, Dmitry A and Dalla Torre, Emanuele G},
  journal={Annals of Physics},
  volume={454},
  pages={169297},
  year={2023},
  publisher={Elsevier},
  doi={10.1016/j.aop.2023.169297}
}

@article{glenn2018high,
  title={High-resolution magnetic resonance spectroscopy using a solid-state spin sensor},
  author={Glenn, David R and Bucher, Dominik B and Lee, Junghyun and Lukin, Mikhail D and Park, Hongkun and Walsworth, Ronald L},
  journal={Nature},
  volume={555},
  number={7696},
  pages={351--354},
  year={2018},
  publisher={Nature Publishing Group UK London},
  doi={10.1038/nature25781}
}

@article{barry2024sensitive,
  title={Sensitive ac and dc magnetometry with nitrogen-vacancy-center ensembles in diamond},
  author={Barry, John F and Steinecker, Matthew H and Alsid, Scott T and Majumder, Jonah and Pham, Linh M and O’Keeffe, Michael F and Braje, Danielle A},
  journal={Physical Review Applied},
  volume={22},
  number={4},
  pages={044069},
  year={2024},
  publisher={APS},
  doi={10.1103/PhysRevApplied.22.044069}
}

@article{taylor2008high,
  title={High-sensitivity diamond magnetometer with nanoscale resolution},
  author={Taylor, Jacob M and Cappellaro, Paola and Childress, Lilian and Jiang, Liang and Budker, Dmitry and Hemmer, Philip Robert and Yacoby, Amir and Walsworth, Ronald and Lukin, Mikhail D},
  journal={Nature Physics},
  volume={4},
  number={10},
  pages={810--816},
  year={2008},
  publisher={Nature Publishing Group UK London},
  doi={10.1038/nphys1075}
}

@article{Degen17,
  title = {Quantum sensing},
  author = {Degen, C. L. and Reinhard, F. and Cappellaro, P.},
  journal = {Rev. Mod. Phys.},
  volume = {89},
  issue = {3},
  pages = {035002},
  numpages = {39},
  year = {2017},
  month = {Jul},
  publisher = {American Physical Society},
  doi = {10.1103/RevModPhys.89.035002},
  url = {https://link.aps.org/doi/10.1103/RevModPhys.89.035002}
}

@article{Ajoy19,
  title={Hyperpolarized relaxometry based nuclear {$T_1$} noise spectroscopy in diamond},
  author={Ajoy, A and Safvati, B and Nazaryan, R and Oon, J T and Han, B and Raghavan, P and Nirodi, R and Aguilar, A and Liu, K and Cai, X and Lv, X and Druga, E and Ramanathan, C and Reimer, J A and Meriles, C A and Suter, D and Pines, A},
  journal={Nature communications},
  volume={10},
  number={1},
  pages={5160},
  year={2019},
  publisher={Nature Publishing Group UK London},
  doi={10.1038/s41467-019-13042-3},
  url={https://www.nature.com/articles/s41467-019-13042-3},
}

@article{Sahin22,
  title={High field magnetometry with hyperpolarized nuclear spins},
  author={Sahin, Ozgur and de Leon Sanchez, Erica and Conti, Sophie and Akkiraju, Amala and Reshetikhin, Paul and Druga, Emanuel and Aggarwal, Aakriti and Gilbert, Benjamin and Bhave, Sunil and Ajoy, Ashok},
  journal={Nature communications},
  volume={13},
  number={1},
  pages={5486},
  year={2022},
  publisher={Nature Publishing Group UK London},
  url={https://www.nature.com/articles/s41467-022-32907-8},
  doi={10.1038/s41467-022-32907-8}
}

@article{sahin22_trajectory,
  title = {Micromotion-based {DC} sensing using continuously tracked trajectories of dipolar coupled nuclear spins},
  author = {Sahin, Ozgur and Al Asadi, Hawraa and Schindler, Paul M. and Pillai, Arjun and Sanchez, Erica and Markham, Matthew and Elo, Mark and McAllister, Maxwell and Druga, Emanuel and Fleckenstein, Christoph and Bukov, Marin and Ajoy, Ashok},
  journal = {Phys. Rev. Res.},
  volume = {7},
  issue = {4},
  pages = {043272},
  numpages = {27},
  year = {2025},
  month = {Dec},
  publisher = {American Physical Society},
  doi = {10.1103/bgdc-rgkd},
  url = {https://link.aps.org/doi/10.1103/bgdc-rgkd}
}

@article{Beatrez21_90s,
	title        = {Floquet Prethermalization with Lifetime Exceeding 90 s in a Bulk Hyperpolarized Solid},
	author       = {Beatrez, William and Janes, Otto and Akkiraju, Amala and Pillai, Arjun and Oddo, Alexander and Reshetikhin, Paul and Druga, Emanuel and McAllister, Maxwell and Elo, Mark and Gilbert, Benjamin and Suter, Dieter and Ajoy, Ashok},
	year         = {2021},
	month        = {Oct},
	journal      = {Phys. Rev. Lett.},
	publisher    = {American Physical Society},
	volume       = {127},
	pages        = {170603},
	doi          = {10.1103/PhysRevLett.127.170603},
	url          = {https://link.aps.org/doi/10.1103/PhysRevLett.127.170603},
	issue        = {17},
	numpages     = {8}
}

@article{DTC_Beatrez2022,
    author={Beatrez, William
    and Fleckenstein, Christoph
    and Pillai, Arjun
    and de Leon Sanchez, Erica
    and Akkiraju, Amala
    and Diaz Alcala, Jesus
    and Conti, Sophie
    and Reshetikhin, Paul
    and Druga, Emanuel
    and Bukov, Marin
    and Ajoy, Ashok},
    title={Critical prethermal discrete time crystal created by two-frequency driving},
    journal={Nature Physics},
    year={2023},
    month={Mar},
    day={01},
    volume={19},
    number={3},
    pages={407-413},
    issn={1745-2481},
    doi={10.1038/s41567-022-01891-7},
    url={https://doi.org/10.1038/s41567-022-01891-7}
}

@article{arradErrorCorrection14,
  title = {Increasing Sensing Resolution with Error Correction},
  author = {Arrad, G. and Vinkler, Y. and Aharonov, D. and Retzker, A.},
  journal = {Phys. Rev. Lett.},
  volume = {112},
  issue = {15},
  pages = {150801},
  numpages = {6},
  year = {2014},
  month = {Apr},
  publisher = {American Physical Society},
  doi = {10.1103/PhysRevLett.112.150801},
  url = {https://link.aps.org/doi/10.1103/PhysRevLett.112.150801}
}

@article{sarkarDroplet24,
author = {Adrisha Sarkar  and Zachary R. Jones  and Madhur Parashar  and Emanuel Druga  and Amala Akkiraju  and Sophie Conti  and Pranav Krishnamoorthi  and Srisai Nachuri  and Parker Aman  and Mohammad Hashemi  and Nicholas Nunn  and Marco D. Torelli  and Benjamin Gilbert  and Kevin R. Wilson  and Olga A. Shenderova  and Deepti Tanjore  and Ashok Ajoy },
title = {High-precision chemical quantum sensing in flowing monodisperse microdroplets},
journal = {Science Advances},
volume = {10},
number = {50},
pages = {eadp4033},
year = {2024},
doi = {10.1126/sciadv.adp4033},
URL = {https://www.science.org/doi/abs/10.1126/sciadv.adp4033}
}

@article{amitonovaSTED15,
author = {L. V. Doronina-Amitonova and I. V. Fedotov and A. M. Zheltikov},
journal = {Opt. Lett.},
number = {5},
pages = {725--728},
publisher = {Optica Publishing Group},
title = {Ultrahigh-contrast imaging by temporally modulated stimulated emission depletion},
volume = {40},
month = {Mar},
year = {2015},
url = {https://opg.optica.org/ol/abstract.cfm?URI=ol-40-5-725},
doi = {10.1364/OL.40.000725}}

@article{acostaNVTemperature10,
  title = {Temperature Dependence of the Nitrogen-Vacancy Magnetic Resonance in Diamond},
  author = {Acosta, V. M. and Bauch, E. and Ledbetter, M. P. and Waxman, A. and Bouchard, L.-S. and Budker, D.},
  journal = {Phys. Rev. Lett.},
  volume = {104},
  issue = {7},
  pages = {070801},
  numpages = {4},
  year = {2010},
  month = {Feb},
  publisher = {American Physical Society},
  doi = {10.1103/PhysRevLett.104.070801},
  url = {https://link.aps.org/doi/10.1103/PhysRevLett.104.070801}
}

@article{fangDQ13,
  title = {High-Sensitivity Magnetometry Based on Quantum Beats in Diamond Nitrogen-Vacancy Centers},
  author = {Fang, Kejie and Acosta, Victor M. and Santori, Charles and Huang, Zhihong and Itoh, Kohei M. and Watanabe, Hideyuki and Shikata, Shinichi and Beausoleil, Raymond G.},
  journal = {Phys. Rev. Lett.},
  volume = {110},
  issue = {13},
  pages = {130802},
  numpages = {5},
  year = {2013},
  month = {Mar},
  publisher = {American Physical Society},
  doi = {10.1103/PhysRevLett.110.130802},
  url = {https://link.aps.org/doi/10.1103/PhysRevLett.110.130802}
}

@article{levittCompositePi79,
title = {{NMR} population inversion using a composite pulse},
journal = {Journal of Magnetic Resonance (1969)},
volume = {33},
number = {2},
pages = {473-476},
year = {1979},
issn = {0022-2364},
doi = {https://doi.org/10.1016/0022-2364(79)90265-8},
url = {https://www.sciencedirect.com/science/article/pii/0022236479902658},
author = {Malcolm H. Levitt and Ray Freeman}
}

@article{gevorgyanCompensatedGates21,
  title = {Ultrahigh-fidelity composite rotational quantum gates},
  author = {Gevorgyan, Hayk L. and Vitanov, Nikolay V.},
  journal = {Phys. Rev. A},
  volume = {104},
  issue = {1},
  pages = {012609},
  numpages = {12},
  year = {2021},
  month = {Jul},
  publisher = {American Physical Society},
  doi = {10.1103/PhysRevA.104.012609},
  url = {https://link.aps.org/doi/10.1103/PhysRevA.104.012609}
}

@article{klein1993young,
  title={Young's modulus and {Poisson's} ratio of {CVD} diamond},
  author={Klein, Claude A and Cardinale, Gregory F},
  journal={Diamond and Related Materials},
  volume={2},
  number={5-7},
  pages={918--923},
  year={1993},
  publisher={Elsevier},
  doi = {10.1016/0925-9635(93)90250-6}
}

@article{steffenPulseShape07,
  title = {Shaped pulses for quantum computing},
  author = {Steffen, Matthias and Koch, Roger H.},
  journal = {Phys. Rev. A},
  volume = {75},
  issue = {6},
  pages = {062326},
  numpages = {4},
  year = {2007},
  month = {Jun},
  publisher = {American Physical Society},
  doi = {10.1103/PhysRevA.75.062326},
  url = {https://link.aps.org/doi/10.1103/PhysRevA.75.062326}
}

@article{daemsPulseShape13,
  title = {Robust Quantum Control by a Single-Shot Shaped Pulse},
  author = {Daems, D. and Ruschhaupt, A. and Sugny, D. and Gu\'erin, S.},
  journal = {Phys. Rev. Lett.},
  volume = {111},
  issue = {5},
  pages = {050404},
  numpages = {5},
  year = {2013},
  month = {Jul},
  publisher = {American Physical Society},
  doi = {10.1103/PhysRevLett.111.050404},
  url = {https://link.aps.org/doi/10.1103/PhysRevLett.111.050404}
}

@article{zopesPulseShape17,
  title = {High-Resolution Quantum Sensing with Shaped Control Pulses},
  author = {Zopes, J. and Sasaki, K. and Cujia, K. S. and Boss, J. M. and Chang, K. and Segawa, T. F. and Itoh, K. M. and Degen, C. L.},
  journal = {Phys. Rev. Lett.},
  volume = {119},
  issue = {26},
  pages = {260501},
  numpages = {5},
  year = {2017},
  month = {Dec},
  publisher = {American Physical Society},
  doi = {10.1103/PhysRevLett.119.260501},
  url = {https://link.aps.org/doi/10.1103/PhysRevLett.119.260501}
}

@article{dualTransitionAtomic23,
title = {Frequency shift compensation for single and dual laser beam pass sensors of a coherent population trapping resonance based coupled dark state magnetometer},
journal = {Measurement: Sensors},
volume = {25},
pages = {100606},
year = {2023},
issn = {2665-9174},
doi = {https://doi.org/10.1016/j.measen.2022.100606},
url = {https://www.sciencedirect.com/science/article/pii/S2665917422002409},
author = {Michaela Ellmeier and Christoph Amtmann and Andreas Pollinger and Werner Magnes and Christian Hagen and Alexander Betzler and Irmgard Jernej and Martín Agú and Laurentius Windholz and Roland Lammegger},
}

@article{dingDualSpecies23,
  title = {Dual-Species All-Optical Magnetometer Based on a {Cs-K} Hybrid Vapor Cell},
  author = {Ding, Yudong and Xiao, Wei and Zhao, Yixin and Wu, Teng and Peng, Xiang and Guo, Hong},
  journal = {Phys. Rev. Appl.},
  volume = {19},
  issue = {3},
  pages = {034066},
  numpages = {8},
  year = {2023},
  month = {Mar},
  publisher = {American Physical Society},
  doi = {10.1103/PhysRevApplied.19.034066},
  url = {https://link.aps.org/doi/10.1103/PhysRevApplied.19.034066}
}

@article{wangDualTransition24,
  title = {Magnetic induction sensor based on a dual-frequency atomic magnetometer},
  author = {Wang, Hengyan and Zugenmaier, Michael and Jensen, Kasper and Zheng, Wenqiang and Polzik, Eugene S.},
  journal = {Phys. Rev. Appl.},
  volume = {22},
  issue = {3},
  pages = {034030},
  numpages = {9},
  year = {2024},
  month = {Sep},
  publisher = {American Physical Society},
  doi = {10.1103/PhysRevApplied.22.034030},
  url = {https://link.aps.org/doi/10.1103/PhysRevApplied.22.034030}
}
\vspace{-1mm}

\clearpage
\appendix
\setcounter{figure}{0}
\renewcommand{\thefigure}{S\arabic{figure}} 

%%%%%%%%%%%%%%%%%%%%%%%%%%%%%%%%
%%% SI
%%%%%%%%%%%%%%%%%%%%%%%%%%%%%%%%
%%%%%%%%%%%%%%%%%%%%%%%%%%%%%%%%%%%%%%%%%%%%%%%%%%%%%%%%%%%%%%%%%%%%%%%%%%%
%                        SI
%%%%%%%%%%%%%%%%%%%%%%%%%%%%%%%%%%%%%%%%%%%%%%%%%%%%%%%%%%%%%%%%%%%%%%%%%%%
%%%%%%%%%%%%%%%%%%%%%%%%%%%%%%%%%%%%%%%%%%%%%%%%%%%%%%%%%%%%%%%%%%%%%%%%%%%

\clearpage
 \onecolumngrid
%\begin{widetext}
\begin{center}
\textbf{\large{\textit{Supplemental Information:} \\ \smallskip Robust Quantum Sensing via Prethermal Spin Orbits}} \\\smallskip
\end{center}

\twocolumngrid

\tableofcontents

\setcounter{table}{0}
\setcounter{figure}{0}

\renewcommand{\thetable}{S\arabic{table}}
\renewcommand{\thefigure}{S\arabic{figure}}

\renewcommand{\theHtable}{S\thetable}
\renewcommand{\theHfigure}{S\thefigure}

%%%%%%%%%%%%%%%%%%%%%%%%%%%%%%%
%            Text
%%%%%%%%%%%%%%%%%%%%%%%%%%%%%%%
%%%%%%%%%%%%%%%%%%%%%%%%%%%%%%%

\section*{Organization of the Supplementary Information}

\noindent
This Supplementary Information provides detailed results, theoretical background, and methods supporting the findings in the main text. We begin by summarizing key performance metrics and robustness benchmarks (App.~\ref{secSI:performancemetrics}), followed by a comparison to other quantum sensing methods (App.~\ref{secSI:existingtechniques}) and a characterization of the sensor's wide and flat operational bandwidth (App.~\ref{secSI:sensorresponse}). We then demonstrate the protocol's capability for dynamic sensing of rapidly varying magnetic fields using a sound wave (App.~\ref{secSI:soundsignal}) and show how prethermal lifetimes are extended under the orbit drive (App.~\ref{secSI:lifetime}). The influence of the pulse angle is explored via an experimental phase diagram (App.~\ref{secSI:phasediagram}), which is followed by a theoretical description of the prethermal spin orbits (App.~\ref{secSI:theorytrajectories}).

Subsequently, we present a comprehensive analysis of the sensor's robustness against a wide range of perturbations. This includes the origin and suppression of transients (App.~\ref{secSI:transients}), a detailed study of immunity to mechanical vibration (App.~\ref{secSI:vibrationNote}), bias field drifts and off-resonant driving (App.~\ref{secSI:RobustnessDCOffRes}), and control pulse imperfections (App.~\ref{SIMethods:pulseangle}). We also document the sensor's performance across wide variations in temperature, from which we infer its robustness to strain (App.~\ref{secSI:tempstrain}). This analysis concludes by characterizing the susceptibility of the RF resonator to external signals (App.~\ref{secSI:RFsusceptibility}) and detailing the differential signal extraction principle that underlies the protocol's robustness to background fields (App.~\ref{secSI:signalextraction}).

Following the robustness analysis, we provide a characterization of the sensor's sensitivity (App.~\ref{secSI:sensitivity}) and discuss pathways to further increase its performance (App.~\ref{secSI:optimization}).

The Supplementary Methods section provides detailed experimental and theoretical procedures, including the reconstruction of the 3D magnetization vector (App.~\ref{secSI:3Dextraction}), the formal definition and evaluation of the suppression factor (App.~\ref{secSI:evalsuppfactor}), the protocol for vibration testing (App.~\ref{SI:MeasurementOfVibration}), details on the pulse sequence timing (App.~\ref{secSI:timingzdrive}), and the numerical methods used for simulating the spin dynamics (App.~\ref{secSI:simspin}).

\section*{\T{Supplementary Notes}}

\section{Performance Metrics and Robustness} \label{secSI:performancemetrics}

    Table~\ref{tab:robustness} summarizes the key performance characteristics of the PRISM measurement protocol, including frequency range, background suppression, 
    and tolerance to various experimental imperfections. The data is presented as typical values achieved under standard conditions, 
    the best values demonstrated in this work, and projections for possible future improvements.

    Notably, the protocol enables broadband, simultaneous detection of all frequencies within the measurement range without retuning, while maintaining strong immunity to bias-field fluctuations, pulse errors, and mechanical vibrations.

    \begin{table*}[t]
        \centering
        \caption{Performance summary of the PRISM measurement protocol.}
        \includegraphics[width=\textwidth]{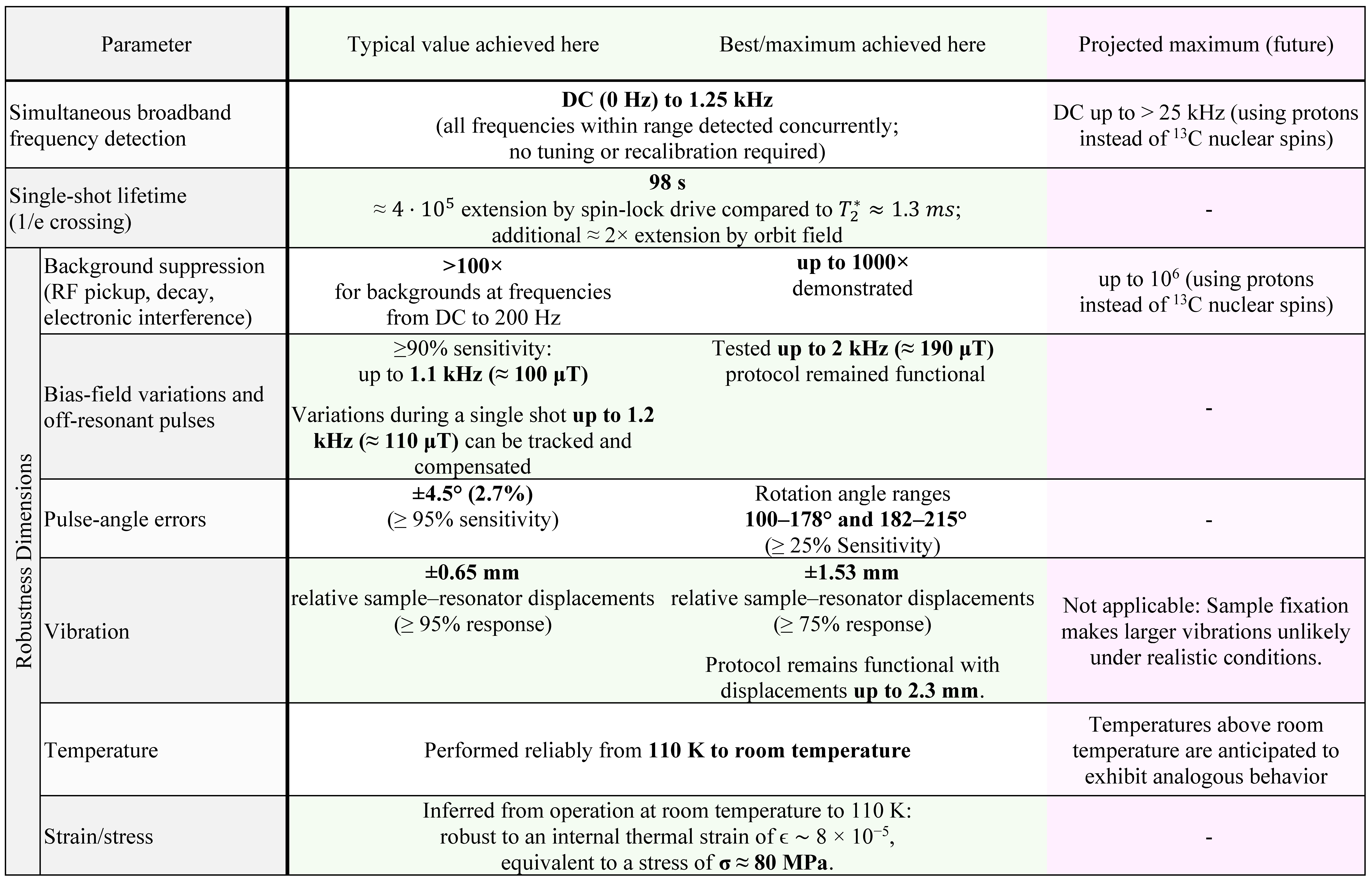}
    
    \label{tab:robustness}
\end{table*}

\begin{table*}[t]
\centering
\small
\setlength{\tabcolsep}{4pt}
\renewcommand{\arraystretch}{1.2}
\caption{\textbf{Comparison to other methods.} 
    Summary of desirable robustness attributes afforded by this and other selected methods, as discussed in above text. The darker green color indicates robustness yielded by toggling the spins between the two axes, while lighter green color indicates robustness inherent to the prethermal nuclear spin platform. Yellow indicates that robustness is dependent on specific realization of the method. Robustness to bias drift for wideband sensing is realized if the sensor can reconstruct the DC field. The robustness of dynamical decoupling sequences to bias offset or pulse angle is subject to precise sequence design. Additionally, micromotion in the dynamical decoupling sequence could provide signal modulation for background discrimination, but this possibility depends on the sequence design as well as phase-sensitivity of the measurement. Referenced techniques are: modulation~\cite{amitonovaSTED15, sarkarDroplet24, haithamNVMWLockIn17}, dual transition sensing~\cite{dingDualSpecies23, dualTransitionAtomic23, wangDualTransition24}, double quantum coherence~\cite{fangDQ13, hart2021n}, wideband sensing \cite{zeng2024wide, aiello2013composite}, dynamical decoupling pulse engineering~\cite{hiroseCDD12, louzonCPDD25, levittCompositePi79, gevorgyanCompensatedGates21, steffenPulseShape07, daemsPulseShape13, zopesPulseShape17},  and quantum error correction~\cite{unden2016quantum, arradErrorCorrection14}.}
    \includegraphics[width=0.97\textwidth]{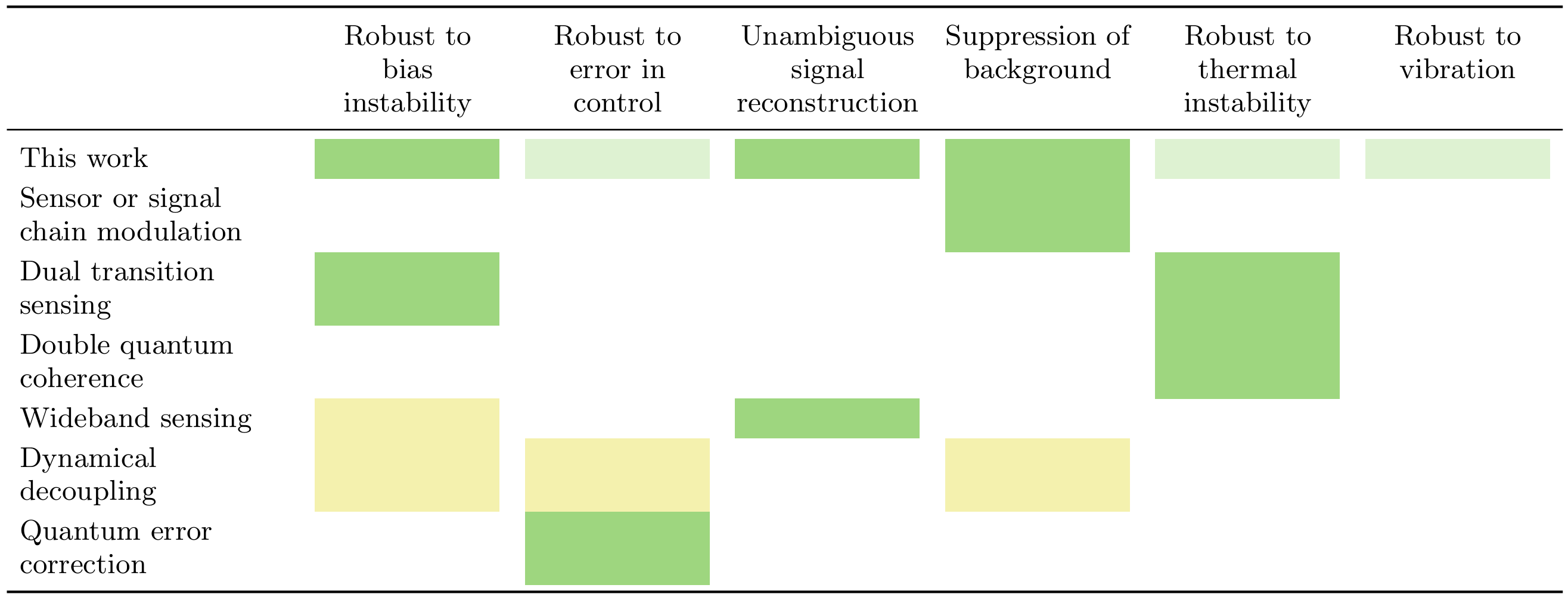}
\label{tab:method-comparison}
\vspace{2em}
\end{table*}

\section{Comparison to Other Robust Quantum Sensing Methods} \label{secSI:existingtechniques}

This paper demonstrates several key attributes of robust sensing, namely unambiguous signal reconstruction, suppression of backgrounds that contaminate spin readout, and robustness to fluctuations in bias, pulse control, temperature, and sensor position. Here, we compare the PRISM method to selected established techniques which achieve subsets of these desirable characteristics. We restrict to methods of deriving robustness instrinsically from control and readout protocols. Methods of external stabilization, by isolating from instabilities or compensating instrumentation with active feedback are not considered. Likewise, we do not further detail gradiometric measurements between two or more sensors, which offer an established means to reject unwanted instabilities, dependent on coupling between the sensors and signal/interference sources. Our technique offers similar attributes of a gradiometer realized in a single sensor.  Comparisons are summarized in Table~\ref{tab:method-comparison}.

Reconstruction of magnetic field signals can be complicated by spurious sensor response, especially in resonant sensing schemes~\cite{loretz2015spurious}. Continuous dynamical decoupling is one method to limit unwanted spectral response away from a narrow band around the target signal frequency~\cite{hiroseCDD12, louzonCPDD25}. Wideband approaches have also been demonstrated that allow robust signal reconstruction \cite{zeng2024wide, aiello2013composite}. A key benefit of PRISM is the flattening of signal response throughout the wide bandwidth of the sensor, including to DC signal. The known trajectory dynamics lift the degeneracy between lower magnetization and greater tilt of the spin-lock axis, allowing for internal calibration of the bias field. Without the trajectory protocol, a static or slowly varying tilt of the prethermal axis by an external field cannot be disentangled from the shot-to-shot variation in the absolute signal intensity and relaxation time. Thus the trajectory protocol mitigates the systematic uncertainty in signal reconstruction throughout the sensing bandwidth due to the bias-dependent response illustrated in \zfr{fig5}a. The technique also disambiguates reconstruction in the high frequency regime (${>}$\SI{1}{\kilo\hertz}) by suppression of transient dynamics, as shown in \zfr{fig4}a. 

The robustness to bias field drift in the wideband sensing approach stems from the capacity to directly measure and thus calibrate out effects of bias offset. Alternatively, dynamical decoupling attains AC sensitivity with echo sequences that rectify the AC signal in the interaction frame of the spins, simultaneously canceling DC signal. In the ideal scenario of instantaneous pulses, this can protect the sensor from bias offset. To negate the effect of offset bias during the finite pulse width, more sophisticated sequence design is required. This is particularly critical in the limit of continuous dynamical decoupling that otherwise conveys advantages of simpler spectral response as mentioned above. Incorporation of dual species or transitions also serves to mitigate bias instability. Demonstrations in atomic magnetometers show that dual transition techniques can suppress performance degradation due to nonlinear magnetic response \cite{dingDualSpecies23, dualTransitionAtomic23, wangDualTransition24}.

The PRISM protocol not only improves the robustness of signal reconstruction through a flattened broadband response, but it also serves to modulate the signal. This allows RF backgrounds and other readout interferences to be distinguished from the spin signal. In principle, a similar modulation effect can be achieved via phase cycling. Our method not only confers the additional benefit over phase cycling of flattened signal response, but is also advantageous in offering rapid modulation in a long-lived sensor. The long-lived prethermal sensing approach is contingent on the preparation of a fixed prethermal axis. Thus, rapid phase cycling would quickly destroy coherence, requiring re-initialization to revive the signal. Phase cycling between multiple shots, which would preserve the long-lived prethermal state, would only modulate the signal across timescales of more than a minute. Only very static backgrounds could be suppressed by such slow modulation. 

The protocol described here allows rapid signal modulation at $\sim$kHz frequencies, corresponding to half the pulse repetition frequency. Other modulation techniques, such as chopping \cite{amitonovaSTED15, sarkarDroplet24}, are common for example in readout of NV center fluorescence. In such an approach any stray backgrounds downstream of the chopper or modulating mechanism remain unmodulated and therefore can be rejected. With PRISM, modulation is internal to the spin system itself, so there is no opportunity for background to enter between the sensor and the modulator. Similar modulation at the sensor level can be achieved via other established techniques, for example in NV centers by modulation of the microwave drive \cite{haithamNVMWLockIn17}.

The described sensor also inherits additional elements of robustness due to the platform method of prethermal spin-lock sensing in a nuclear spin ensemble. Thermal response can mimic magnetic signals in other sensors. For example, the zero-field splitting of NV centers   is temperature-dependent, rendering NV measurements based on single quantum coherence simultaneously sensitive to both magnetic field and temperature \cite{acostaNVTemperature10}. Double quantum coherence techniques have been developed to produce NV magnetometers that are insensitive to temperature \cite{fangDQ13, hart2021n}. Similarly, compensation of temperature effects in atomic magnetometers has been demonstrated by measurement of multiple transitions \cite{dualTransitionAtomic23}. In our case, the sensor is inherently robust to temperature because the Zeeman splitting of the two-level $^{13}$C nuclei is insensitive to temperature, as shown in \zfr{fig5}e.  

Errors in the control sequence also can affect performance of quantum sensors, by causing decoherence or producing systematic error in signal reconstruction. Approaches such as quantum error correction~\cite{unden2016quantum, arradErrorCorrection14} or pulse engineering in dynamically decoupled sequences (e.g. using composite ~\cite{levittCompositePi79, gevorgyanCompensatedGates21} or shaped pulses \cite{steffenPulseShape07, daemsPulseShape13, zopesPulseShape17}) can mitigate control error. Phase cycling is also useful for error suppression - for example, it has been demonstrated for suppressing residual single quantum coherence in double quantum magnetometry with NV centers~\cite{hart2021n}. In our sensor, static detuning of the resonance frequency used for RF control of spins manifests equally to a static bias field, and thus can be internally calibrated as discussed above. The coherence of the sensor remains long-lived for any spin-lock pulse angle away from $\pi$. Pulse angle error affects the splitting of the trajectory manifolds, as illustrated in \zfr{fig5}c, but does not produce a net tilt that mimics a magnetic signal. The splitting of the trajectories does affect the response of the sensor, but we observe a regime with flat dependence of trajectory splitting on pulse angle, indicating the effect of error is absent at first order. The robustness to control error also produces robustness to vibration in our sensor, as displacement of the sensor in the inhomogeneity of $B_0$ and $B_1$ fields manifests as detuning or pulse angle error respectively.

\section{Wide and Flat Sensor Bandwidth} \label{secSI:sensorresponse}
        The PRISM sensing protocol exhibits a very broad frequency response, and therefore its response must be characterized across the entire frequency spectrum.  

        \begin{figure}[t]
            \centering
            \includegraphics[width=0.49\textwidth]{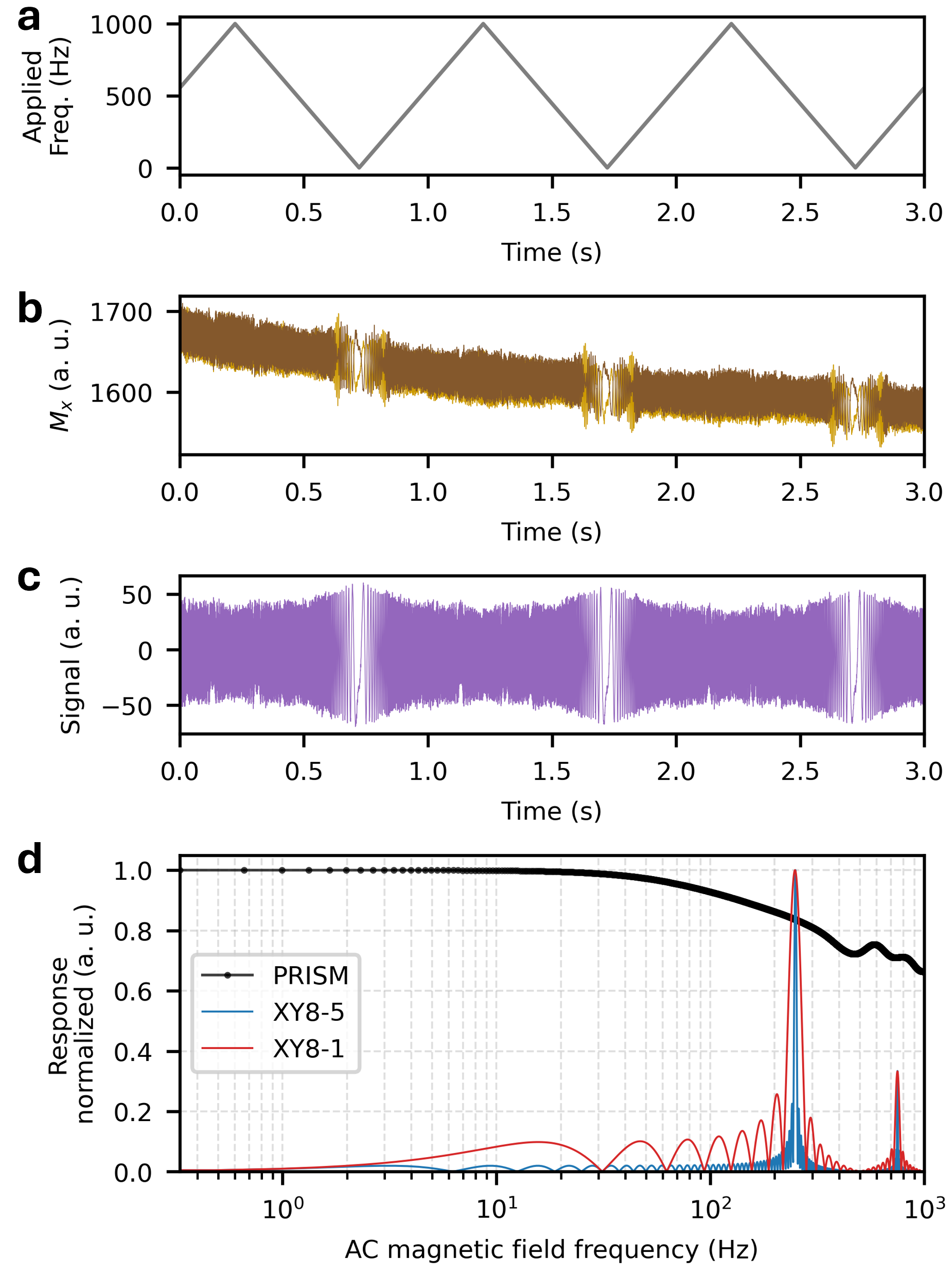}
            \caption{\textbf{Frequency‑dependent response of the sensor.}
                (\textbf{a}) Frequency of the modulated signal supplied to the auxiliary AC coil. 
                (\textbf{b}) Raw time‑domain data acquired using the PRISM protocol. Alternating points are colored gold/brown.
                (\textbf{c}) Normalized differential signal extracted from data in panel b.  
                (\textbf{d}) Comparison of the response function (filter function) obtained with the PRISM method presented in this work to that of the XY8 dynamical decoupling sequence.  
                Shown is the Fourier spectrum from panel~c (PRISM), smoothed with a Gaussian filter ($\sigma = \SI{50}{\hertz}$) to transform discrete peaks into a spectral‑density‑like representation (black dots connected by a black line).  
                For comparison, the calculated response functions of a single‑cycle (XY8‑1) and a five‑cycle (XY8‑5) AC measurement using the XY8 protocol are also shown, assuming $\delta$‑pulses and a spacing of \SI{2}{\milli\second} between pulses.
            }
            \label{figSI:response}
        \end{figure}

        Determining the frequency or phase of an unknown AC magnetic field requires no prior calibration, as these quantities can be read directly from the recorded time‑domain signal.  
        In contrast, quantitative measurement of the field amplitude does require calibration to convert the measured signal strength from arbitrary units (a. u.) into physical units such as nT.
        
        Several calibration strategies are possible.
        One approach is to inject a reference signal of known amplitude and frequency into the sensor simultaneously with the measurement.  
        The ratio between the known amplitude and the measured signal strength yields a calibration factor that can be used to convert a. u. values into nT.
        
        Alternatively, the sensor can be calibrated prior to the measurement by determining its response function.  
        To obtain this response function, an AC magnetic field was applied during a single shot while its frequency was swept from \SIrange{0}{1000}{\hertz}, using a triangular modulation function at \SI{1}{\hertz} (Fig.~\ref{figSI:response}a-b).  
        From the time‑domain signal (Fig.~\ref{figSI:response}c) obtained via the normalized differential signal method (App.~\ref{secSI:effectofdecay}), the magnitude of the FFT peaks is extracted and smoothed using a Gaussian filter ($\sigma = \SI{50}{\hertz}$) to turn the individual peaks into a spectral density type plot (Fig.~\ref{figSI:response}d).  
        
        The sensor’s response is relatively uniform over the measured frequency range, with a slight decrease at high frequencies, most likely due to transients (cf. App.~\ref{secSI:theorytrajectories}), which have a mildly damping effect at higher frequencies.
        
        From the measured response, a frequency‑dependent calibration factor can be derived and subsequently applied to convert a.~u. values into nT.  
        By additionally extracting the elevation angle at the start of each experiment, as described in App.~\ref{secSI:3Dextraction}, small shot‑to‑shot deviations in the elevation angle can be incorporated into the calibration factor.  
        In the shown data, this correction was omitted, as the observed shot‑to‑shot fluctuations in the elevation angle were negligible.

\section{Dynamic Sensing of Rapidly Varying Magnetic Fields} \label{secSI:soundsignal}
        \begin{figure*}[t]
            \centering
            \includegraphics[width=0.9\textwidth]{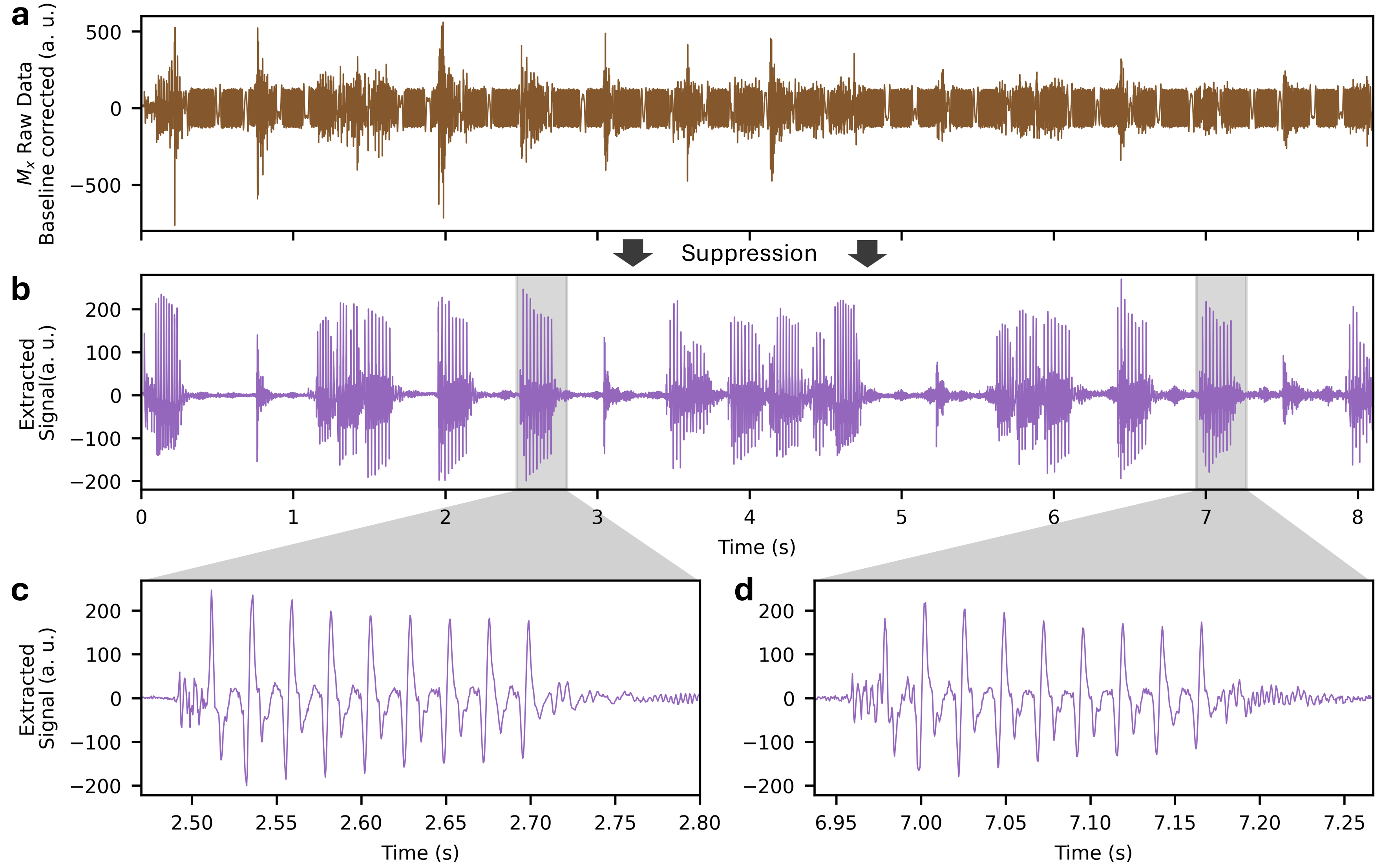}
            \caption{
                \textbf{Single-shot acquisition of an acoustic signal in the presence of a strong radio-frequency background and subsequent signal extraction.}
                Acquisition time per sample: \SI{76}{\micro\second}; sampling rate: \SI{2453.9}{\hertz}; pulse duration: \SI{103.73}{\micro\second}; pulse separation: \SI{100.03}{\micro\second}.
                (\textbf{a}) $M_x$ component of the magnetization of one of the two manifolds obtained in trajectory sensing, equivalent to a conventional measurement without trajectory. 
                (\textbf{b}) Reconstructed signal after suppression of the background via manifold subtraction and baseline normalization. 
                (\textbf{c}) Close-up view of the extracted acoustic waveform between $2.47$ and \SI{2.80}{\second}. 
                (\textbf{d}) Close-up of the same waveform as in (c) at a later time segment between $6.94$ and \SI{7.27}{\second}.
            }
            \label{figSI:soundmeasurementlong}
        \end{figure*}
        
        To evaluate the capability of the proposed PRISM detection scheme for tracking rapidly varying magnetic fields, we measured a time-dependent signal corresponding to a \SI{\approx 4.5}{\second} long acoustic waveform. The waveform was periodically applied as an AC magnetic field and recorded in a single-shot measurement. The peak amplitude of the AC magnetic signal was \SI{37}{\micro\tesla}. The pulse length was $\tau_{\mathrm{p}}=\SI{103.73}{\micro\second}$ and the interpulse delay $\tau_s = \SI{100.03}{\micro\second}$, corresponding to an effective sampling rate of \SI{2.454}{\kilo\hertz} (Nyquist limit \SI{1.227}{\kilo\hertz}). Each acquisition window (\SI{76}{\micro\second}) began \SI{12}{\micro\second} after the $\xhat$-pulse.
        
        For testing robustness against strong backgrounds, the signal was superimposed with an artificially generated radio-frequency (RF) background. The background consisted of a sinusoidal waveform with a triangular frequency modulation at \SI{2.5}{\hertz}, with a deviation of $\pm \SI{400}{\hertz}$ around the Larmor precession frequency of $\Cs$. After the measurement, the signal of one of the two axes was subtracted by its baseline to save it as a sound file (Fig.~\ref{figSI:soundmeasurementlong}a). This raw signal, containing the strong background contamination, is provided as \hyperref[SIaudio1]{Supplementary Audio 1}.
        
        Following manifold subtraction and normalization to the baseline, the original acoustic waveform can be reconstructed with high fidelity (Fig.~\ref{figSI:soundmeasurementlong}b), even under strong background contamination. This cleaned signal, demonstrating the high-fidelity waveform recovery, is available as \hyperref[SIaudio2]{Supplementary Audio 2}. The magnified views in Fig.~\ref{figSI:soundmeasurementlong}c,d illustrate the recovery of fine temporal features at two distinct intervals in the acquisition. Weak residual background artefacts remain visible at the end of the shown intervals. Notably, transient high-frequency components—such as the initial drum hit—cannot be fully recovered due to their Fourier content exceeding the Nyquist limit. Increasing the sampling rate (App.~\ref{secSI:increasesamplingrate}) would mitigate this limitation.

\section{Extension of Prethermal Lifetimes Under Orbit Drives} \label{secSI:lifetime}
            \begin{figure}[t]
            \centering
            \includegraphics[width=0.49\textwidth]{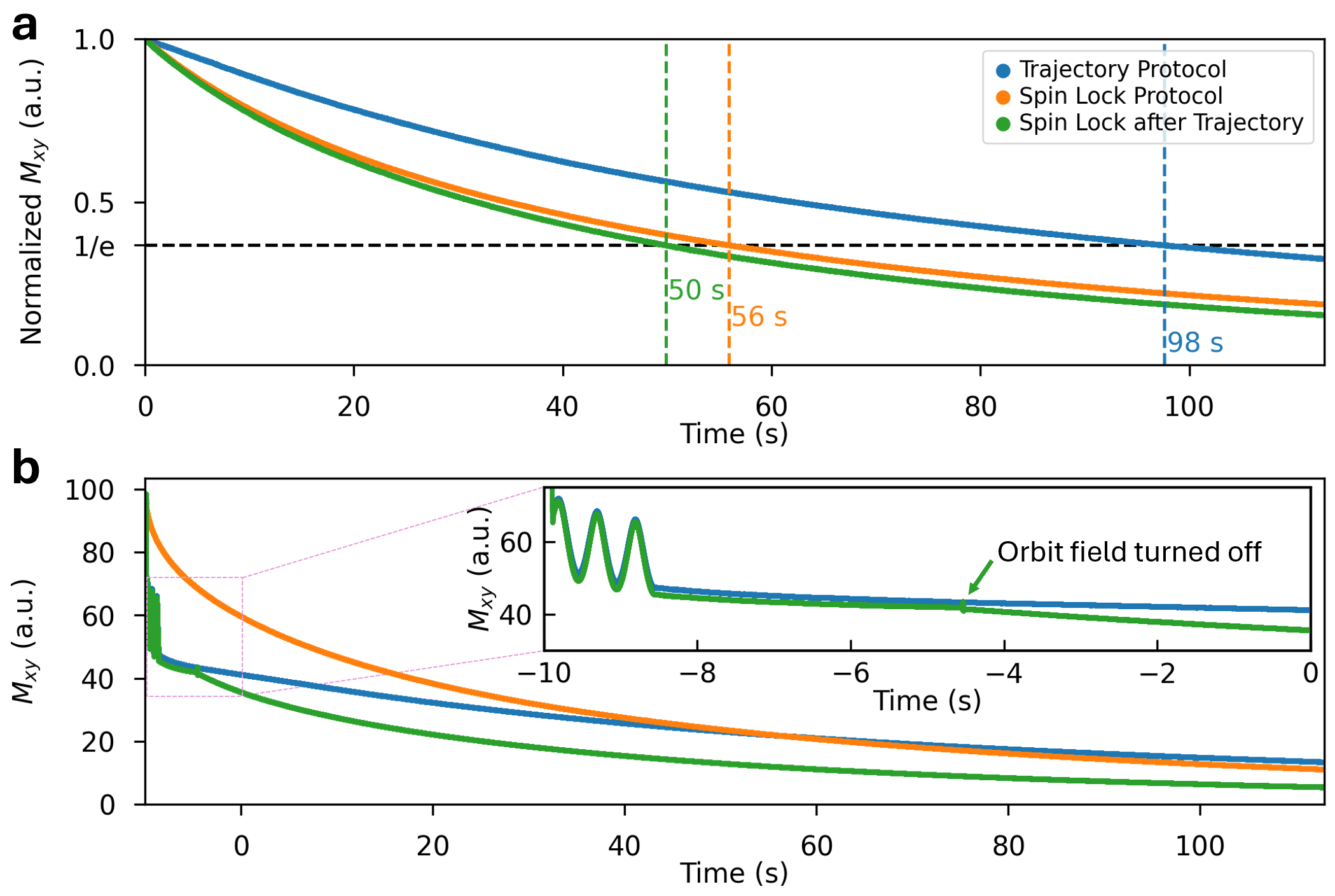}
            \caption{
                \textbf{Lifetime of the magnetization.}  
                (\textbf{a}) Decay of the normalized signal amplitude for continuous spin lock without $\zhat$-drive (orange), for spin lock following brief trajectory activation (green), and for driven spin trajectories under the orbit field ($\zhat$-drive, blue). Lifetimes $T'_2$ are defined as the time at which the amplitude falls to $1/e$ of its initial value. The time axis origin is set to \SI{10}{\second} after the initial excitation pulse. Activation of the orbit field increases $T'_2$ by a factor of 1.75 relative to pure spin lock.  
                (\textbf{b}) Raw (unnormalized) decay signals for the same datasets and color scheme as in panel~a. The initial signal loss upon trajectory activation is visible at $t=\SI{-10}{\second}$. Due to the extended lifetime provided by the orbit field, the trajectory-driven signal surpasses the spin-lock signal after ${\sim}\SI{70}{\second}$.  
                \textbf{Inset:} Initialization of the trajectory by a short calibration rotation from ${t=}\SI{-10}{\second}$ to ${t}\SI{\approx-8.5}{\second}$, followed by positioning both prethermal eigenstates near the $\xhat\zhat$-plane with maximal elevation. For the green dataset, the orbit field was switched off at ${t}\SI{\approx-4.5}{\second}$, resulting in a subsequent decay rate matching that of pure spin lock.
            }
            \label{figSI:lifetime}
        \end{figure}
        
        The magnetization lifetime, $T'_2$, characterizes the rate at which the measured signal decays and sets the maximum continuous measurement interval before reinitialization is required. Longer lifetimes directly enable extended sensing durations and improve achievable sensitivity at later times during a single shot.
        
        To isolate the influence of the orbit field ($\zhat$-drive), which generates the spin trajectory, we compared its performance with a conventional spin-lock protocol in which the orbit field was disabled. Here, $T'_2$ is defined as the time at which the detected amplitude drops to $1/e$ of its initial value. All measurements were performed sequentially under identical experimental conditions, including matching the spin-lock pulse angle of the $\xhat$-drive across protocols.
        
        Activating the orbit field extended $T'_2$ by a factor of $1.75$ relative to pure spin lock (Fig.~\ref{figSI:lifetime}a, blue vs.\ orange). In a control test, the orbit field was applied only briefly to initialize the spin trajectory and then switched off after a few seconds (Fig.~\ref{figSI:lifetime}b, inset). The primary purpose of this test was to determine whether the observed extension of $T'_2$ in the orbit-field-driven protocol might result from site-selective relaxation, i.e., certain ${}^{13}\mathrm{C}$ nuclear spin sites decaying more rapidly upon trajectory activation and thereby altering the overall decay profile. The control data shows that this effect, if present, is negligible: the measured lifetime immediately reduced to ${\sim}\SI{50}{\second}$, in close agreement with the pure spin-lock value of \SI{56}{\second} (green vs.\ orange in Fig.~\ref{figSI:lifetime}a). This confirms that the lifetime enhancement arises from the dynamics induced by the sustained orbit field, rather than from preferential decay of specific spin sites.
        
        Under the application of weak sinusoidal AC magnetic fields, the measured trajectory lifetime was indistinguishable from the orbit-field-driven case without sensing (blue traces in Fig.~\ref{figSI:lifetime}a), confirming that such weak fields do not measurably perturb the decay dynamics.
        
        Orbit-field activation produces a rapid transient drop in magnetization, arising from two effects: (i) the sudden modification of the effective Hamiltonian, and (ii) elevation of the magnetization vector, which reduces its transverse ($\xhat\tm\yhat$) component. However, the extended $T'_2$ compensates for this initial loss, with the orbit-field-driven signal exceeding that of pure spin lock after approximately \SI{70}{\second} (Fig.~\ref{figSI:lifetime}b). This demonstrates that the prolonged lifetime more than offsets short-term amplitude penalties, thereby improving long-term sensing performance.        

\section{Phase Diagram of $M_z$ vs Flip Angle} \label{secSI:phasediagram}

    \begin{figure}[t]
                \centering
                \includegraphics[width=0.49\textwidth]{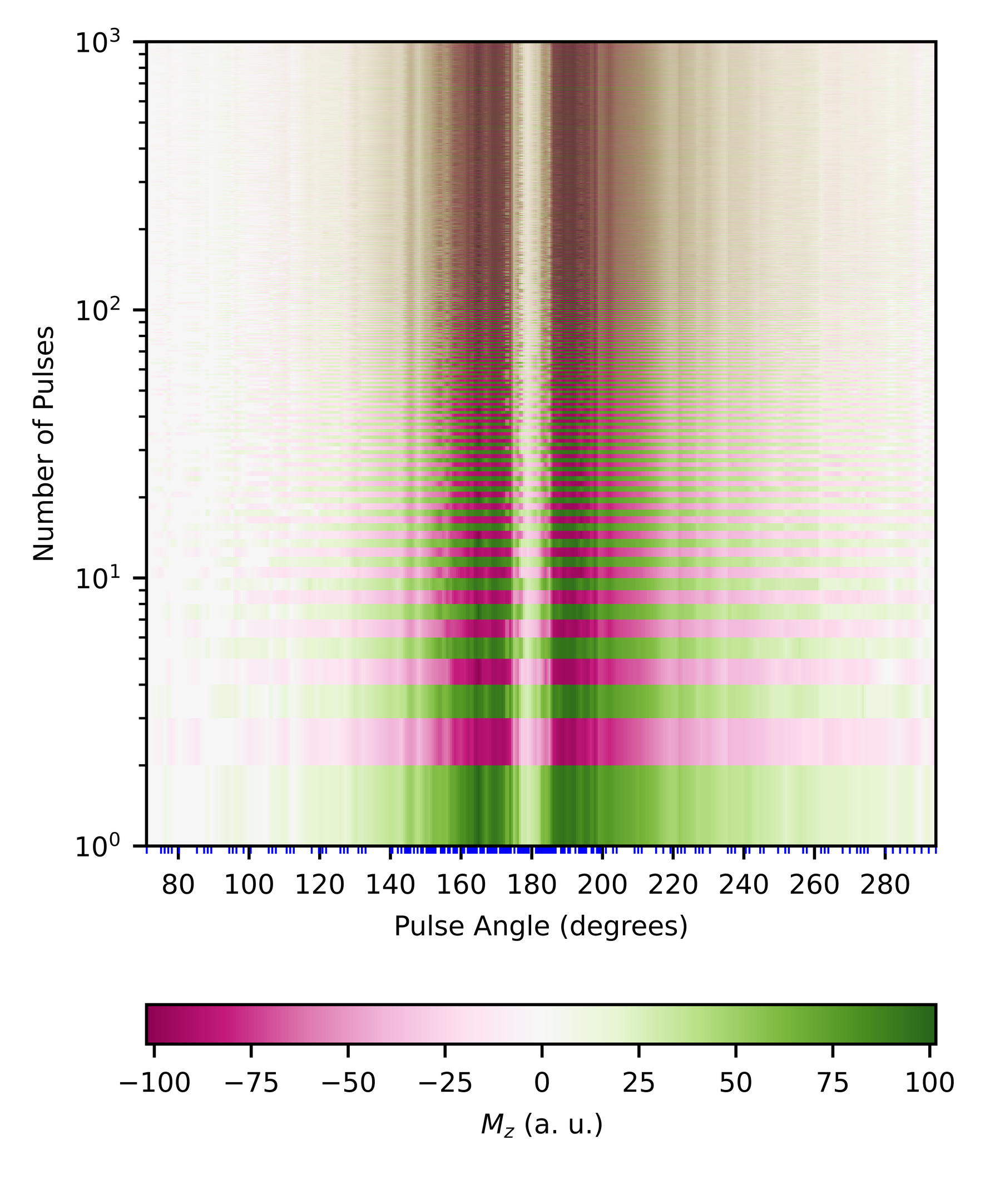}
                \caption{
                    \T{Colormap of the $\zhat$-component of the magnetization vector} ($M_z$) for each single shot experiment over 1000 x-pulses, plotted as a function of pulse angle. Minor tick marks on the $\xhat$-axis (blue) indicate pulse angles at which individual single-shot experiments were performed (188 total), with values between these angles interpolated. The color scale highlights transitions of the magnetization vector between the upper (green) and lower (purple) hemispheres of the Bloch sphere. Two broad stable intervals, close to a pulse angle of \SI{180}{\degree}, are evident. At exactly \SI{180}{\degree}, $M_z$ exhibits a pronounced drop to near-zero, corresponding to a markedly reduced trajectory quality.
                }
                \label{figSI:pulseangle_phasediagram}
            \end{figure}

        Figure \ref{figSI:pulseangle_phasediagram} reveals the evolution of the $\zhat$-component of the magnetization vector over the initial 1000 $\xhat$-pulses as a function of the pulse angle. The data clearly shows two broad, stable bands surrounding the \SI{180}{\degree} rotation. Within these regions, the large magnitude of the $\zhat$-component corresponds to high sensitivity and therefore robust performance of the protocol, with the color alternation indicating the vector's switching between the upper and lower prethermal state axes. Conversely, the $\zhat$-component decreases for pulse angles far from \SI{180}{\degree} and also drops sharply at this exact value. This loss of signal is attributed to both a reduced elevation angle and a general decrease in the norm of the magnetization vector (App. \ref{SIMethods:pulseangle}).

\section{Theoretical Description of Prethermal Spin Orbits} \label{secSI:theorytrajectories}
        Beyond its role in facilitating robust, noise-rejected quantum sensing, the Floquet prethermal physics governing the sensor’s operation is of intrinsic interest in the field of physics, owing to the emergence of multi-axial prethermal orbits. Our analysis builds upon the theoretical treatment of periodically driven Floquet systems in Ref.~\cite{sahin22_trajectory}. In the following, we outline the key elements of this theory that are essential to the operation of our sensor.

        When the orbit field is commensurate with the spin-lock cycle (\(T_{\mathrm{orbit}} = 2\tau\)), the dynamics are captured by an effective Hamiltonian consisting of a spin-spin interaction (\(H_\mathrm{int}\)) term and a non-commuting emergent field ($\boldsymbol{w}_k(\varphi)$)
        \begin{equation}
        H_{\mathrm{eff},k} = H_{\mathrm{int}} + \bm{w}_k(\varphi) \cdot \bm{I}.
        \label{eq:effective_hamiltonian}
        \end{equation}
        In Equation~\eqref{eq:effective_hamiltonian}, \(\varphi\) is the phase of the orbit field, \(\bm{I}\) is a vector of spin 1/2 operators, and \(k = 1,2\) indexes the stroboscopic frame, corresponding to periods defined relative to the first or second spin-lock pulse.
        
        Magnetization is read out after each pulse, resulting in two distinct manifolds in the amplitude and phase, which together define the sensor’s operational basis. The spins rapidly prethermalize into a state well approximated by the canonical Gibbs form \(\rho_{\beta} \propto e^{-\beta H_{\mathrm{eff},k}}\). Importantly, \(\beta\propto\epsilon\), where \(\epsilon\) is the deviation of the spin-lock pulse angle from \(\pi\). Deviation in the pulse angle is therefore critical for the spin system's prethermalization to a finite temperature state, yielding a lifetime that is significantly enhanced in the driving frequency~\cite{Beatrez21_90s,sahin22_trajectory}. For this reason, our sensing scheme is robust to errors in the pulse sequence (\ref{SIMethods:pulseangle}), separating it from traditional sensing protocols.
        
        In the high-temperature limit (\(\beta \ll 1\)), the magnetization per spin during the prethermal plateau is
        \begin{equation}
        \frac{\langle \bm{I}_{k} \rangle}{L} \equiv \frac{\mathrm{Tr}\!\left[ \rho_{\beta} \bm{I}_{k} \right]}{L} \approx -\beta(\varphi)\,\bm{w}_{k}(\varphi),
        \label{eq:magnetization}
        \end{equation}
        where the magnetization direction is dependent on the phase of the orbit field, \(\varphi\). Consequently, the readout reveals a robust oscillation between \(\langle \bm{I}_{1} \rangle\) and \(\langle \bm{I}_{2} \rangle\) (Fig.~\ref{fig:fig2}b(ii)), related by a \(180^\circ\) rotation about the spin-lock axis.
        
        For on-resonance pulses, the orbit is centered along \(\mathbf{x}\). Off-resonance driving tilts the spin-lock axis toward \(\pm\mathbf{z}\) by an angle
        \(\alpha = \arctan(\Delta\omega / \Omega)\),
        where \(\Delta\omega\) denotes the detuning and \(\Omega\) the on-resonance Rabi frequency. The spin-lock axis is then mapped as
        \begin{equation}
        \mathbf{x} \longrightarrow \cos\alpha \,\mathbf{x} + \sin\alpha \,\mathbf{z}.
        \label{eq:spinlock_axis}
        \end{equation}

        We now assess the perturbation introduced by the target field \(B_{\mathrm{sense}}\). In the regime \(\omega_{\mathrm{orbit}} \gg \omega_{\mathrm{sense}}\), the sensed field appears as a quasi-static bias analogous to that induced by off-resonant driving (\(\Delta\omega\mathbf{z}\)). The effective spin-lock (SL) axis then evolves slowly in time,
        \begin{equation}
        \mathbf{n}_{\mathrm{SL}}\big(\alpha(t)\big) = \cos\alpha(t)\,\mathbf{x} + \sin\alpha(t)\,\mathbf{z},
        \end{equation}
        with
        \begin{equation}
        \alpha(t) = \arctan\left[ \frac{\Delta\omega + B_{\mathrm{sense}}(t)}{\Omega} \right].
        \end{equation}
        
        This adiabatic variation causes \(\langle \bm{I}_{1} \rangle\) and \(\langle \bm{I}_{2} \rangle\) to oscillate in the \(\pm\mathbf{z}\) direction (Fig.~\ref{fig:fig2}), modulating their transverse-plane projections. This modulation establishes a direct quantitative link between \(B_{\mathrm{sense}}\) and the measured signal in the experiments.

        However, our sensing scheme remains robust even as the frequency of \(B_{\mathrm{sense}}\) approaches \(\omega_\mathrm{{orbit}}/2\) and the variation of the magnetization is no longer adiabatic. In this regime, the effect of the external field in tilting the spin-lock axis persists, but additional evolution is observed due to the spin ensemble being far from a prethermal state. This manifests in a "transient" signal, with dynamics that are reflected in the two manifolds, and can thus be suppressed. The transient dynamics are explored further in ~\ref{secSI:transients}.

\section{Origin of Transients and their Suppression} \label{secSI:transients}   
\begin{figure}[t]
            \centering
            \includegraphics[width=0.49\textwidth]{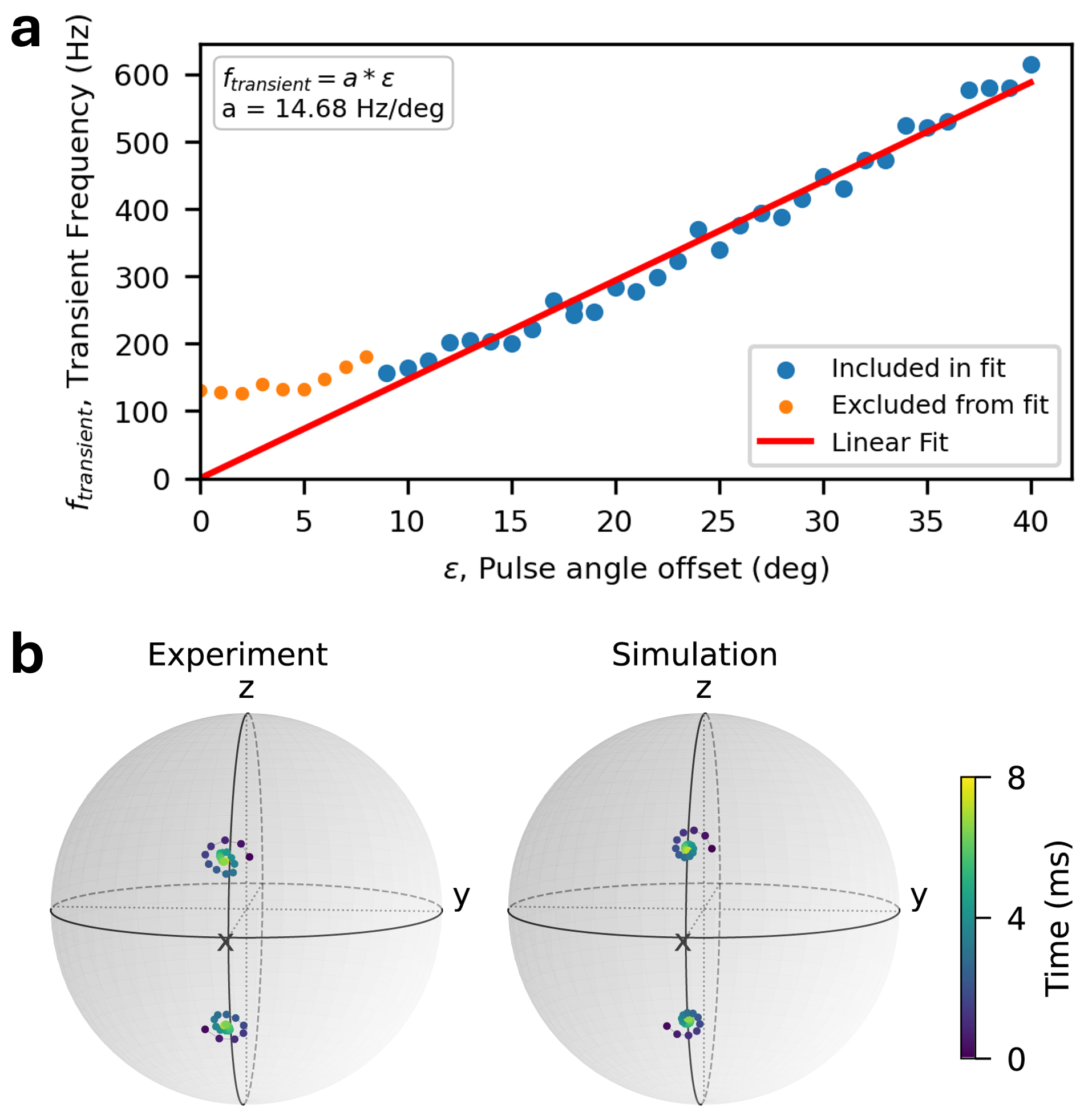}
            \caption{
                \textbf{Oscillation frequency of transients during square-waveform detection.}  
                At each rising and falling edge of the square wave, transient oscillations occur as the magnetization precesses towards the new prethermal states defined by the updated effective Hamiltonian.
                (\textbf{a}) For each experimental shot, a \SI{12}{\milli\second} time window of the measured data was extracted following the waveform edge. A fast Fourier transform was applied to this window, and the dominant peak was fitted with a pseudo-Voigt function to determine the transient oscillation frequency. Measurements were repeated in random order for a range of $\xhat$-pulse rotation angles. Small pulse angle offsets were masked due to poor fit reliability in this regime. Data for pulse-angle offsets above $9^\circ$ were fitted with a linear model, yielding a slope of $14.68~\mathrm{Hz/deg}$ in the transient oscillation frequency.
                (\textbf{b}) Comparison of experimental data with numerical simulation based on the phenomenological model. The measured transient waveform is shown alongside the results of a numerical simulation, demonstrating strong agreement with our model. Specifically, the simulation correctly reproduces key features of the transient, including its oscillation frequency and handedness.
            }
            \label{figSI:transientfrequency}
        \end{figure}

        When the magnetization of the spin ensemble does not commute with the effective Floquet Hamiltonian, rapid evolution in the stroboscopically observed magnetization can occur. This transient behavior decays on the $1-\SI{10}{\milli\second}$ timescale until the quasi-equilibrium prethermal state is reached. In the case of a perturbation evolving more quickly than this timescale, such as a rapidly oscillating sensed AC field, this motion can dominate the dynamics of the spin ensemble. 
        In the case of spins driven in an orbit between two axes by periodic pulses at angle $\theta=\pi+\epsilon$, the deviations $\epsilon$ in the pulse angle imprint directly on stroboscopic observations of the magnetization. In particular, the magnetization is rotated about the perturbed quasi-equilibrium axis at a frequency $\epsilon/2 T$=\SI{14}{\hertz\per deg}, as shown in  Fig.~\ref{figSI:transientfrequency}. Any deviation from the prethermal axis in one hemisphere is rotated by the combined angle of the x-pulse and z-drive rotation (${\approx}\pi$ around $\xhat$) into the other hemisphere. Thus the transients manifest in spirals within each hemisphere that are phase-shifted by \SI{180}{\degree} with respect to each other, as in Fig~\ref{fig:fig4}a. To first order in $\epsilon$, the transient imprint is canceled by summing between the trajectories. However, the measured signal is preserved due to motion of the prethermal axis itself. Hence this method flattens the sensor response at high frequencies by suppressing the additional transient dynamics.

        For clarity, we now demonstrate the mechanism of transient suppression explicitly.  
        Consider an initial state 
        \[
        \rho_0 \propto U(\yhat,\phi)\, I^x\, U^{\dagger}(\yhat,\phi),
        \]
        whose magnetization is rotated by an elevation angle $\phi$ about the $\yhat$-axis (we use axis-angle notation, e.g., $U(\yhat,\phi)=e^{-i\phi I^y}$).  
        Such a state is representative of one of the magnetization vectors appearing in our sensing protocol.  
        During the transient period, we retain only the effect of the $\tau$-periodic spin-lock pulses, modeled as
        \begin{equation}
            H_{\mathrm{transient}}
            = (\pi + \epsilon)\, I^x \sum_{l} \delta(t - l\tau),
            \label{eq:SI:transient_hamiltonian}
        \end{equation}
        where the pulses are taken to be instantaneous and include a small flip-angle error $\epsilon$.
        
        After the $n^{\mathrm{th}}$ pulse, the expectation value of the magnetization is
        \begin{equation}
        \begin{split}
        \langle \bm{I} \rangle(n\tau^+)
        &= \mathrm{Tr}\!\left[
            U(\xhat,n\epsilon)\,
            U(\xhat,n\pi)\,
            U(\yhat,\phi)\, I^x\, U^\dagger(\yhat,\phi) \right. \\[4pt]
        &\qquad\left.
            U^\dagger(\xhat,n\pi)\,
            U^\dagger(\xhat,n\epsilon)\,
            \bm{I}
        \right],
        \end{split}
        \label{eq:transient_magnetization}
        \end{equation}
        which evaluates to
        \begin{equation}
        \begin{split}
        \langle \bm{I} \rangle(n\tau^+)
        &= \cos(\phi)\,\hat{\mathbf{x}}
           + (-1)^n \sin(\phi)\Big[
                \sin(n\epsilon)\,\hat{\mathbf{y}}
                - \cos(n\epsilon)\,\hat{\mathbf{z}}
             \Big].
        \end{split}
        \label{eq:transient_magnetization_2}
        \end{equation}
        Equation~\eqref{eq:transient_magnetization_2} shows that the magnetization precesses about the $\xhat$-axis at an angular frequency $\epsilon/\tau$, with successive even and odd points differing by a phase of $\pi$.
        
        This simple picture neglects the fact that the spins relax toward a new (quasi-)equilibrium configuration over a finite prethermalization time,
        \[
        T_{\mathrm{eq}} = n_{\mathrm{eq}} \tau.
        \]
        To incorporate this, we rewrite the rotation in the $\yhat\tm\zhat$-plane from \eqref{eq:transient_magnetization_2} as
        \begin{equation}
            \langle I^y \rangle + i\langle I^z \rangle
            = -i\,(-1)^n \sin(\phi)\,e^{in\epsilon}
            \equiv g(n).
        \end{equation}

        In the toggling frame defined by the $\pi$ pulses about $\xhat$, the alternating factor $(-1)^n$ is removed. In this frame, we assume that $g(n)$ relaxes exponentially from its initial value $g(0)$ toward a new equilibrium value $g_{\mathrm{eq}}$, while also undergoing coherent rotation about $\xhat$.  
        Transforming back to the original rotating frame gives
        \begin{equation}
            g(n)
            = (-1)^n\!\left[
                g_{\mathrm{eq}}
                + \big(g(0) - g_{\mathrm{eq}}\big)
                  \exp\!\left[(i\epsilon - 1/n_{\mathrm{eq}})\,n\right]
              \right].
            \label{eq:transient_magnetization_3}
        \end{equation}
        The $\yhat$- and $\zhat$-components of the magnetization are then obtained as
        \[
        \langle I^y \rangle = \mathrm{Re}[g(n)],
        \qquad
        \langle I^z \rangle = \mathrm{Im}[g(n)].
        \]
        
        Equation~\eqref{eq:transient_magnetization_3} explains why the transient trajectories appear as spirals in the upper and lower hemispheres of the Bloch sphere, both exhibiting the same handedness: although the equilibrium value $g_{\mathrm{eq}}$ is fixed in the toggling frame, it acquires opposite signs at even and odd stroboscopic points when mapped back to the original rotating frame.
        
        To obtain the $\xhat$-component of the magnetization, we assume that small perturbations to the system Hamiltonian change the magnitude of the magnetization vector only weakly.  
        Thus the relaxation is primarily \emph{directional} rather than a decay of the magnetization vector norm, allowing us to approximate
        \[
        \langle I^x \rangle(n\tau^+)
        \approx
        \sqrt{\big|\langle \bm{I} \rangle(0)\big|^2 - \big|g(n)\big|^2}.
        \]
        Because $|g(n)|$ is independent of the factor $(-1)^n$, the transient contribution to $\langle I^x \rangle$ is nearly identical for both even and odd stroboscopic points.  
        Consequently, it cancels to first order in $\epsilon$ in the differential signal, while the desired motion of the prethermal axis is preserved.

\section{Detailed Study of Robustness to Mechanical Vibration} \label{secSI:vibrationNote}
            \begin{figure*}[t]
                \centering
                \includegraphics[width=0.5\textwidth]{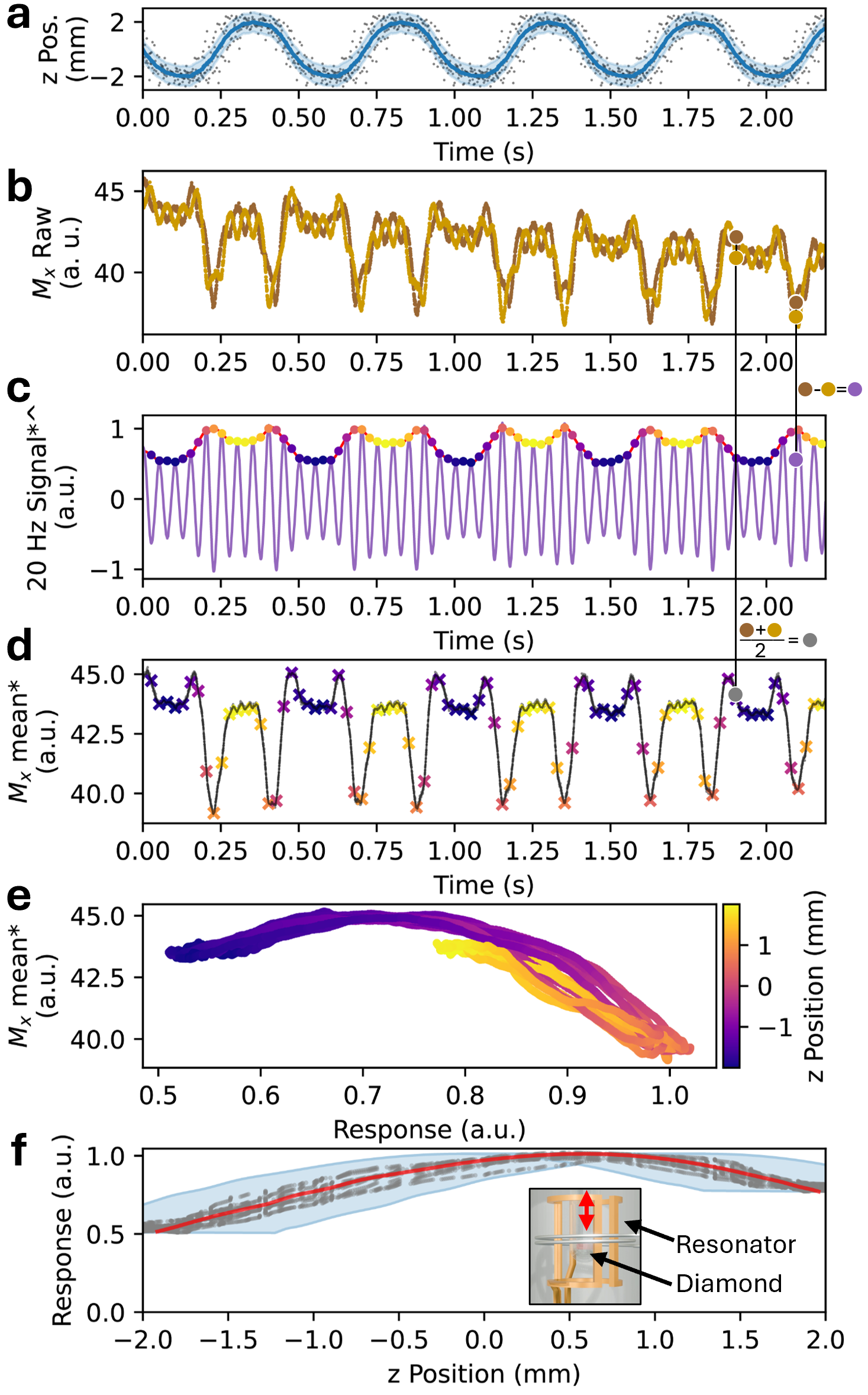}
                \caption{
                    \textbf{Vibration tolerance of the PRISM protocol.}  
                    (\textbf{a}) Sample displacement along $B_0$ generated by the shuttler, with raw position data (grey points), smoothed position (blue), and single-period standard deviation (light blue).  
                    (\textbf{b}) Raw $\xhat$ magnetization data, shown separately for even-indexed points (“A”, brown) and odd-indexed points (“B”, gold).  
                    (\textbf{c}) Extracted \SI{20}{\hertz} AC-field response from the coil, obtained as $(A-B)$ after correction via Biot–Savart modelling of the coil’s spatial field, and of static magnetic field variation with $\zhat$. Envelope smoothed using Gaussian filtering. Z position marked as colored points. 
                    (\textbf{d}) Mean $M_x$ without AC imprint, given by $(A+B)/2$. Colored crosses mark z position.  
                    (\textbf{e}) Mean $M_x$ versus AC-field response magnitude, illustrating that off-center displacement lowers response by reducing RF pulse power and flip angle, which increases $M_x$ projection at low displacements. Large offsets cause overall magnetization reduction, reducing mean $M_x$.  
                    (\textbf{f}) AC-field response versus z position (gray points), with spline fit (red) and position uncertainty (blue). Uncertainty mainly stems from imperfect periodicity in shuttler motion; additional contributions arise from weak external fields unrelated to the coil and residual Hilbert transform artifacts. Inset: 3D rendering of the RF resonator with the AC coil surrounding it and the diamond at the center.
                    Symbols in labels: ``*'' indicates decay removed by division by a linear baseline; ``\^{}'' indicates quantities relative to expected magnetic field, after removing DC components and components correlated with shuttler motion (${\approx}\SI{2}{\hertz}$).
                }
                \label{figSI:vibration}
            \end{figure*} 
            
            The PRISM protocol exhibits a high tolerance to mechanical vibrations, defined here as the displacement of the diamond containing $^{13}$C nuclear spins relative to the RF resonator. This resonator both delivers RF pulses and detects the emitted nuclear magnetization. Relative motion produces two primary effects:
            
            \begin{enumerate}
                \item Variation in RF pulse power received by the $^{13}$C ensemble due to spatial inhomogeneity of the resonator's RF field, leading to changes in the $\xhat$-pulse flip angle.
                \item Reduced efficiency in detecting RF signals from precessing spins, as optimal sensitivity is achieved at the resonators's coil center (“sweet spot”).
            \end{enumerate}
            
            To quantify robustness, we deliberately oscillated the sample at $\pm \SI{2}{\milli\meter}$ amplitude and $\approx \SI{2.1}{\hertz}$ along the static field axis ($B_0$, $\zhat$-axis; see Fig.~\ref{figSI:vibration}a and Fig.~\ref{figSI:vibration}f inset) using the method outlined in App.~\ref{SI:MeasurementOfVibration}. The $\xhat$ and $\yhat$ components of magnetization were continuously recorded (Fig.~\ref{figSI:vibration}b) and decomposed into:  
            (i) the phase-shifted component (“response”, see Fig. \ref{figSI:vibration}c), and  
            (ii) an in-phase component (“baseline”, see Fig.~\ref{figSI:vibration}d).

            The extracted response validated detection of the intentionally injected \SI{20}{\hertz} AC magnetic field (Fig.~\ref{figSI:vibration}c). Processing involved the following steps:
            
            \begin{enumerate}[itemsep=0pt]
                \item DC offset removal to suppress bias fields,
                \item Removal of a weak $\pm \SI{2.1}{\hertz}$ imprint from $B_0$ inhomogeneity in combination with shuttler movement,
                \item Savitzky–Golay smoothing (window = 41 points, polynomial order = 2),
                \item Decay compensation via division by the fitted decay function,
                \item AC coil inhomogeneity correction using Biot–Savart modelling,
                \item Hilbert transform to obtain the analytic signal, and
                \item Gaussian filtering ($\sigma$ = 40 samples) of the analytic signal envelope (red trace in Fig.~\ref{figSI:vibration}c).
            \end{enumerate}
            To determine the true response function, the extracted amplitude was compared to the known, deliberately injected AC magnetic field. 
            The performed steps do not bias this comparison; in particular, steps 1, 2, and 5 are necessary as they remove artifacts that would bias this comparison.
            
            The baseline reflects changes in $M_x$ due to vibration (Fig.~\ref{figSI:vibration}d). This trace was obtained by averaging the even-indexed with the corresponding odd-indexed data points, normalizing by the decay function, and applying Savitzky–Golay smoothing (window = 15, polynomial order = 2).
            
            Plotting the baseline (mean $M_x$) against the response (Fig.~\ref{figSI:vibration}e) reveals that displacement reduces the measured response, yet initially increases $M_x$. This counterintuitive effect results from reduced RF power at non-central resonator positions, decreasing the flip angle. In Bloch sphere terms, the reduced excitation lowers the elevation angle of the two prethermal eigenstates (cf. Fig.~\ref{figSI:pulseangle}), increasing the $\xhat$ projection length despite a decrease in total detectable magnetization. For displacements $>\SI{1}{\milli\meter}$, overall magnetization loss dominates and $M_x$ decreases.
            
            The main uncertainty in the data stems from position-tracking inaccuracies. Motion reconstruction assumes perfectly periodic shuttler oscillations (App.~\ref{SI:MeasurementOfVibration}). In practice, small deviations in acceleration during reversal lead to discrepancies between cycles. Furthermore, $\zhat$-axis motion can induce secondary vibrations in the $\xhat\tm\yhat$ plane, subtly affecting the response.

            Fig.~\ref{figSI:vibration}f shows the response versus $\zhat$ position, peaking at \SI{0.58}{\milli\meter}. This offset arises because the coil’s detection optimum is offset from the mechanical zero. The response decays symmetrically about this peak, and within the interval $\SI{-0.33}{\milli\meter} \leq z \leq \SI{1.03}{\milli\meter}$ the response remains above $95\%$ of its maximum—far exceeding realistic vibration amplitudes in operational conditions.

        \begin{figure*}[t!]
            \centering
            \includegraphics[width=0.95\textwidth]{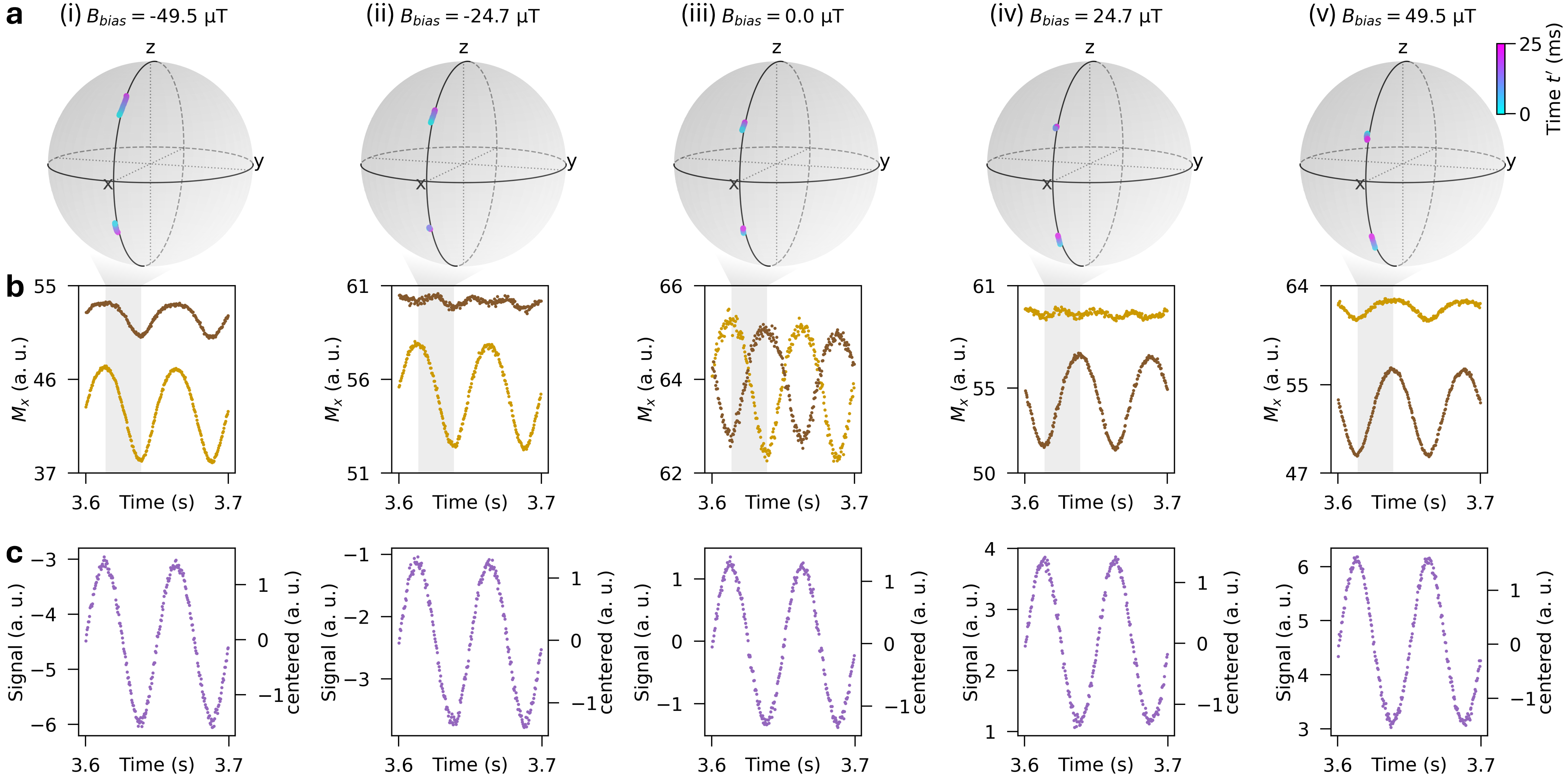}
            \caption{
            \textbf{Sensing an AC field in the presence of a DC bias field.}
            (\textbf{a}) 3D trajectories of the tracked magnetization vector during a sensing experiment under different external bias fields. The data shows a segment corresponding to half a wavelength (\SI{25}{\milli\second}) of the target AC magnetic field (\SI{1.8}{\micro\tesla}). Each column shows data for different applied bias fields ranging from \SIrange{-49.5}{49.5}{\micro\tesla} (i to v). The AC magnetic field induces arcs on the $\xhat\zhat$ unit circle. At zero bias field (iii), two arcs of equal length are traced in the same rotational direction. At approximately $\pm\SI{25}{\micro\tesla}$ (ii and iv), one arc's length becomes nearly zero while the other's is roughly doubled. For even larger bias fields ($\pm\SI{50}{\micro\tesla}$, i and v), the two arcs have opposite directions of rotation and continue to have unequal lengths.
            (\textbf{b}) The raw measurement data for the $M_x$ component (gold/brown traces) clearly shows the inversion of the signal imprint at zero bias field (iii). As a bias field is applied, the amplitude and shape of the imprints change significantly, consistent with the arcs shown in (a).
            (\textbf{c}) Despite the changes in the raw data, the target signal can still be reliably extracted. The extracted signal's amplitude shows no significant change (ii to iv) or only a very slight change (i and v). The bias field introduces an offset in the extracted signal that depends monotonically on the field strength. This effect can be harnessed for measuring DC magnetic fields or for correcting the minor amplitude dependence of the AC signal on the bias field.
            }
            \label{figSI:biasfielddata}
        \end{figure*}

\section{Robustness to Bias Field Drifts and Off-Resonant Driving} \label{secSI:RobustnessDCOffRes}

            \begin{figure}[t]
                \centering
                \includegraphics[width=0.49\textwidth]{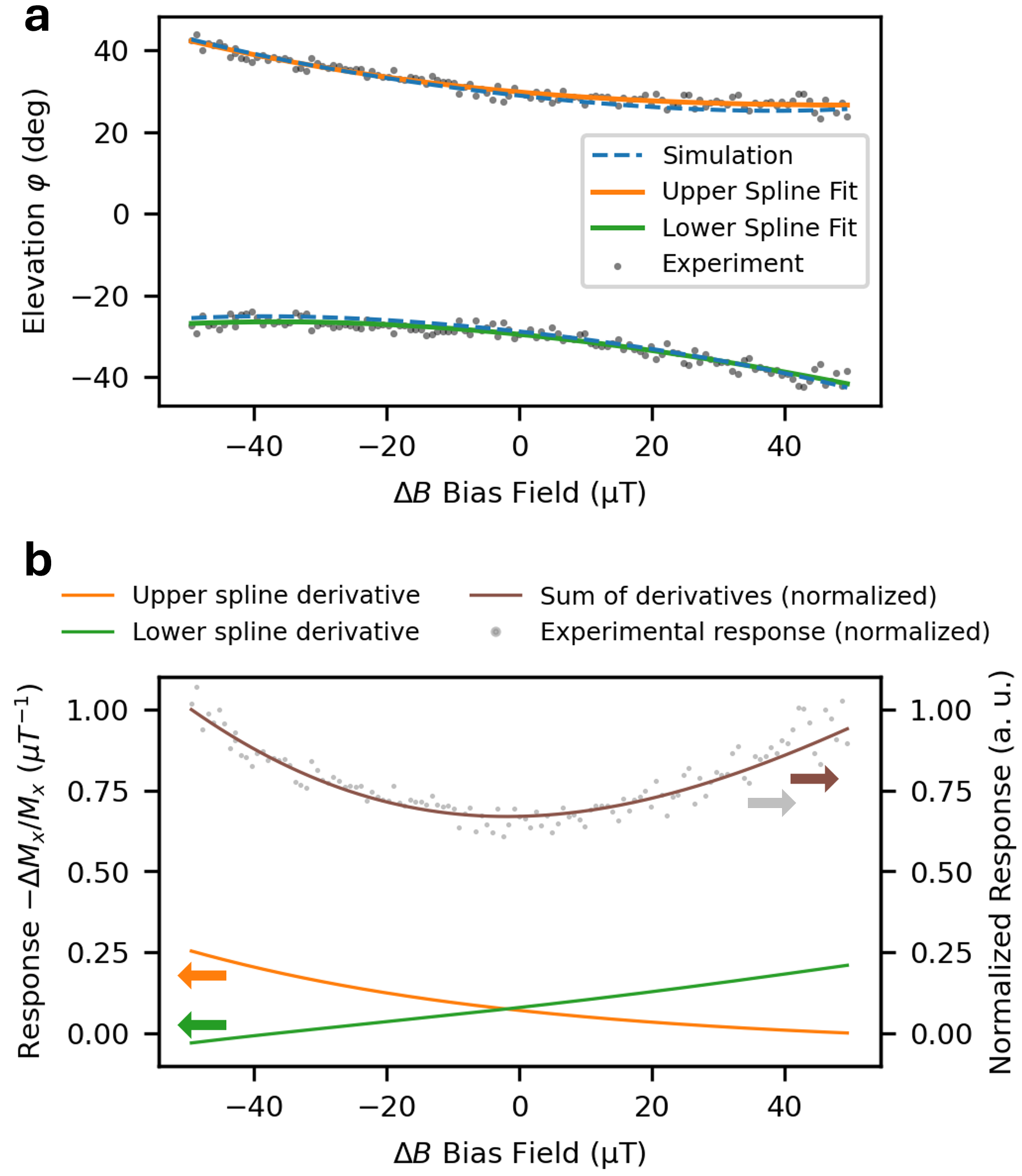}
                \caption{
                    \textbf{Bias-field dependence of elevation angle and response function.}  
                    (\textbf{a}) Elevation angle of the two magnetization states as a function of applied DC bias field.  
                    A total of 122 distinct bias-field values were measured in random order. The elevation angle from each shot was extracted for both eigenstates (grey points) and spline-fitted (orange and green curves). The expected elevation angle was independently calculated using the method in App.~\ref{secSI:SimulationRotationMatricesBRAYDEN} (blue dashed lines). In simulations, all parameters were fixed except for the pulse angle, which differed by only $\sim 3\%$ from its nominal value—within the calibration accuracy of the experiment.
                    (\textbf{b}) Response function obtained by normalizing the measured test-signal imprint amplitude for each shot (grey points). To evaluate the predicted link between elevation angle (panel~a) and response, the derivative of the two spline fits from~(a) was calculated, converted to the expected response using Eq.~\ref{eq:ResponseFunction}, divided by $M_x$ to remove the influence of transients during trajectory initialization, and finally normalized (brown curve). Excellent agreement between prediction and measured response validates Eq.~\ref{eq:ResponseFunction} experimentally.
                }
                \label{figSI:biasfield}
            \end{figure}

            In addition to AC-field sensing, the protocol can quantify DC bias fields. In realistic conditions, both weak AC/DC perturbations and larger magnetic-field drifts may occur. We evaluate robustness against such strong bias fields, as well as against RF pulses detuned from the ${}^{13}\mathrm{C}$ Larmor frequency (\emph{off-resonance}).  
            Detuning can arise from imperfect RF frequency calibration or from DC bias fields shifting the Larmor frequency, related via  
            \[
            \Delta \omega = \gamma\,\Delta B,
            \]  
            where $\Delta \omega$ is the RF frequency detuning, $\gamma$ is the gyromagnetic ratio and $\Delta B$ is the magnetic bias field.

            \paragraph{\T{Effects of a bias field on prethermal states and sensing:}}
                To investigate the effect of a bias field on the sensing protocol, we performed the same experiment repeatedly while applying an external DC bias field of varying strength. A sinusoidal AC test field ($f=$\SI{20}{\hertz}, $B_{AC}=\SI{1.8}{\micro\tesla}$) was used as the target signal. As shown in Fig. \ref{figSI:biasfielddata}, the applied bias field significantly alters the trajectory of the magnetization vector during the measurement. Nevertheless, the AC signal can still be reliably extracted with minimal amplitude distortion. Notably, the bias field introduces an offset in the extracted signal that depends monotonically on the field strength, a feature that can be utilized for DC magnetometry.

            \paragraph{\T{Theory linking bias field, elevation angle, and response:}}  
            A DC bias field alters the elevation angle $\phi$ of the magnetization vectors on the Bloch sphere. Differentiating $\phi$ with respect to $\Delta B$ yields the elevation-angle change per unit bias field. Multiplying by $\sin\phi$ converts this change into the corresponding variation in $M_x$, the $\xhat$-component of the magnetization vector, which represents the extracted signal response for a weak signal amplitude $\Delta B$:
            \begin{equation}
                \frac{dM_x}{d(\Delta B)} 
                = \frac{dM_x}{d\phi} \cdot \frac{d\phi (\Delta B)}{d(\Delta B)} 
                = \sin(\phi(\Delta B)) \, \frac{d\phi (\Delta B)}{d(\Delta B)}.
                \label{eq:ResponseFunction}
            \end{equation}
            For off-resonant pulses, $\Delta B$ is equivalently expressed via the frequency detuning as $\Delta B = \Delta \omega / \gamma$.
            
            \paragraph{\T{Experimental protocol:}}  
            We determined both the response function and $\phi$ under two conditions:  
            (i) applying a DC bias field via an external coil;  
            (ii) varying the RF pulse frequency on a shot-by-shot basis.  
            The bias field in condition (i) was switched on during the measurement sequence, whereas the frequency detuning for condition (ii) was present for the entire shot, beginning with the initialization pulse.
            In each case, the full three-dimensional magnetization vector was reconstructed, $\phi$ extracted, and a calibrated AC test field injected to measure the response strength for the given bias field or detuning.
            
            \paragraph{\T{Simulation:}}  
            Using the experimental parameters exactly, we simulated $\phi$ and magnetization vector orientations via methods in App.~\ref{secSI:SimulationRotationMatricesBRAYDEN}. 
            
            For the bias field scan simulation, all known parameters—pulse length, inter-pulse spacing, and $\zhat$-drive amplitude—were fixed; the $\xhat$-pulse power (rotation angle per $\xhat$-pulse) remained the only free parameter. Optimal agreement with experiment was found for an $\xhat$-rotation of \SI{160.2}{\degree}, versus a pre-estimate of \SI{165.6}{\degree}, yielding excellent fidelity between simulated and measured elevation angles (Fig.~\ref{figSI:biasfield}a). These simulations were also augmented with an analysis from the perspective of Floquet theory as outlined in section App.~\ref{secSI:SimulationFloquet}.

            \begin{figure}[t]
                \centering
                \includegraphics[width=0.45\textwidth]{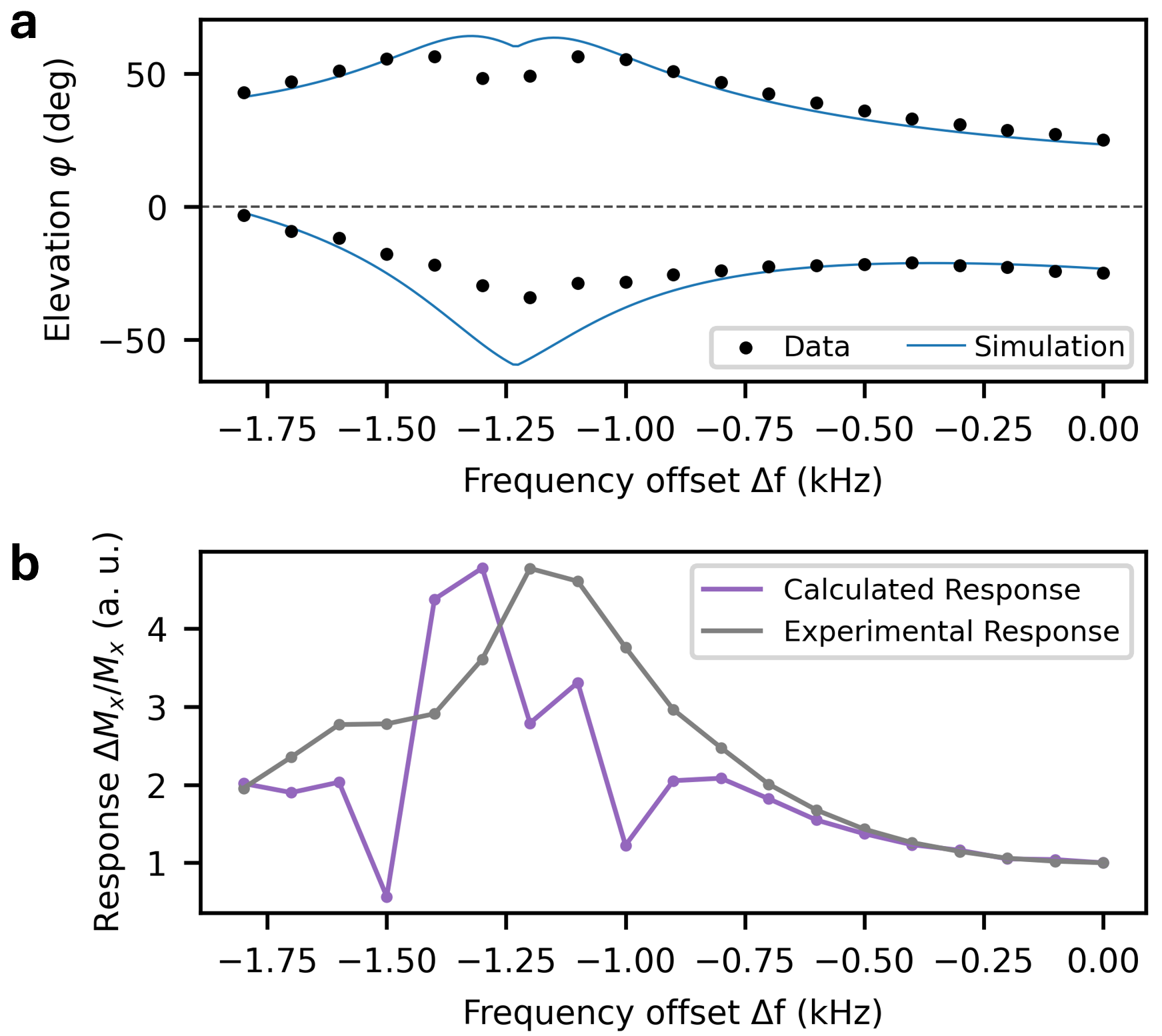}
                \caption{
                \textbf{Dependence of elevation angle and response function on RF pulse detuning.}  
                (\textbf{a}) Elevation angles of the two prethermal magnetization eigenstates as a function of RF pulse frequency offset. Data was acquired at 19 distinct offsets in \SI{100}{\hertz} increments, with the angles for both eigenstates extracted from each dataset (black). Simulations (App. \ref{secSI:SimulationFloquet}) reproduce the overall trend and show excellent quantitative agreement for detunings up to approximately \SI{-0.75}{\kilo\hertz} (blue). The sharp dips at ${\app}\SI{1.2}{\kilo\hertz}$ occur as the inter-vector angle separating the two prethermal axes approaches its maximum value for the given pulse length.
                (\textbf{b}) Response function obtained out of the measured test-field imprint for each shot (grey points). To validate the predicted link between elevation angle and response, the curve in (a) was differentiated, converted to the expected response using Eq.~\ref{eq:ResponseFunction}, divided by $M_x$ to remove the influence of transients during trajectory initialization, and finally normalized (purple curve). The data points at \SI{0}{\kilo\hertz} are used to normalize both graphs. Excellent agreement is observed for detunings up to \SI{-0.5}{\kilo\hertz}. Deviations at larger frequency offsets are likely due to the coarse sampling of frequency points.
                }                
                \label{figSI:offresonance}
            \end{figure}
            
            \paragraph{\T{Validation and explanation:}} 
            The theoretically derived response function relation (Eq.~\ref{eq:ResponseFunction}) agrees closely with experimental data across the full range of tested bias fields and large parts of the investigated detuning range (Fig.~\ref{figSI:biasfield}b, Fig.~\ref{figSI:offresonance}b). Deviations appear in the off-resonance data below \SI{-0.5}{\kilo\hertz} detuning (${\approx}\SI{-47}{\micro\tesla}$), where the geometry of the magnetization vectors on the Bloch sphere becomes more intricate. For on-resonance pulses, the angle between the two magnetization vectors is simply \(2|\phi|\). Under detuning, however, the response is not captured by a simple tilting of both vectors toward one hemisphere, because detuning changes both the effective rotation axis and the effective flip angle of the pulse. As shown in the next section, the flip angle is a predominant factor in determining the angle between the two vectors (we highlight the distinction between the elevation angle \(\phi\) and the \emph{inter-vector angle}). Consequently, below \SI{-0.5}{\kilo\hertz}, this inter-vector angle rapidly approaches the maximum value permitted by the pulse duration (occurs around \SI{-1.2}{\kilo\hertz} in Fig.~\ref{figSI:offresonance}a) and eventually attains this maximum, after which it decreases. This leads to the distinct elevation angle traces in Fig.~\ref{figSI:offresonance}a.

            This geometric interplay also influences the imprint of the sensed field. The signal imprint---normally \SI{180}{\degree} phase-shifted at small detunings---becomes in phase between \SI{0.4}{\kilo\hertz} and \SI{2}{\kilo\hertz} (Fig.~\ref{fig:fig5}b(iii), top), originating from changes in the gradients of the \emph{elevation-angle} traces in Fig.~\ref{figSI:offresonance}a. Importantly, the imprint remains extractable throughout this range because the gradient differences between the upper (positive-$\zhat$) and lower (negative-$\zhat$) components persist (Fig.~\ref{fig:fig5}b(iii), bottom).

            \paragraph{Relationship between response and sensitivity}
            \begin{figure}[t]
                \centering
                \includegraphics[width=0.45\textwidth]{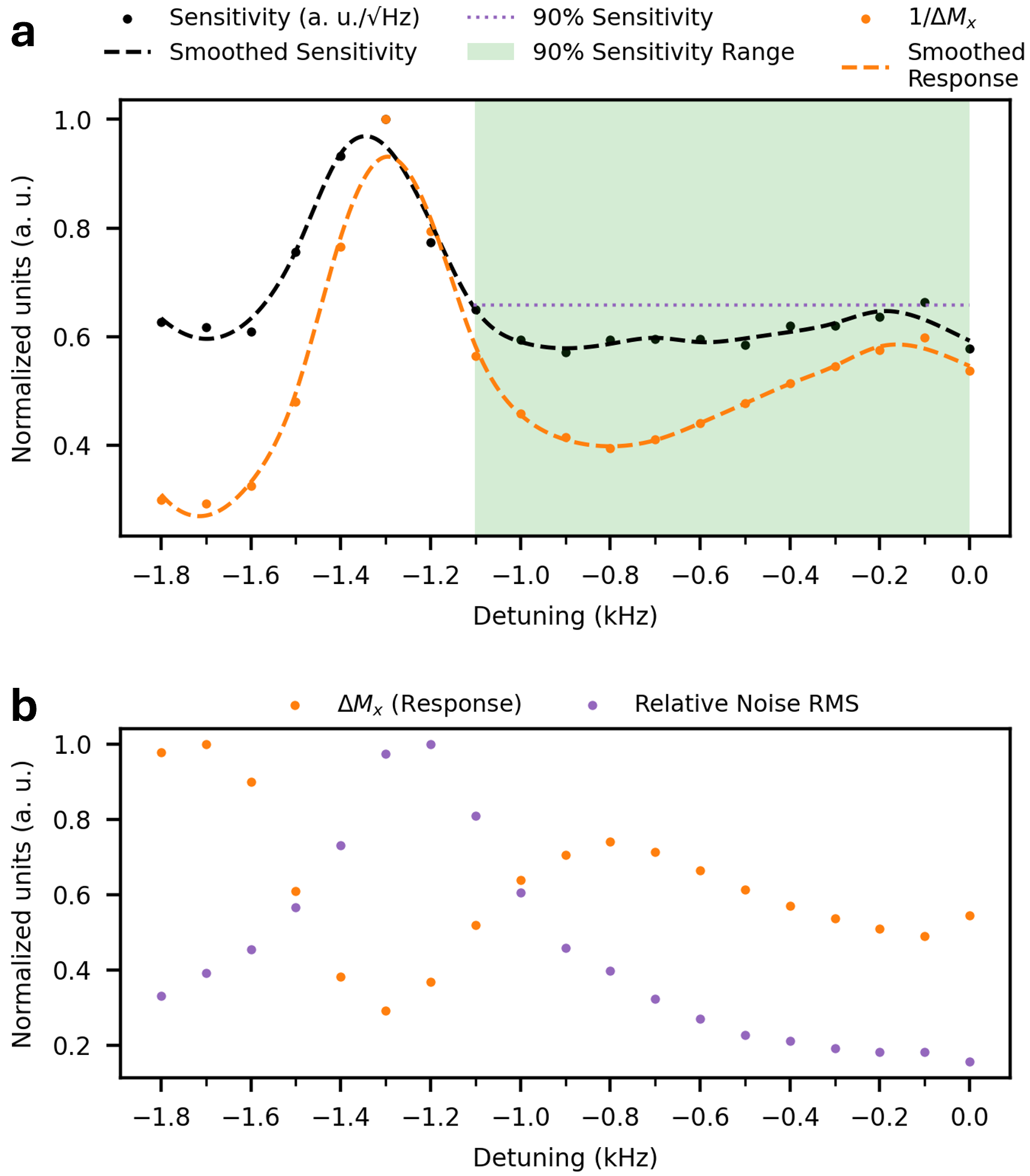}
                \caption{
                \textbf{Relation between sensitivity and response at different detuning values.}
                (\textbf{a})~Comparison of the sensitivity (black) and the response magnitude (orange) as a function of detuning $\Delta f$. The raw data (points) was smoothed for this analysis using a Savitzky-Golay filter (window length 5 samples, polynomial order 2, dashed lines). A threshold corresponding to 90\% of the peak sensitivity (purple dotted line) is indicated, highlighting the operational range where this performance level is maintained (green shaded area).
                (\textbf{b})~Comparison of the response, $\Delta M_x$ (orange), with the root-mean-square (RMS) noise (purple). The noise is calculated and plotted relative to both the maximum signal amplitude, $M_x$, and the response magnitude, showing its dependence on the detuning.
                }       
                \label{figSI:responsevssensitivity}
            \end{figure}
            
            The measurements, which are also analyzed in Fig.~\ref{figSI:offresonance}, involved applying a synthetic AC signal to characterize both the sensor's response strength ($\Delta M_x$) and its sensitivity. For a constant noise floor, the sensitivity is expected to be inversely proportional to the absolute response ($\propto 1/\Delta M_x$). 
            Figure~\ref{figSI:responsevssensitivity}a shows that while the two curves follow a similar trend, they exhibit a clear deviation from a simple inverse relationship. This discrepancy arises because the noise floor is not constant, but varies with the detuning, as shown in Fig.~\ref{figSI:responsevssensitivity}b. This variation can be attributed to changes in the spin dynamics; for instance, the effect of RF pulse inhomogeneities on the nuclear spins is altered by the detuning.
            Despite this effect, the sensor maintained at least 90\% of its peak sensitivity for detuning values up to \SI{1.1}{\kilo\hertz}.

\section{Robustness to Control Pulse Imperfections} \label{SIMethods:pulseangle}
        \begin{figure*}[t]
                    \centering
                    \includegraphics[width=0.95\textwidth]{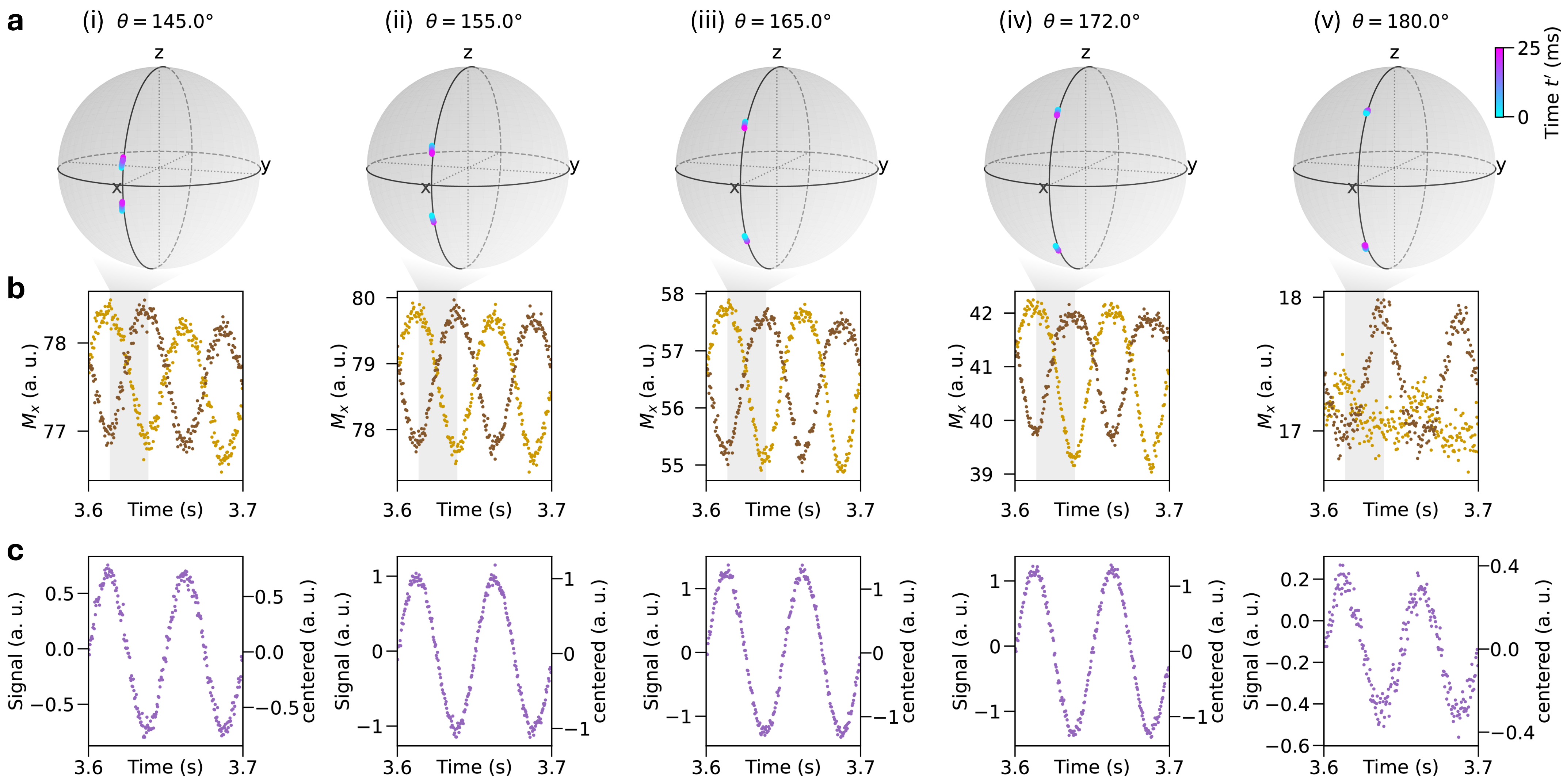}
                    \caption{
                    \textbf{Robustness of AC field sensing to pulse angle deviations.}
                    (a) 3D trajectories of the tracked magnetization vector for different $\xhat$-pulse rotation angles. The data shows a segment corresponding to half a wavelength (\SI{25}{\milli\second}) of the target AC magnetic field (\SI{1.8}{\micro\tesla}). Each column corresponds to a different pulse angle, from \SI{145}{\degree} (i) to \SI{180}{\degree} (v). The AC magnetic field induces arcs on the $\xhat\zhat$ unit circle. Near the optimal angle of \SI{165}{\degree} (iii), the response is maximized. As the angle deviates from this optimum (\SI{155}{\degree} in ii and \SI{172}{\degree} in iv), the elevation angle of the prethermal axes changes but the trajectory is still symmetric. Even for a large deviation to \SI{145}{\degree} (i) and \SI{180}{\degree} (v), a clear trajectory is maintained.
                    (b) The raw measurement data for the $M_x$ component (brown and gold traces) confirms the behavior seen in the trajectories. The signal imprint is clearly visible over a broad pulse angle range. As the pulse angle deviates from the optimum at \SI{165}{\degree}, the amplitude of $M_x$ changes. At \SI{180}{\degree} (v), the signal amplitude is sharply reduced.
                    (c) Despite the variations in the raw data, the target signal shape is reliably extracted across the entire range by subtracting the traces in (b). The response remains nearly constant across a broad plateau around the optimum (ii to iv), with only a slight reduction outside of this plateau at \SI{145}{\degree} deviation (i). This demonstrates the protocol's exceptional resilience to calibration errors in the control pulses.
                    }
                    \label{figSI:pulseangledata}
                \end{figure*}

                \begin{figure*}[t]
                \centering
                \includegraphics[width=0.94\textwidth]{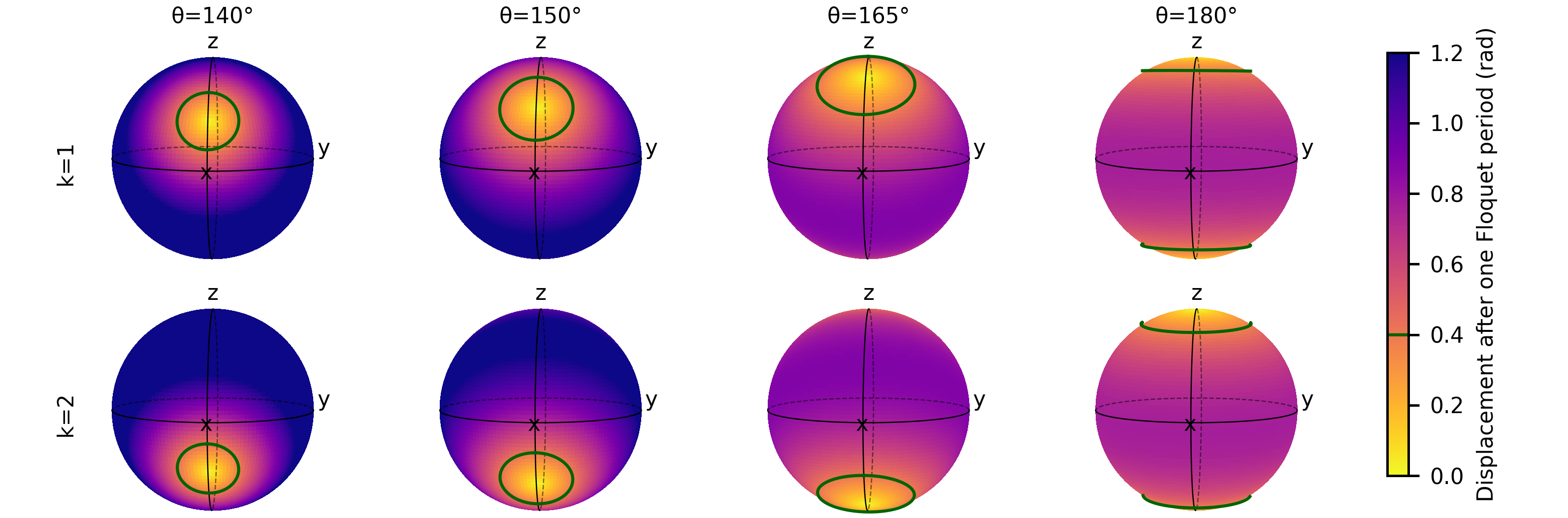}
                \caption{
                    \textbf{Numerical simulation of Bloch sphere dynamics under a Floquet cycle.} The rotation matrix (cf. App. \ref{secSI:SimulationRotationMatricesBRAYDEN}) for a single Floquet cycle was calculated using parameters analogous to the experiments presented in the main text: a pulse duration of \SI{100}{\micro\second}, an inter-pulse spacing of \SI{100}{\micro\second}, and a peak z-drive amplitude corresponding to a rotation of \SI{18}{\degree} per \SI{100}{\micro\second}. Each point on the surface of the Bloch sphere was then rotated by this matrix, and the resulting arc distance (displacement) from its initial point is displayed using the color scale. To guide the eye, a green contour line is drawn at a threshold displacement of \SI{0.4}{rad}. The columns represent different $\xhat$-drive pulse flip angles ($\vartheta$), while the rows show the state of the system at the two different stroboscopic frames: at the center of the inter-pulse spacing after the first (k=1) and second (k=2) spin-locking pulse.
                    These simulations reveal the location and extent of the prethermal orbits' stability basin. The size of this region (enclosed by the green contour) is a direct proxy of the system's resilience to driving field inhomogeneities. Under inhomogeneous driving fields, a broad stability basin allows different spatial subsets of the spin ensemble to lock into broadly spread prethermal orientations. This orientational dephasing leads to a partial cancellation of the net magnetization, resulting in a weaker signal. Conversely, a more focused region suppresses this dephasing and preserves the signal. This mechanism culminates at $\vartheta = \SI{180}{\degree}$, where the stability region splits into two distinct, nearly opposite orientations along $\pm \zhat$, causing the near-complete signal cancellation observed experimentally.
                }
                \label{figSI:delocalizedeigenstates}
            \end{figure*}
            
            \begin{figure}[t]
                \centering
                \includegraphics[width=0.49\textwidth]{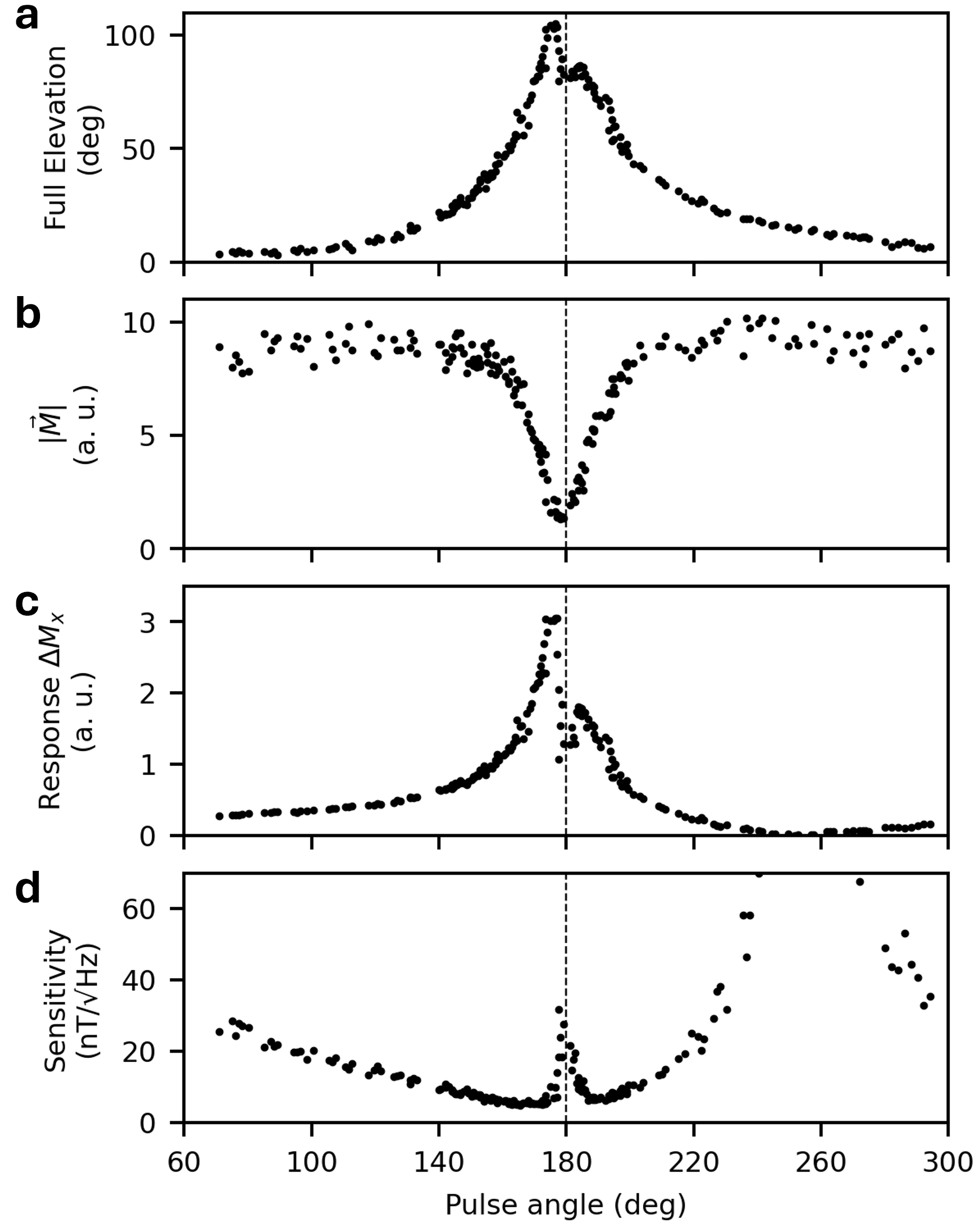}
                \caption{
                \textbf{Robustness of the PRISM protocol against pulse-angle variations.}
                    Evaluation of a 188-shot scan in which the $\xhat$-pulse rotation angle was varied in random order to test the robustness.
                    \textbf{(a)} Sum of the elevation angles of the two prethermal eigenstates of the magnetization vector. The elevation angle increases as the $\xhat$-pulse rotation approaches \SI{180}{\degree}, with a slight drop observed exactly at \SI{180}{\degree}.  
                    \textbf{(b)} Reconstructed absolute length of the magnetization vector from full 3D reconstruction. A pronounced drop is visible near $180^\circ$, where the net magnetization approaches zero.  
                    \textbf{(c)} Measured response $\Delta M_x$ to a test signal, indicating the magnitude of the imprint on the $\xhat$-projection of the magnetization vector. The response is maximal for $\xhat$-pulse rotations near \SI{175}{\degree}.  
                    \textbf{(d)} Sensitivity extracted from the scan, normalized to the maximum measured sensitivity determined in App.~\ref{secSI:sensitivity}. A distinct peak in sensitivity degradation appears between \SI{174}{\degree} and \SI{186}{\degree} due to the drop in net magnetization. Regions outside this interval exhibit consistently high sensitivity, underscoring the protocol's robustness over a broad range of pulse angles.
                }
                \label{figSI:pulseangle}
            \end{figure}

            A notable feature of the presented PRISM protocol is its intrinsic robustness to inaccuracies in rotation-angle calibration: deviations of up to $\pm \SI{5}{\degree}$ in the applied pulse angle do not significantly degrade performance. This robustness also means that inhomogeneity of the RF resonator is not a significant concern, further enhancing the protocol’s reliability. Such tolerance is exceptional given that in many nuclear spin and other qubit manipulation schemes, even errors on the order of $1^\circ$ can measurably reduce state fidelity. In conventional implementations, precise control over the rotation angle is critical for high-fidelity state preparation and gate operations, and substantial effort is typically devoted to calibrating pulse durations and amplitudes---for example via iterative Rabi oscillation measurements or error-compensating composite pulse schemes---to minimize over- or under-rotation. The robustness of our approach to such calibration errors can therefore markedly simplify experimental procedures and enhance resilience against both systematic and random drifts in control parameters. Furthermore, the pulse-angle robustness is intimately connected to the Floquet prethermal physics underlying the sensor's operation and is explored in App.~\ref{secSI:theorytrajectories}.

            Figure \ref{figSI:pulseangledata} presents experimental data acquired using a range of rotation angles $\vartheta$ for the $\xhat$-pulses, demonstrating that signal extraction is successful across the entire investigated range. The response amplitude exhibits a slight dependence on the pulse angle, reaching its maximum near \SI{165}{\degree} (Fig. \ref{figSI:pulseangledata}(iii)). This optimal value is situated within a broad plateau, leading to only a marginal reduction in response at angles of \SI{155}{\degree} (Fig. \ref{figSI:pulseangledata}(ii)) and \SI{172}{\degree} (Fig. \ref{figSI:pulseangledata}(iv)). Notably, even for a substantial deviation to \SI{145}{\degree}, the measurement protocol remains robust, and the sinusoidal test signal can be unambiguously reconstructed (Fig. \ref{figSI:pulseangledata}(i)). As the angle approaches \SI{180}{\degree}, the signal amplitude diminishes sharply (Fig. \ref{figSI:pulseangledata}(v)), which can be attributed to a significant delocalization of the pulse sequence's eigenstates due to small inherent driving field inhomogeneities (Fig.~\ref{figSI:delocalizedeigenstates}); nevertheless, a trajectory can still be resolved, which can be partially attributed to slight spatial inhomogeneities in the pulse's $B_1$ field and a bias towards eigenvectors with a positive x-component.

            For maximum sensitivity, a large elevation angle is desirable, as it maximizes the responsiveness of the transverse ($\xhat$-axis) projection of the magnetization vector to small changes induced by external magnetic fields (Fig.~\ref{figSI:pulseangle}a). The sensitivity peaks near \SI{180}{\degree}; however, at precisely \SI{180}{\degree} the net magnetization approaches zero (Fig.~\ref{figSI:pulseangle}b), meaning most of the initial spin polarization is lost upon activation of the $\zhat$-drive, resulting in a decreased response and therefore decreased sensitivity (Fig.~\ref{figSI:pulseangle}c). This loss shortens the transverse magnetization vector and consequently diminishes the signal-to-noise ratio (SNR). Since both the elevation and the net amplitude ($M_x$) affect the measurement quality, we investigated sensitivity as the primary metric for quantifying robustness across the plateau regions (Fig.~\ref{figSI:pulseangle}d).
            
            The insensitivity to pulse parameter variations and the broad plateau of high sensitivity make the PRISM protocol particularly well-suited for environments where long-term calibration stability is difficult to maintain, where control hardware resources are limited, or where resilience to environmental noise is essential for sustained high-precision measurements.

    \section{Robustness to Temperature and Strain Variations} \label{secSI:tempstrain}
        \begin{figure}[t]
            \centering
            \includegraphics[width=0.49\textwidth]{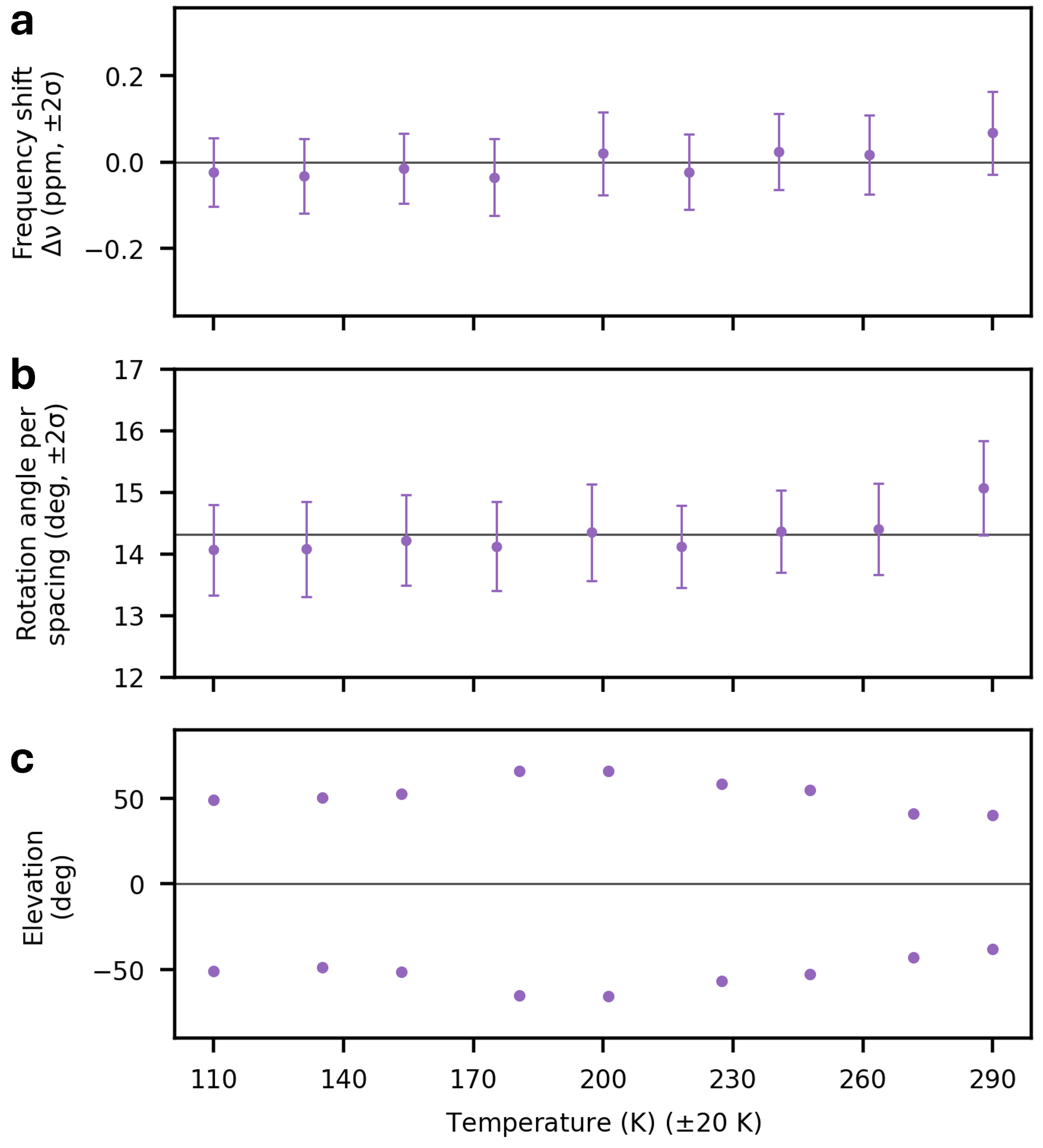}
            \caption{
                \textbf{Temperature dependence of the PRISM protocol measured on ${}^{13}\mathrm{C}$ nuclear spins in diamond.} 
                (\textbf{a}) Frequency shift of the Larmor precession frequency determined from free induction decay (FID) measurements at temperatures ranging from ${\approx}\SI{110}{\kelvin}$ to room temperature. As expected for nuclear spins, no significant temperature dependence is observed.  
                (\textbf{b}) Strength of the $\zhat$-drive, tracked in the presence of a DC bias field, remains close to the optimal value (cf. Fig. \ref{figSI:elevationanglevszdrive}) across the entire temperature range, with only negligible deviation near room temperature. The specified value is the maximum angular velocity relative to the spacing length of \SI{74}{\micro\second}. The reported standard error reflects the uncertainty from fitting the FFT peaks used to determine the $\zhat$-drive amplitude.
                (\textbf{c}) Elevation angle extracted shortly after the start of each measurement confirms consistent protocol operation over the full range. A small variation in elevation angle is observed, likely due to temperature-dependent changes in the RF resonator and its capacitors, which can affect the emitted pulse power and thus the effective rotation angle. Owing to the protocol's intrinsic tolerance to pulse-angle deviations, no appreciable sensitivity loss is expected.
            }
            \label{figSI:temperature}
        \end{figure}
        
        The presented method was tested on ${}^{13}\mathrm{C}$ nuclear spins in diamond. Nuclear spins offer the advantage of being generally insensitive to temperature: the Larmor precession frequency $\omega_\mathrm{L} = \gamma B_0$
        depends only on the constant gyromagnetic ratio $\gamma$ and the applied magnetic field $B_0$. For a constant $B_0$—or within the bias-field plateau region where the protocol is mostly insensitive to small $B_0$ changes (App.~\ref{secSI:RobustnessDCOffRes})—no temperature dependence is expected.

        To verify this experimentally, we first cooled the full setup, i.e. the diamond, the RF resonator with its capacitors, the AC coil, and the microwave coil for hyperpolarization, to \SI{110}{\kelvin}, and then evaluated the protocol over a broad temperature range from approximately \SI{110}{\kelvin} to room temperature, progressively heating these components. At each temperature, without recalibration, the Larmor precession frequency was determined via free induction decay (FID) measurements. As predicted, no significant change in $\omega_\mathrm{L}$ was observed (Fig.~\ref{figSI:temperature}a).
        
        During these measurements, the FID procedure was repeated in the presence of a small DC bias field to track the $\zhat$-drive strength. Without additional adjustments, the $\zhat$-drive remained close to the optimal amplitude throughout the temperature range (cf. Fig.~\ref{figSI:elevationanglevszdrive}), with only a very slight reduction at room temperature (Fig.~\ref{figSI:temperature}b).
        
        We further tested the full PRISM protocol by extracting the elevation angle shortly after the start of the acquisition at each temperature. The PRISM protocol was operated with a \SI{74}{\micro\second} pulse length, \SI{74}{\micro\second} spacing between pulses, and an $\xhat$-pulse angle of \SI{{\approx}165}{\degree}. The protocol operated flawlessly across the entire range, although a small, systematic variation in elevation angle was noted (Fig.~\ref{figSI:temperature}c). Since both the Larmor frequency (Fig.~\ref{figSI:temperature}a) and $\zhat$-drive amplitude (Fig.~\ref{figSI:temperature}b) are essentially unchanged within the robustness plateaus, we attribute the observed variation to temperature-dependent effects of the RF resonator and its capacitors. Such changes can slightly alter the emitted RF pulse power and thereby the effective $\xhat$-pulse rotation angle.
        
        Due to the intrinsic pulse-angle tolerance of the protocol, these variations produce negligible changes in sensitivity. In cases of major temperature drifts, they can be further compensated either by post-processing using the full 3D magnetization reconstruction or via periodic Rabi calibrations to maintain optimal rotation angles. This resilience further reduces the need for active environmental control, highlighting the practicality of the protocol in cryogenic and variable-temperature experiments.

        Finally, we can quantify the internal strain induced within the diamond lattice by this temperature change. The thermal contraction upon cooling from room temperature ($T_\mathrm{RT} \approx \SI{295}{\kelvin}$) to $T_\mathrm{low} = \SI{110}{\kelvin}$ generates a mechanical strain, $\varepsilon$. This strain is not a linear function of the temperature drop, as the coefficient of linear thermal expansion for diamond, $\alpha(T)$, is itself strongly temperature-dependent, decreasing rapidly at cryogenic temperatures~\cite{jacobson2019thermal}. The total strain is therefore given by the integral over this range:
        \begin{equation}
            \varepsilon = \int_{T_\mathrm{low}}^{T_\mathrm{RT}} \alpha(T) \, \mathrm{d}T.
        \end{equation}
        Using reference data for the thermal expansion of diamond~\cite{jacobson2019thermal}, we estimate the integrated contraction over this temperature range to be at the order of $\varepsilon \sim \SI{8e-5}{}$. While seemingly small, this strain corresponds to a significant internal stress ($\sigma$) due to diamond's exceptionally high Young's modulus ($E \sim \SI{1000}{\giga\pascal}$)~\cite{klein1993young}. The resulting stress is:
        \begin{equation}
            \sigma = E \varepsilon \approx \SI{1000}{\giga\pascal} \times \SI{8e-5}{} =\SI{80}{\mega\pascal}.
        \end{equation}
        The protocol's flawless operation under these conditions demonstrates its profound insensitivity to the intrinsic mechanical stress within the sensor material itself. This further validates our conclusion that the minor observed variations are attributable to hardware components, not a fundamental limitation of the sensing protocol.

\section{Susceptibility of the RF resonator to external RF signals} \label{secSI:RFsusceptibility}
        \begin{figure}[t]
            \centering
            \includegraphics[width=0.47\textwidth]{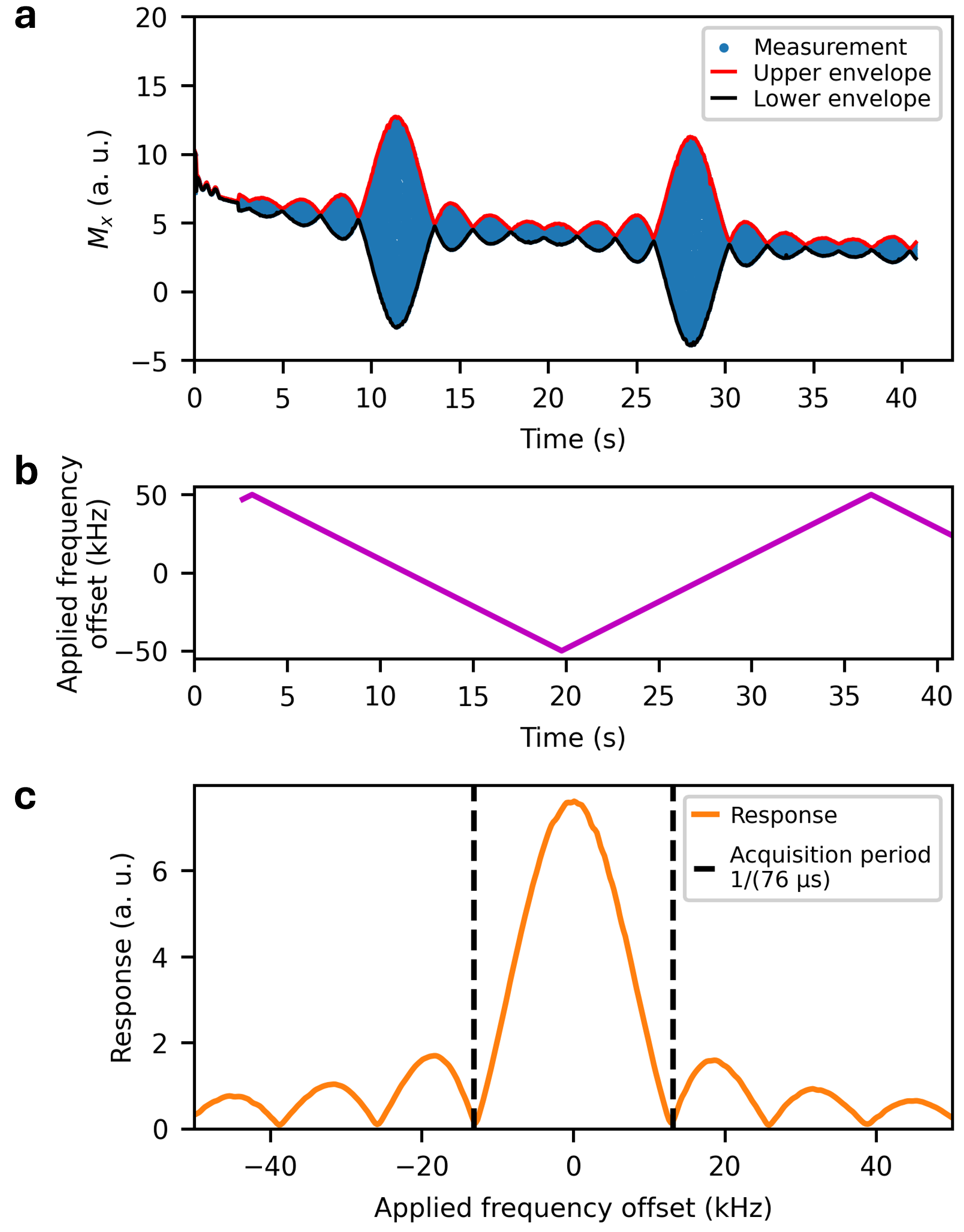}
            \caption{
                \textbf{RF susceptibility of the ${}^{13}\mathrm{C}$ RF resonator under PRISM operation.}  
                (\textbf{a}) Raw measured PRISM trajectory signal with the externally applied RF imprint (blue trace). The upper (red) and lower (black) envelopes mark the instantaneous amplitude modulation due to the injected RF background.  
                (\textbf{b}) Applied RF background frequency modulation relative to the Larmor frequency (\SI{\approx75.4}{\mega\hertz}), generated by a triangular waveform at \SI{0.03}{\hertz} spanning \SI{\pm 50}{\kilo\hertz}. The RF background was emitted via a slightly tilted AC coil, producing a finite transverse ($\xhat\tm\yhat$) component and inhomogeneous coupling into the RF resonator.  
                (\textbf{c}) Extracted frequency-response function from envelope analysis in (a), plotted versus the instantaneous frequency offset in (b). The band-like structure arises from the finite bandwidth of the RF resonator resonance and periodic signal cancellation intrinsic to the acquisition scheme. Magnetization is sampled in discrete \SI{76}{\micro\second} windows, and the Larmor peak is extracted by averaging the real and imaginary ($I$, $Q$) components from dual-quadrature acquisition. If the external RF field completes an integer number of periods within the acquisition window, the averaged imprint vanishes, producing minima at integer multiples of $1/(\SI{76}{\micro\second}) \approx \SI{13.2}{\kilo\hertz}$ offset from the Larmor frequency (black dashed lines). These periodic nulls are superimposed on the underlying frequency-dependent resonator response determined by its quality factor $Q$.
                }
            \label{figSI:rfsusceptibility}
        \end{figure} 
        
        The RF resonator detects the total magnetization vector of the ${}^{13}\mathrm{C}$ nuclear spins by monitoring their Larmor precession in the radio-frequency (RF) domain. In the present setup, the Larmor frequency is approximately \SI{75.4}{\mega\hertz}. Because the detection electronics and the RF resonator are resonant at this frequency, they are also sensitive to RF signals originating from external sources in the same spectral range. We experimentally quantified this susceptibility.
        
        For this purpose, the nuclear spins were polarized and then aligned and stabilized using the PRISM protocol. The parameters (pulse power, pulse duration, spacing between pulses) were identical to those used in the PRISM protocol measurements in other experiments. Following initialization, a controlled artificial RF background was emitted via the AC coil (Fig. \ref{figSI:rfsusceptibility}a). The AC coil was mounted at a slight tilt, such that the resulting RF radiation had a finite transverse ($\xhat\tm\yhat$) component. Inhomogeneity of the AC field additionally induces coupling to the RF resonator. The artificial RF background was injected exactly at the ${}^{13}\mathrm{C}$ Larmor frequency, and its frequency was modulated in time with a triangular waveform at \SI{0.03}{\hertz}, spanning \SI{\pm 50}{\kilo\hertz} (Fig. \ref{figSI:rfsusceptibility}b).
        
        The imprint amplitude was determined by extracting the envelope of the detected RF signal at each instantaneous modulation frequency. The resulting intensity map exhibits a band-like structure (Fig. \ref{figSI:rfsusceptibility}c). This arises from two factors: the finite bandwidth of the RF resonator resonance and the periodic signal cancellation inherent to the acquisition scheme. Magnetization is recorded in discrete windows of \SI{76}{\micro\second}, and from each window the spectral peak at the Larmor frequency is extracted by averaging the real and imaginary components ($I$ and $Q$) from the digitizer (dual-quadrature acquisition). If the externally applied RF field completes exactly one period within the detection window, the averaged imprint is zero. This occurs when the period length equals \SI{76}{\micro\second}, corresponding to a frequency offset of \SI{13.2}{\kilo\hertz}. Consequently, the frequency-response function exhibits its first minima at \SI{\pm 13.2}{\kilo\hertz} relative to the Larmor frequency. This effect repeats periodically with frequency and is superimposed on the intrinsic frequency-dependent response of the RF resonator, which is governed by its quality factor $Q$.

   \section{Robustness to Background Fields: Principle of Differential Signal Extraction} \label{secSI:signalextraction}
        \begin{figure}[t]
            \centering
            \includegraphics[width=0.49\textwidth]{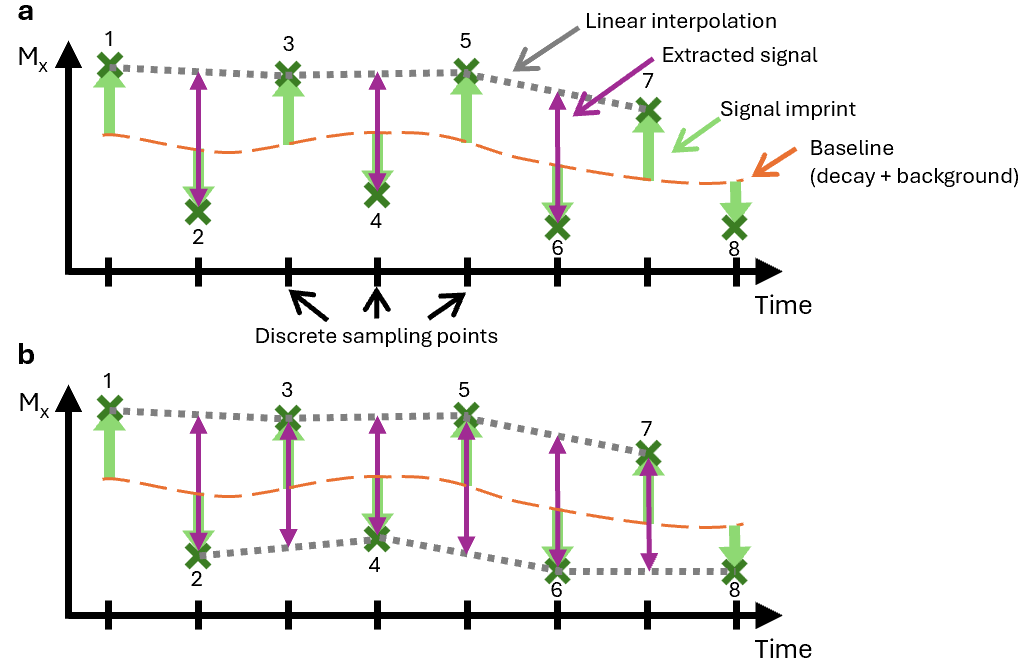}
            \caption{\textbf{Signal extraction method.} 
            (\textbf{a}) Schematic illustration of the reconstruction procedure. 
            Measured data points (green crosses) are separated into even- and odd-indexed points, which correspond to two magnetization vector orientations in the Floquet eigenbasis. 
            Background components (orange dashed line), such as slow signal decay or backgrounds, affect both indices equally, whereas the target DC or AC magnetic field induces opposite changes in $M_x$ for the two orientations, producing a 180° phase-shifted imprint (green arrows). 
            Subtraction of even from odd points isolates the desired signal while suppressing the background. 
            A temporal offset between odd and even datapoints is compensated by linear interpolation (gray dotted lines).
            (\textbf{b}) Schematic illustrating interpolation applied to both even- and odd-indexed data points, which increases the effective sampling rate of the reconstructed signal to the full measurement repetition rate.
            }
            \label{figSI:signal_extraction}
        \end{figure}

        \begin{figure}[t]
                \centering
                \includegraphics[width=0.49\textwidth]{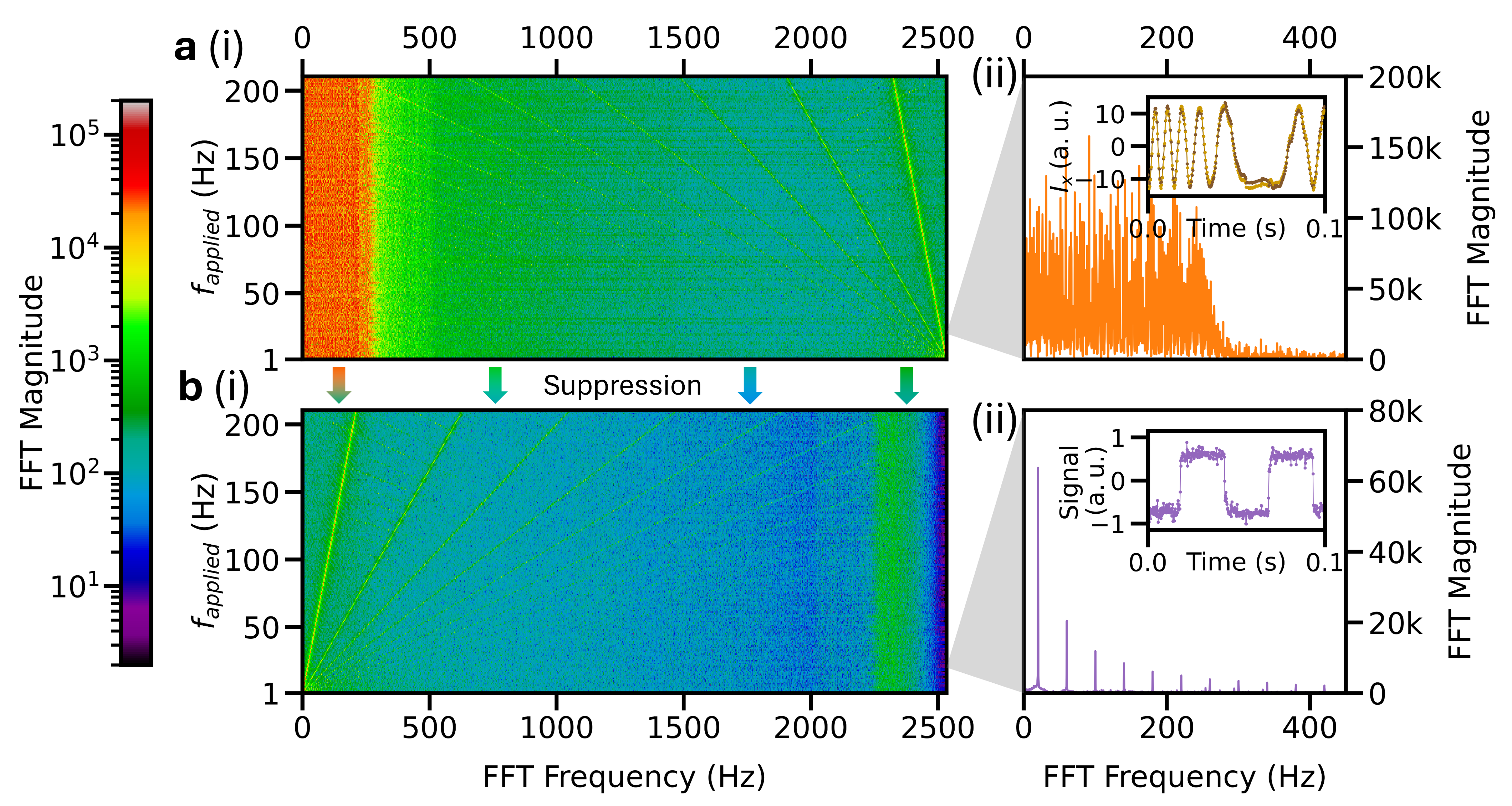}
                \caption{\textbf{Extended differential signal extraction with experimental data.} 
                Measurement of an AC magnetic field with a square waveform in the presence of a broadband RF background (“swish”-type waveform) spanning 0–\SI{250}{\hertz}.  
                (\textbf{a})~(i) Each row of the colormap shows the FFT magnitude spectrum from a single-shot acquisition. The square-wave frequency was varied between shots, scanned from 1 to \SI{210}{\hertz}, producing harmonic features that appear as diagonal lines. Data shown is from the raw signal without background suppression, revealing strong broadband interference from the RF background. Linear interpolation of odd- and even-indexed data points results in mirroring of signal peaks according to $f_\mathrm{background} \rightarrow f_\mathrm{Nyquist} - f_\mathrm{background}$.  
                (ii) FFT spectrum for a \SI{20}{\hertz} square-wave measurement, showing the dominant RF background across a wide frequency range. The inset shows a segment of the time-domain raw data, highlighting the background dominance.  
                (\textbf{b})~(i) Single-shot spectra after applying the background suppression technique (same experimental data as a(i)). Signal peaks retain their original amplitude, while the background is greatly reduced and residual components are mirrored into the higher-frequency range.  
                (ii) Example FFT spectrum for the \SI{20}{\hertz} square wave after suppression, with peaks at odd harmonics of the base frequency clearly resolved. The inset shows a segment of the reconstructed square wave in the time domain.
                }
                \label{figSI:extended_signal_extraction}
            \end{figure}

        Fig.~\ref{figSI:signal_extraction}a summarizes the procedure used to reconstruct DC and AC magnetic signals from the measured transverse magnetization. Because NMR measurements probe the transverse component of the magnetization vector, we analyze the $M_x$ component of the magnetization vector oriented along the $\xhat$-axis. The acquired stream $M_n \equiv M_x(t_n)$ at repetition times $t_n = n\,\Delta t$ is partitioned into two subsequences at even ($k=2n$) and odd ($k=2n+1$) indices. Consecutive samples probe the two stable orientations of the magnetization vector defined by the periodic drive. Because these orientations are related by a \SI{{\approx}180}{\degree} rotation about the spin-lock axis, a small sensing-field-induced tilt of the effective rotation axis—and with it both stable magnetization vector states—produces opposite-sign changes in the measured projection $M_x$ for the two parities, while static or slowly varying backgrounds (relative to the sampling rate) and the slow signal decay imprint identically on both. Consequently, as the sensing field amplitude increases, the $M_x$ values of odd-indexed points increase while those of even-indexed points decrease, creating a mirrored imprint equivalent to a \SI{180}{\degree} phase flip (green arrows in Fig.~\ref{figSI:signal_extraction}a).

        To isolate the signal, we account for the half-cycle temporal offset between odd and even samples by linearly interpolating the odd-indexed sequence onto the even-index time grid (gray dotted lines in Fig.~\ref{figSI:signal_extraction}a), and then form the differential signal $d_{2n}$ on the even grid as
        \begin{equation}
            d_{2n} \equiv 
            M_{2n} - \frac{M_{2n-1}+M_{2n+1}}{2} \, .
        \end{equation}
        This operation cancels common-mode backgrounds while retaining the phase-flipped component, yielding a background-suppressed, time-compensated series. We refer to $d_{2n}$ as the \textit{differential signal}.
        
        The measured magnetic field alters the elevation angles of the two magnetization-vector orientations on the Bloch sphere; the resulting $M_x$ variations are slightly influenced by the sphere’s curvature. At low field strengths (below a few \si{\micro\tesla} in our experiments), this curvature-induced distortion is negligible and largely removed by the subtraction procedure. At larger fields, geometry-based post-processing can be used to account for the curvature. A secondary distortion source is the mild non-linearity between elevation angle and DC bias field (Fig.~\ref{fig:fig5}a). PRISM simultaneously measures DC bias fields directly from the reconstructed signal, enabling calibration of this relationship and removal via post-processing. We observe a monotonic dependence between the reconstructed DC component and the true bias field up to $\sim$\SI{100}{\micro\tesla}, providing a large dynamic range (Fig.~\ref{fig:fig5}b).
        
        It is noteworthy that this sensing scheme can detect magnetic fields stronger than the periodic $\zhat$-drive used to generate the stroboscopic orbit. The orbit itself arises from two stable orientations (rotation eigenvectors of the driven evolution) induced by resonant $\zhat$-rotation, which produces a large splitting between the two magnetization positions on the sphere; magnetic sensing fields are then extracted from changes in the effective axis direction.

        \paragraph{\T{Effect of Signal Decay:}}  \label{secSI:effectofdecay}
        The $^{13}\mathrm{C}$ nuclear spins in diamond exhibit a very long $T'_2$ lifetime under our pulse sequence—more than $70,000$ times longer than $T_2^\ast$~ (cf. Fig. \ref{figSI:sensitivity_ramsey}a and Fig. \ref{figSI:lifetime}a)—ensuring a stable signal over timescales of several seconds and enabling continuous interrogation over minutes.  
        Over these longer durations, however, a gradual decay of the net magnetization is observed (Fig.~\ref{figSI:lifetime}), dependent on sample properties and potentially accelerated by transients in case of strong sudden changes in the external field.  
        
        For such measurements, the differential signal is normalized to a baseline that follows the amplitude decay.  
        This baseline is extracted by computing the mean of interpolated even and odd values (rather than their difference), followed by smoothing over neighboring points.  
        The smoothing window length can be tuned according to background intensity and spectral content.  
        We term the decay-compensated trace the \textit{normalized differential signal}.
        
        \paragraph{\T{Sampling Rate and Frequency Range:}} \label{secSI:samplingrate}
        Because the reconstructed signal is obtained from the difference between odd- and even-indexed data points, its effective sampling rate is half of the raw measurement rate (i.e., the $\xhat$-pulse repetition rate).  
        
        Consequently, the highest measurable signal frequency is limited to half the Nyquist frequency of the original acquisition.  
        For example, with $\xhat$-pulses of $\SI{{\approx}100}{\micro\second}$ duration and \SI{100}{\micro\second} spacing, the raw sampling rate is \SI{5}{\kilo\hertz}, yielding a maximum measurable signal frequency of \SI{1.25}{\kilo\hertz}.  
        
        Increasing pulse power shortens the pulse duration, thereby increasing both the sampling rate and the maximum measurable signal frequency.  
        Similarly, reducing the inter-pulse spacing increases sampling rate but slightly reduces SNR unless the pulse duration is scaled accordingly.

        \paragraph{\T{Extended Signal Extraction Method:}}
            As an extension of the standard reconstruction scheme described above, linear interpolation can be applied to \emph{both} the even-indexed and odd-indexed data points (Fig.~\ref{figSI:signal_extraction}b).  
            In the frequency domain, this procedure results in mirrored signal peaks: any signal frequency $f_\mathrm{signal}$ is mapped to $f_\mathrm{Nyquist}-f_\mathrm{signal}$ (Fig.~\ref{figSI:extended_signal_extraction}a).  
            
            Because even- \textit{and} odd-indexed data points are temporally interpolated, the effective sampling rate of the extracted signal becomes equal to the full measurement sampling rate (i.e., the $\xhat$-pulse repetition rate).  
            However, this increased rate does not yield a higher signal-to-noise ratio, as the interpolation does not introduce additional independent measurements.
            
            Applying a Fourier transform to the interpolated differential signal allows the identification of signal peaks and their separation from background components (Fig.~\ref{figSI:extended_signal_extraction}b).  
            This approach is particularly advantageous when background noise occupies only a limited spectral range and for static or quasi-static signals in which frequency and amplitude remain constant throughout the acquisition.  
            Under such conditions, the signal can be cleanly extracted from the Fourier spectrum, enabling complete suppression of broadband but spectrally limited background interference.

\section{Characterization of Sensitivity} \label{secSI:sensitivity}
        We quantified the magnetic field sensitivity using a calibrated test signal generated by a small coil positioned adjacent to the sensor. The coil was driven by a function generator producing a sinusoidal field of known amplitude and frequency along the $\zhat$–axis. For sensitivity determination, a \SI{2.47}{\micro\tesla} signal at \SI{20}{\hertz} was applied.
        
        \begin{figure}[t]
            \centering
            \includegraphics[width=0.49\textwidth]{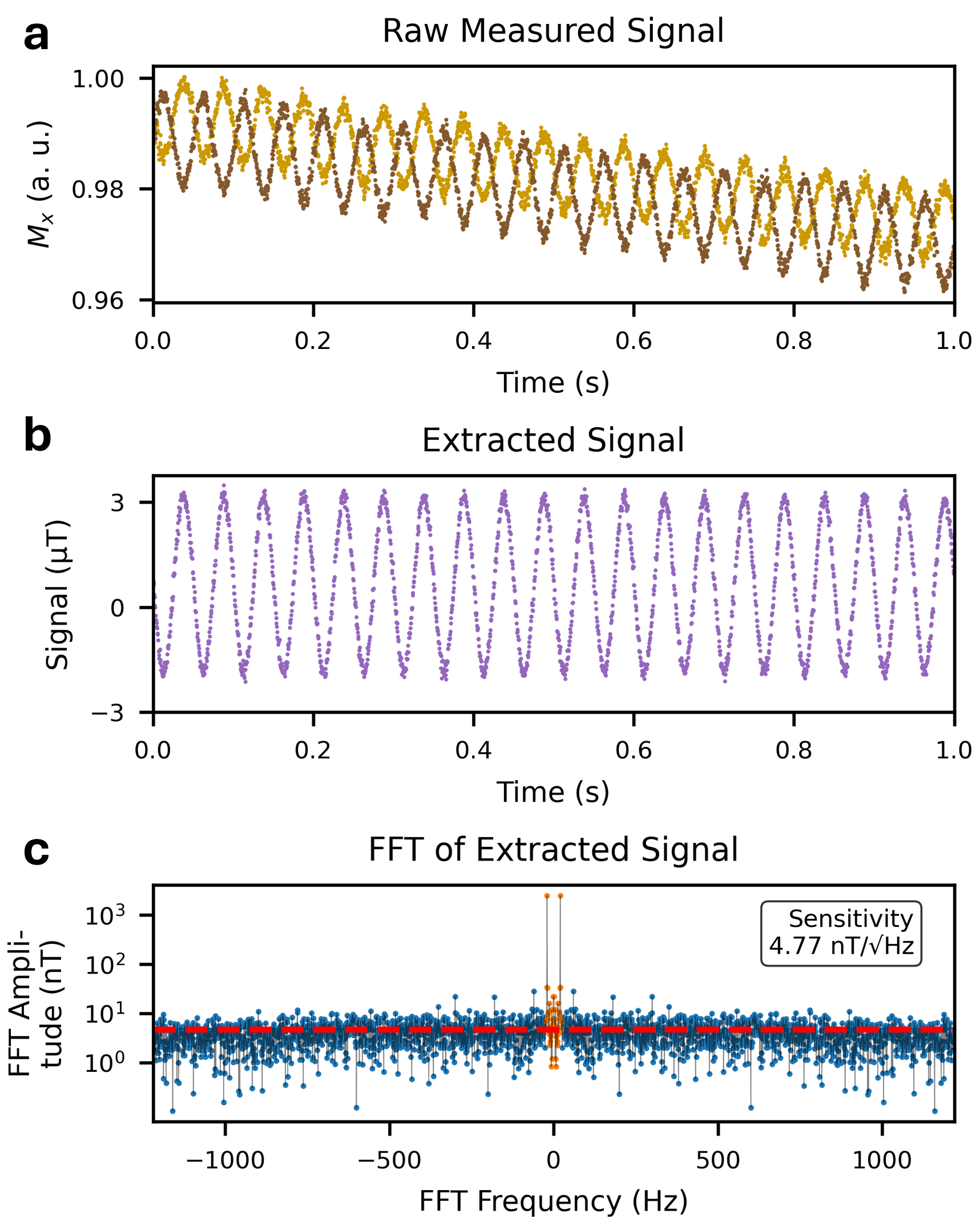}
            \caption{\textbf{Sensitivity measurement.}
                (\textbf{a}) Raw $M_x$ signal showing the imprint of the injected sinusoidal test field (\SI{2.47}{\micro\tesla}, \SI{20}{\hertz}). Data points are plotted alternately in brown and gold, highlighting the two manifolds.
                (\textbf{b}) Differential signal extracted from the data in (a), scaled to the magnetic field strength obtained from auxiliary coil calibration. 
                (\textbf{c}) Fourier spectrum (blue) of the extracted signal in (b). The main signal peaks and the DC offset have been masked (orange); the RMS noise floor (red dashed line) was calculated from the remaining peaks, yielding a sensitivity of \SI{4.77}{\nano\tesla\per\sqrt{\hertz}}.
            }
            \label{figSI:sensitivity_heatmap}
        \end{figure}
        
        From the continuous measurement, a \SI{1}{\second} segment was taken at a predefined interval after the start of the measurement (Fig.~\ref{figSI:sensitivity_heatmap}a-b). The extracted signal was Fourier-transformed, and the amplitude of the injected signal peak was used to scale the spectral magnitudes. The signal and low-frequency components (e.~g. due to decay and bias field) were masked, leaving only the noise-dominated portion of the spectrum (Fig.~\ref{figSI:sensitivity_heatmap}c).
        
        The sensitivity $\mathcal{S}$, reported in \si{\nano\tesla\per\sqrt{\hertz}}, was obtained from the root-mean-square (RMS) level of the residual noise floor, normalized by the square root of the spectral bin width $\Delta f$:
        \begin{equation}
            \mathcal{S} 
            = \frac{\mathrm{RMS}(\text{noise floor})}{\sqrt{\Delta f}},
        \end{equation}
        with
        \begin{equation}
            \mathrm{RMS}(\text{noise floor}) 
            = \sqrt{\frac{1}{N} \sum_{i=1}^{N} \left( \frac{2|X_i|}{L} \right)^2 },
        \end{equation}
        and
        \begin{equation}
            \Delta f = \frac{f_s}{L},
        \end{equation}
        where $X_i$ are the Fourier amplitudes in the noise-only spectral region, $L$ is the total sample length, $N$ is the number of bins used for noise evaluation, and $f_s$ is the sampling rate.
        
        Using this procedure, the measured sensitivity was \SI{4.77}{\nano\tesla\per\sqrt{\hertz}}. This estimate assumes that no additional AC magnetic fields are present beyond the injected test field. In practice, weak external signals can elevate the apparent noise floor; for instance, the amplifier driving the coil introduced a \SI{60}{\hertz} magnetic field, which was correctly detected by the sensor since it is indeed a magnetic field and not interference, but artificially degraded the calculated sensitivity. The intrinsic sensitivity of the shown measurement is therefore expected to be slightly better than the stated value.
        
        \paragraph{\T{Comparison with Ramsey Sensing:}}
            The experimental configuration presented here was not specifically optimized for best sensitivity. Therefore, to contextualize our sensitivity, we compare them to a standard Ramsey sensing protocol, which is widely used in precision magnetometry and can achieve high sensitivities under optimal conditions. Despite its prevalence, Ramsey sensing differs from—and in certain respects is disadvantageous to—our continuous measurement protocol.
            
            Ramsey sensing relies on the free precession of spins after an initial $\pi/2$ excitation, and is therefore limited by the transverse relaxation time $T_2^*$. In systems with short $T_2^*$, such as $^{13}\mathrm{C}$ in diamond, the signal lifetime can be a few \si{\milli\second}, in contrast to the ${>}\SI{1}{\minute}$ continuous acquisition achievable with our method. Consequently, Ramsey measurements must be reinitialized repeatedly. For the purposes of this comparison and consistency with literature values, we neglect the reinitialization overhead, although in practical continuous-sensing scenarios it significantly reduces the effective sensitivity by several orders of magnitude compared to our proposed PRISM sequence.
            
            To determine the Ramsey sensitivity, we performed a free induction decay (FID) measurement. $^{13}\mathrm{C}$ spins were hyperpolarized using the same procedure as in our protocol. A $\pi/2$ pulse along the $\yhat$–axis was then applied, placing the spins along the $\xhat$–axis. The ensuing FID was acquired over \SI{4}{\milli\second}, sufficient for the polarization to decay completely given the short $T_2^*$ (${\approx}\SI{1.3}{\milli\second}$).
            
            For analysis, we computed the Fourier spectrum of the same data set but masked over varying intervals $\tau_\mathrm{acq}$, each starting immediately after the $\pi/2$ pulse. This allowed determination of the acquisition interval yielding maximum sensitivity. For each spectrum, the central peak was fit using a Pseudo-Voigt function, and the standard error of the peak center was extracted. This error was then normalized by $\sqrt{\tau_\mathrm{acq}}$ to yield the sensitivity in $\mathrm{nT}/\sqrt{\mathrm{Hz}}$.
            
\begin{figure}[t]
                \centering
                \includegraphics[width=0.49\textwidth]{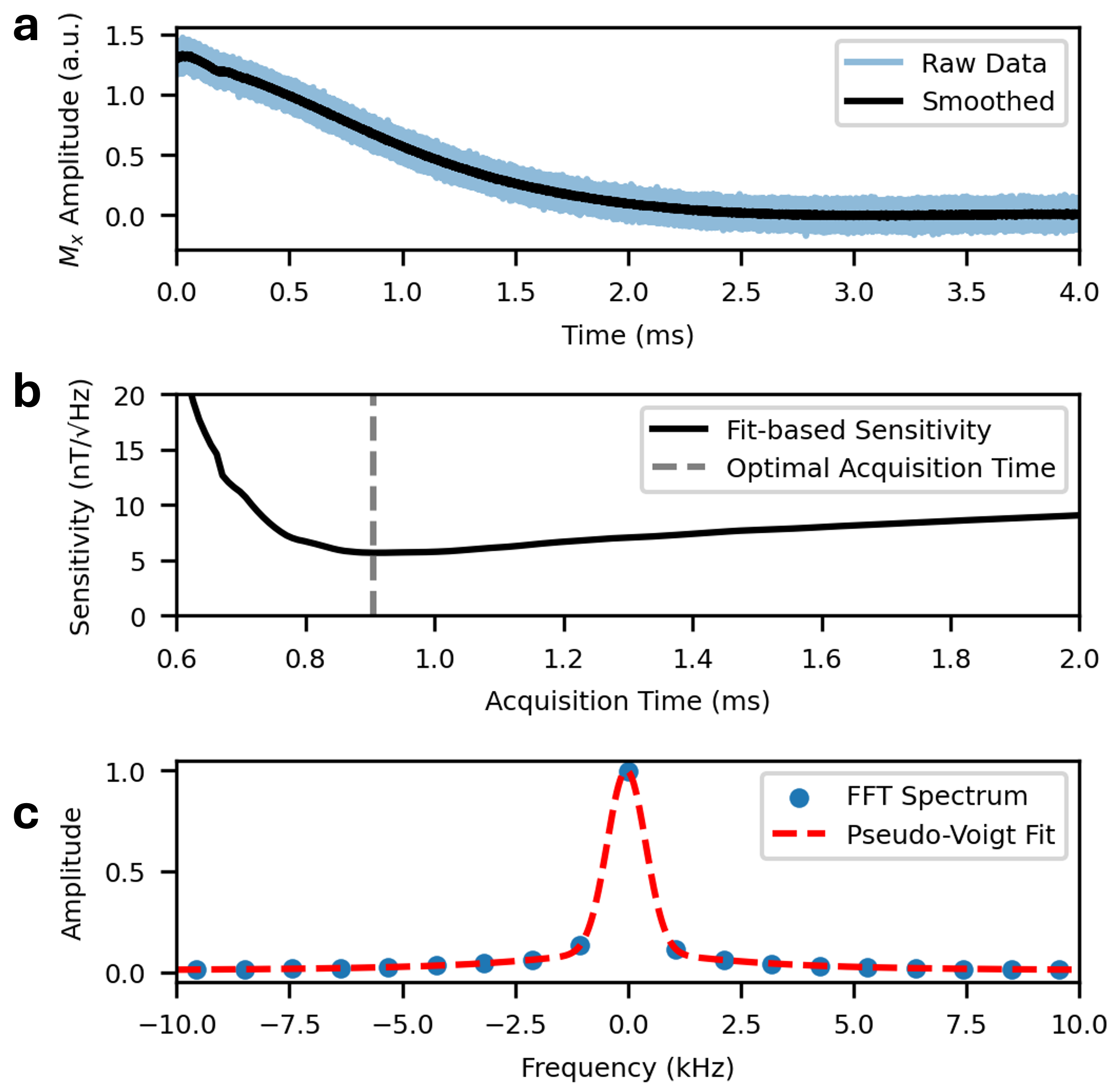}
                \caption{\textbf{Sensitivity of Ramsey sensing for comparison with PRISM.}
                    (\textbf{a}) $M_x$ amplitude during the FID, shown as raw data (light blue) and smoothed trace (black).
                    (\textbf{b}) Sensitivity as a function of acquisition interval $\tau_\mathrm{acq}$ (black line), with optimum reached at $\tau_\mathrm{acq}=\SI{0.906}{\milli\second}$ (gray dashed line).
                    (\textbf{c}) Fourier spectrum for the optimum $\tau_\mathrm{acq}$ (blue scatter), with fitted Pseudo-Voigt profile overlaid (red dashed line).
                }
                \label{figSI:sensitivity_ramsey}
            \end{figure}

            Using this method, the optimum sensitivity was found to be \SI{5.69}{\nano\tesla\per\sqrt{\hertz}} for an acquisition window of $\tau_\mathrm{acq} = \SI{0.906}{\milli\second}$ (Fig.~\ref{figSI:sensitivity_ramsey})—approximately $20\,\%$ worse than that achieved with our newly proposed robust protocol. Under the constraint of the short acquisition window, the best achievable resolution with the Ramsey protocol was limited to \SI{182}{\nano\tesla}, compared to the sub-\si{\nano\tesla} regime achieved with PRISM when considering the long continuous measurement time.

\section{Pathways to Increase Sensitivity} \label{secSI:optimization}
        The sensitivity of the system can be enhanced by tuning key control parameters. We discuss a few critical factors below.
        
        \paragraph{\T{Maximizing Elevation Angle:}} \label{secSI:maximizingelevationangle}
            The measurement protocol relies on detecting changes in the $\xhat$‑component of the magnetization, which directly follows the applied external magnetic field. Sensitivity therefore increases with the magnetization’s deflection towards the $\zhat$-axis. This deflection is quantified by the \emph{elevation angle}, defined as the angle between the $\xhat$‑axis and each of the two distinct orientations (axes) of the magnetization vector.
            
            The elevation angle can be increased either by adjusting the pulse angle (Fig.~\ref{fig:fig5} and App. \ref{SIMethods:pulseangle}) or by varying the amplitude of the $\zhat$‑drive field. To quantify this dependence, the $\zhat$‑drive strength was scanned in random order to suppress effects from slow temporal experimental drifts. Two competing effects are observed:
            \begin{itemize}
                \item A stronger $\zhat$‑drive increases the elevation angle, thereby enhancing sensitivity.
                \item Conversely, a stronger $\zhat$‑drive reduces the overlap between the system’s Hamiltonian before and after activation of the $\zhat$‑drive, leading to increased signal loss.
            \end{itemize}

            A useful single metric capturing both effects is the $\zhat$‑component of the magnetization, $M_z$, which is therefore employed to determine the optimal operating point.
            
\begin{figure}[h]
                \centering
                \includegraphics[width=0.49\textwidth]{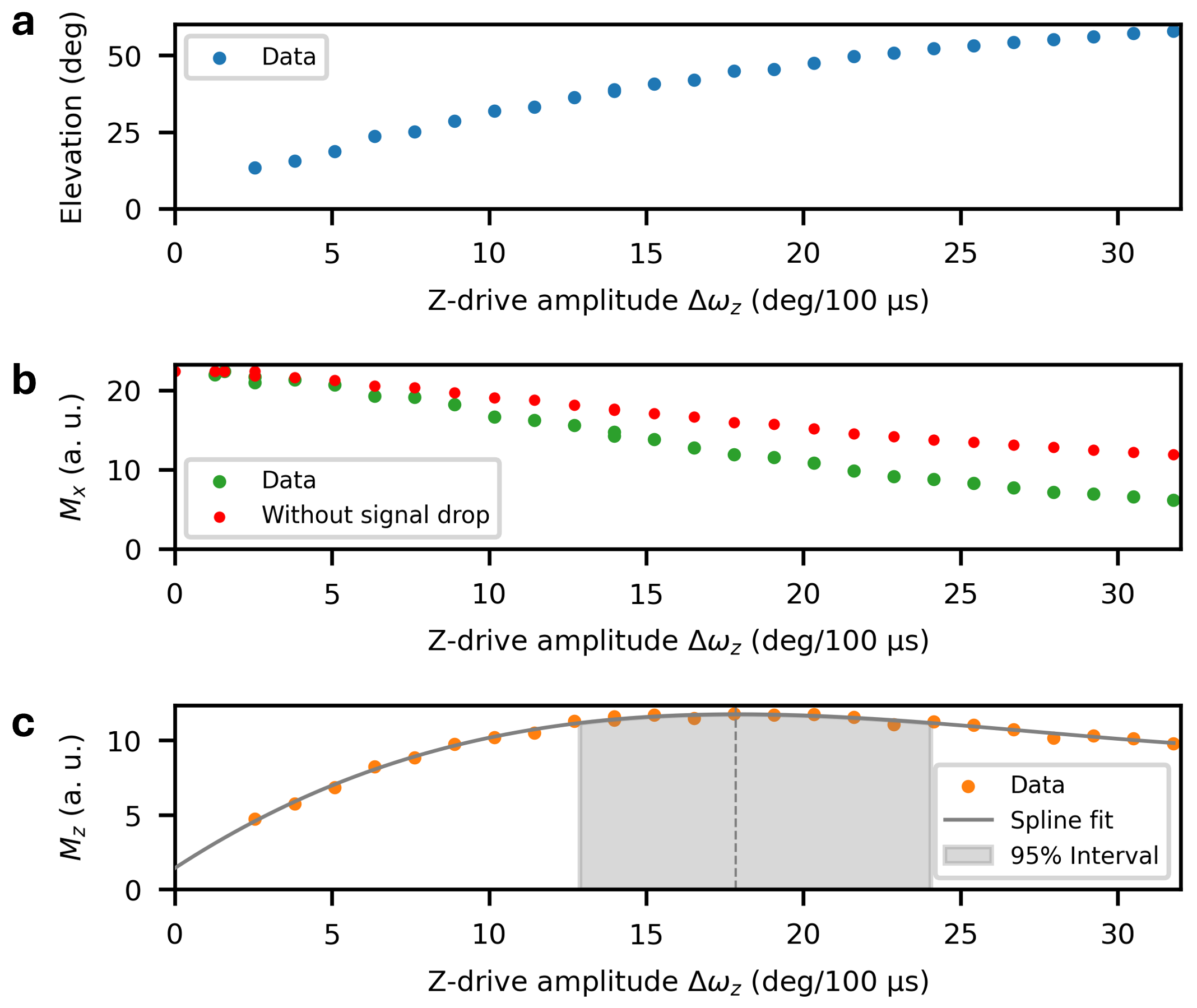}
                \caption{
                    \textbf{Dependence of the magnetization vector position on $\zhat$-drive strength.}
                    (\textbf{a}) Measured elevation angle of the magnetization vector as a function of the $\zhat$-drive amplitude. A stronger $\zhat$-drive deflects the magnetization further towards the $\zhat$-axis, producing a monotonic increase in elevation angle and enhancing the relative response to external fields.  
                    (\textbf{b}) Geometrically predicted $M_x$ reduction (red) due solely to the change in elevation angle, compared to the experimentally measured $M_x$ reduction (green). The additional loss beyond the geometric prediction is attributed to decreased overlap between the system’s Hamiltonians before and after activation of the orbit field, leading to transient signal loss.  
                    (\textbf{c}) The $\zhat$‑component of the magnetization, $M_z$, serves as a single metric capturing both sensitivity gain from elevation-angle increase and sensitivity loss from reduced Hamiltonian overlap. The resulting curve exhibits a broad maximum, with the optimum for a spacing length of \SI{100}{\micro\second} occurring at a rotation rate of $18^\circ/\SI{100}{\micro\second} $. A 95\% performance plateau extends from $13^\circ/\SI{100}{\micro\second}$ to $24^\circ/\SI{100}{\micro\second} $, indicating that precise $\zhat$‑drive tuning is unnecessary and coarse adjustment near the optimum suffices for robust operation.
                }
                \label{figSI:elevationanglevszdrive}
            \end{figure}

            The experimental data reveals that the elevation angle increases monotonically with $\zhat$‑drive strength (Fig.~\ref{figSI:elevationanglevszdrive}a). Geometrically, a larger elevation angle necessarily reduces the $\xhat$‑projection of the magnetization vector (Fig.~\ref{figSI:elevationanglevszdrive}b, red points). However, the measured reduction in the $\xhat$‑component (Fig.~\ref{figSI:elevationanglevszdrive}b, green points) is greater than predicted by geometry alone, indicating an additional signal‑loss mechanism, namely the reduced overlap of the system's Hamiltonians at orbit field activation. As these two effects counteract in their influence on $M_z$, the resulting dependence exhibits a maximum but with a broad plateau (Fig.~\ref{figSI:elevationanglevszdrive}c).
            
            For a spacing length of \SI{100}{\micro\second} the optimal performance occurs at a rotation rate of $18^\circ/\SI{100}{\micro\second}$, with a $95\%$ plateau spanning $13^\circ/\SI{100}{\micro\second}$ to $24^\circ/\SI{100}{\micro\second}$ (Fig.~\ref{figSI:elevationanglevszdrive}c). This wide tolerance implies that precise tuning of the $\zhat$‑drive amplitude is unnecessary; coarse adjustment near the optimum is sufficient.

        \paragraph{\T{Increasing Sampling Rate:}} \label{secSI:increasesamplingrate}
            Increasing the sampling rate broadens the sensor’s accessible frequency bandwidth and significantly enhances rejection of spurious background signals. The latter improvement arises because linear interpolation between discrete measurement points (cf. App.~\ref{secSI:samplingrate}) incurs progressively smaller errors at higher sampling frequencies.
            
            Three practical strategies can increase the sampling rate:
            
            \begin{itemize}
                \item \textbf{Shorter $\xhat$‑pulses:} Increasing $\xhat$‑pulse power reduces pulse duration, directly expanding the sensor’s frequency bandwidth while increasing the suppression factor.
                \item \textbf{Reduced pulse spacing:} Shorter intervals between $\xhat$‑pulses similarly broaden the bandwidth and strengthen background suppression. This approach slightly reduces the acquisition time within each cycle, lowering sensitivity if the ratio of acquisition time to total cycle period decreases. In practice, sensitivity losses can be mitigated by shortening the $\xhat$‑pulse duration in concert, thereby maintaining the optimal ratio.
                \item \textbf{Alternative spin systems:} Systems with inherently faster spin dynamics extend the operating range of the protocol. Protons ($^1$H), for example, have a gyromagnetic ratio roughly four times larger than that of $^{13}$C, permitting significantly shorter control pulses and thus higher sampling rates without loss of signal integrity.
            \end{itemize}
            
\begin{figure}[h]
                \centering
                \includegraphics[width=0.49\textwidth]{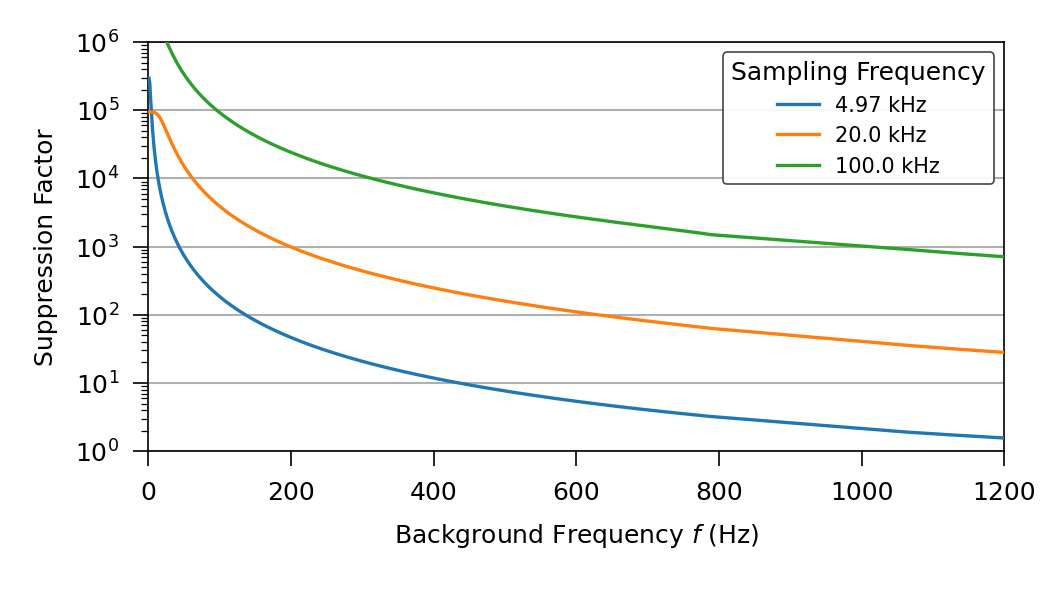}
                \caption{
                    \textbf{Effect of sampling rate on background suppression.}  
                    Calculated suppression factor versus the frequency of interfering background signals for three sampling‑rate scenarios:  
                    (i) pulse/spacing durations of \SI{101.2}{\micro\second}  / \SI{100}{\micro\second}  (${^{13}\mathrm{C}}$, present work, blue);  
                    (ii) \SI{15}{\micro\second}  / \SI{35}{\micro\second}  (orange); and  
                    (iii) \SI{5}{\micro\second}  / \SI{5}{\micro\second} (${^{1}\mathrm{H}}$‑compatible regime, green).  
                    Higher sampling rates improve suppression by up to three orders of magnitude while proportionally extending the usable AC sensing bandwidth; at a \SI{100}{\kilo\hertz} sampling rate, the Nyquist limit of \SI{25}{\kilo\hertz} defines the upper bound of detectable AC fields. The substantial gain in noise rejection demonstrates that modest modifications to pulse duration and spacing can yield dramatic improvements in background suppression performance.
                    }
                \label{figSI:samplingratecomparison}
            \end{figure}

            We calculated the influence of sampling rate on the suppression factor, quantifying the degree of background rejection, for several experimentally relevant scenarios (Fig.~\ref{figSI:samplingratecomparison}):  $\xhat$‑pulse and spacing durations of
            (i) \SI{101.2}{\micro\second} and \SI{100.0}{\micro\second}, as used in the present ${}^{13}\mathrm{C}$ measurements;  
            (ii) \SI{15}{\micro\second} and \SI{35}{\micro\second}, representing stronger pulses and reduced spacing;  
            (iii) \SI{5}{\micro\second} and \SI{5}{\micro\second}, representative of possible performance with ${}^{1}\mathrm{H}$ nuclear spins.
            
            Across these scenarios, the suppression factor increases by up to three orders of magnitude, while the measurable AC‑field bandwidth at a \SI{100}{\kilo\hertz} sampling rate extends from DC to \SI{25}{\kilo\hertz}. These findings illustrate that relatively modest adjustments to control parameters can substantially boost both background resilience and dynamic range, without sacrificing sensitivity when protocol parameters are appropriately changed.

        \paragraph{\T{Optimizing Diamond Sample:}}
            The experiments reported in this study were performed using a natural-abundance $^{13}$C (1.1\%) single-crystal CVD diamond sample and can be reproduced with other commercially available diamonds. The signal amplitude, and consequently the sensitivity, can be enhanced through the use of $^{13}$C-enriched diamond material. 

            The diamond used in this work is significantly smaller than the RF resonator, resulting in a filling factor of $\sim$4\%. An increase in the filling factor, combined with the intrinsic robustness of the protocol, is expected to yield up to a fourfold sensitivity enhancement. The quality factor ($Q$) of the RF resonator in the present setup is approximately ${\sim}77$ and offers further potential for optimization.

        \paragraph{\T{Enhancing Hyperpolarization:}} \label{secSI:enhancinghyperpolarisation}
            The initial hyperpolarization of $^{13}$C nuclear spins at the start of the experiment determines the number of spins subsequently available for detection. A higher degree of nuclear spin polarization directly translates to stronger NMR signal intensity. Optimization can be achieved through variation of the microwave control parameters (e.g., sweep bandwidth, power), as well as by improving the spin polarization of the NV$^-$ centers via higher power optical pumping with green laser light (e.g., increased laser power, more homogeneous illumination across the sample volume)~\cite{Sarkar22}. 

            NV$^-$ centers in the diamond lattice act as a polarization source. The polarization is transferred to proximal $^{13}$C nuclei. Increasing the NV$^-$ center density enhances the spatial polarization distribution. However, higher NV$^-$ concentrations typically entail increased P1 center (substitutional nitrogen) content due to growth-related constraints. Both NV$^-$ and P1 centers serve as relaxation channels for $^{13}$C nuclear spins, reducing their longitudinal relaxation times ($T_1$). Consequently, optimal matching of NV$^-$ concentration and N:NV$^-$ ratio is crucial for maximizing sensitivity~\cite{plotzki2025defect}.

        \paragraph{\T{Pulse-assisted Initialization of Trajectory:}}
            Activation of the $\zhat$-drive induces a transient evolution of the nuclear spins from the spin-lock axis (aligned along the $\xhat$-axis) toward the new prethermal eigenstates of the driven pulse sequence. During this non-adiabatic transition the net magnetization is reduced, resulting in a ${\sim}25\%$ drop in signal amplitude—and hence sensitivity—in our experiments (Fig.~\ref{figSI:elevationanglevszdrive}b, red vs.~green points at $18^\circ/\SI{100}{\micro\second}$).
            
            This loss can be mitigated by applying a short pre-alignment $\yhat$-pulse at the onset of the $\zhat$-drive. The pulse rotates the spins close to the expected eigenstate configuration, reducing the spin rotation required and thereby suppressing non-adiabatic transitions. 
            The approach is conceptually analogous to counterdiabatic driving in quantum control, serving to keep the system close to its instantaneous eigenstates during parameter changes and thereby suppressing non-adiabatic transitions. In practice, such pulse‑assisted initialization can recover much of the transiently lost magnetization, enhancing signal amplitude and sensitivity.

\section*{\T{Supplementary Methods}}
    \section{Reconstruction of the 3D Magnetization Vector} \label{secSI:3Dextraction}
        In conventional NMR experiments, only the components of the magnetization \emph{perpendicular} to the external magnetic field (i.e., perpendicular to the $\zhat$-axis) can be measured directly.  
        These correspond to the transverse $\xhat$ and $\yhat$ components, while the longitudinal $\zhat$-component is not directly accessible.
        For background suppression, we therefore use only the projection of the magnetization onto the $\xhat\yhat$-plane, i.e. the ones we can measure directly.  
        However, for understanding and verifying the operating principle of our sensor, as well as for measuring the elevation angle dependency on other parameters and for calibrating the sensor, it is advantageous to also determine the $\zhat$-component of the magnetization.  
        To this end, we developed a method to extract $M_z$ in every shot using a tailored protocol.

\begin{figure}[h]
            \centering
            \includegraphics[width=0.49\textwidth]{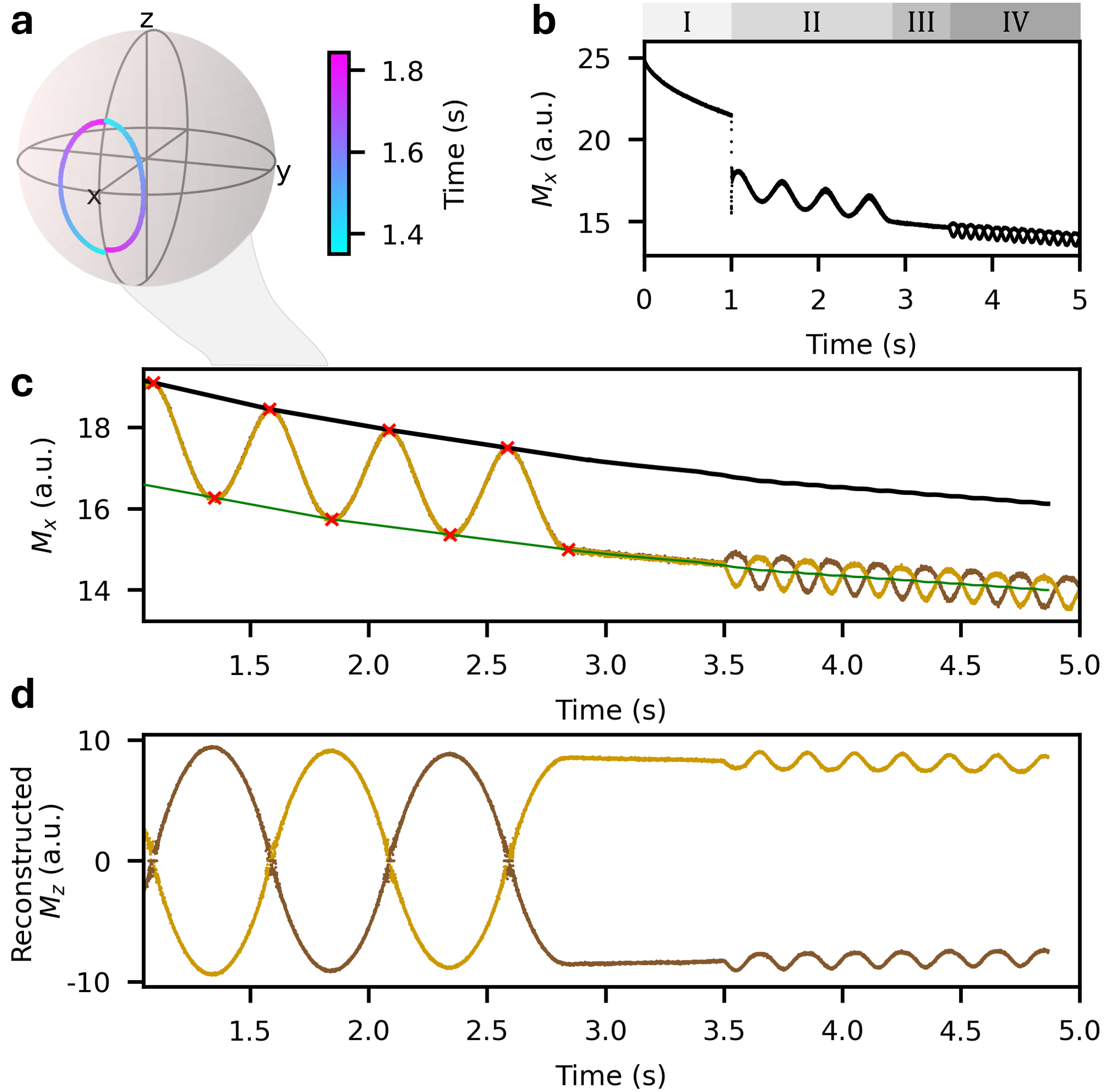}
            \caption{
                \textbf{Rotation of two magnetization vector states via phase modulation of the $\zhat$‑drive.}  
                (\textbf{a}) Plot of experimental data of the rotation of two magnetization eigenstates from the PRISM trajectory when the phase of the $\zhat$‑drive (orbit field) is continuously varied. Both states undergo coherent rotation about the $\xhat$‑axis.  
                (\textbf{b}) Experimental sequence:  
                (I) spin‑lock pulses only;  
                (II) activation of the $\zhat$‑drive to generate the trajectory, with slow variation of the orbit‑field phase to induce $\xhat$‑axis rotation of both eigenstates;  
                (III) rotation halted at the point where the elevation angles of both states are maximized;  
                (IV) sensing: introduction of an external AC magnetic field, subsequently detected by the protocol.  
                (\textbf{c}) Magnified view of the same measurement, showing the time‑resolved trajectory signal. Minima and maxima are identified (red crosses), and the maxima are referenced to a baseline level (green) to extrapolate the norm of the full 3D magnetization vector (black).  
                (\textbf{d}) $\zhat$‑component of the magnetization, $M_z$, determined from the total magnetization norm in (c) for the same experiment.
            }
            \label{figSI:zextractionexplained}
        \end{figure}

        Our approach exploits the fact that the norm of the magnetization vector changes only minimally over short time intervals (e.g. one second), being affected primarily by the intrinsic decay (characterized by $T_2'$) and thermal noise in the readout circuit.
        We also utilize the ability to control the rotation of the magnetization vector about the $\xhat$-axis by shifting the phase of the $\zhat$-drive $B_{\mathrm{orbit}}$.  
        At the beginning of the experiment, the $\zhat$-drive is switched on and its phase is continuously varied.  
        This is achieved by introducing a small frequency offset of approximately $0.005\%\!-\!0.01\%$ to the $\zhat$-drive frequency for a duration of about one second.  
        This offset corresponds to a rotation frequency of the two measured magnetization vector axes about the $\xhat$-axis of roughly $1$–\SI{2}{\hertz} (Fig.~\ref{figSI:zextractionexplained}).

        While the two points on the Bloch sphere rotate about the $\xhat$-axis, we continuously measure the $\xhat\yhat$-projection of the magnetization.  
        Whenever the points cross the $\xhat\yhat$-plane, the $\zhat$-component is zero and the $\xhat\yhat$-projection reaches its maximum.  
        At these instants, we determine the full norm of the magnetization $|\vec{M}|$ (Fig.~\ref{figSI:zextractionexplained}c, upper red crosses).  
        By recording several complete rotations, we obtain multiple such maxima and linearly interpolate the values between them (Fig.~\ref{figSI:zextractionexplained}c, black curve).  
        This allows us to calculate the $\zhat$-component over the entire interval via:
        \begin{equation}
            M_z = \pm\sqrt{|\vec{M}|^2 - |\vec{M}_{xy}|^2}.
            \label{eq:SI:3D_extraction}
        \end{equation} 
        To determine the sign, we assume that the magnetization alternates between configurations in which one of the two vectors lies above and the other below the effective pulse axis, i.e. the $\xhat\yhat$-plane without bias fields. 
        This assumption is justified by both experimental observations and numerical simulations (cf. App. \ref{secSI:SimulationRotationMatricesBRAYDEN}). 

        To reduce the influence of readout noise, the measured values for finding the envelope are smoothed at each manifold by convolution over windows of length $100$~samples. This value is adjusted depending on the specific experiment.

        We stop the rotation at the moment when the $\xhat\yhat$-projection is minimal, i.e., when $M_z$ is maximal.  
        At this instant, the $\yhat$-component is nearly zero because the points lie in the $\xhat\zhat$-plane, and we use just $M_x$ as the readout component for the AC magnetic field detection.  
        The $\zhat$-component during the subsequent measurement is then determined by extrapolating the magnetization norm along with the smoothed baseline. This baseline is shown as a green line and the extrapolated norm as a black line in the right half of Fig.~\ref{figSI:zextractionexplained}c.  
        The baseline is obtained using convolution with a window size adapted to the specific experiment (typically ${\sim}2000$ samples) or with Savitzky–Golay filtering.  
        The $\zhat$-component is then calculated using Eq. \ref{eq:SI:3D_extraction} again.
        When the baseline smoothing window size is sufficiently large, it is unaffected by the measured signal, ensuring that $M_z$ is determined correctly even in the presence of strong imprints due to external AC magnetic fields.

        While not required in the present work, this procedure can, in principle, be repeated multiple times during a single shot for recalibration, potentially minimizing errors in extended measurements lasting several minutes.

        If, during the measurement, the coil picks up a strong background in the $\xhat\yhat$-plane, it adds up to the RF field of the spins and therefore the measured magnetization norm will change, although the true magnetization norm of the $^{13}$C nuclear spins remains constant.  
        To account for this, we fixed $M_z$ when showing RF background effects to the previously determined $\zhat$-component and represented the variations in the $\xhat\yhat$-plane (Fig.~\ref{fig:fig2}b(iv)), in accordance with theoretical expectations.

        \paragraph{\T{Vibration Measurement:}} \label{secSI:vibration3Dextraction}
            In the vibration measurement protocol, the sample was rapidly oscillated along the vertical ($\zhat$) axis during acquisition to quantify the effect of positional modulation on the sensor readout. In contrast to measurements without vibration, the assumption of a slowly varying magnetization magnitude does not hold here: movement within the RF resonator also alters the coupling efficiency, directly affecting the fraction of signal received by the RF resonator.
            
            We therefore make the assumption, based on experiments, that vertical vibration predominantly modulates the effective pulse angle of the spin-lock $\xhat$-drive. This occurs through changes in the relative position between the diamond sample and the RF resonator, which in turn directly influences the measured response function. Potential variations in the static bias field, arising from small vertical displacements, were neglected as a contributor to changes in the trajectory elevation angle. Such bias-field shifts would be directly detected by the sensor; no significant variations were observed in the acquired data.
            
            The transverse magnetization components, \(M_x\) and \(M_y\), were extracted directly from the measurement data by averaging over both manifolds (corresponding to the two eigenstates of the driven spin trajectory). In contrast, the longitudinal component, \(M_z\), is not directly accessible in this protocol and was instead inferred from \(M_x\) and \(M_y\) combined with the estimated trajectory elevation angle.
            
            Determination of the elevation angle relied on the calibrated dependence of the response function on the $\xhat$-pulse angle, as described in App.~\ref{SIMethods:pulseangle}. From the calibration scan, mapping functions were constructed to convert measured test signal response magnitude to corresponding pulse angles, and subsequently to trajectory elevation angles. The analysis was restricted to pulse angles below \(175^\circ\), and the calibration data was smoothed to yield well-defined, monotonic functions suitable for quantitative analysis.

    \section{Suppression Factor: Definition and Experimental Evaluation} \label{secSI:evalsuppfactor}
        The suppression factor $\eta$ quantifies the ability of the protocol to remove a background signal while preserving the amplitude of the desired target signal. It is defined as the ratio of the background amplitude before and after signal extraction, normalized by the change in the target signal amplitude:
        
        \begin{equation} \label{eqSI:SuppFacCalculation}
            \eta =
            \frac{\mathrm{FFT}_{\mathrm{RF,\ before}}}{\mathrm{FFT}_{\mathrm{RF,\ after}}}
            \times
            \frac{\mathrm{FFT}_{\mathrm{Signal,\ after}}}{\mathrm{FFT}_{\mathrm{Signal,\ before}}}
        \end{equation} 

        where $FFT$ stands for the respective peak magnitude in the FFT spectrum.
        
        To determine the suppression factor over the full accessible frequency spectrum, we applied a sinusoidal RF background of known frequency and high amplitude during the measurement of a fixed‑frequency target signal (chosen to avoid spectral overlap with the injected background, e.g. \SI{10}{\hertz} and \SI{50}{\hertz}). The RF background frequencies are scanned in random order  to preemptively mitigate slow drifts. After signal extraction, the FFT of the target signal is computed both before and after extraction, and Eq.~\ref{eqSI:SuppFacCalculation} is used to quantify the suppression factor at each background frequency.
        
        For comparison, we computed a simulated suppression factor by generating an idealized synthetic dataset consisting of the sum of a sinusoidal target and a sinusoidal background. The same extraction procedure is applied, and FFTs before and after extraction yield the suppression factor across the scanned frequencies.
        
        The extraction method assumes that the background remains approximately constant between consecutive sampling points. At higher background frequencies, this condition breaks down as the phase and amplitude vary more rapidly between samples, leading to an observed reduction in suppression factor in the high‑frequency regime. This trend is reproduced by the simulation, confirming the origin of the suppression‑factor roll‑off.

    \section{Experimental Procedure for Vibrational Robustness Testing} \label{SI:MeasurementOfVibration}
        To assess the influence of mechanical vibrations to the measurement method, the sample was moved along the $\zhat$-axis using a shuttler during a single-shot measurement.  
        The shuttler displaced the sample by \SI{\pm 2}{\milli\meter} at a frequency of approximately \SI{2.1}{\hertz}, moving it up and down inside the RF resonator.  
        This displacement is up to an order of magnitude larger than what would be expected under realistic experimental conditions since in a typical implementation, the sensor could be fixed inside the resonator allowing only minor displacement. Nevertheless, this extreme case was chosen to demonstrate the robustness of the measurement method with respect to vibrations. 

        The shuttler motion consists of an acceleration phase and a deceleration phase.  
        To track the motion as accurately as possible, the movement was recorded with a camera.  
        Colored markers were placed on the shuttler and fixed in the background to determine the sample position in each video frame.  
        The shuttler was set to maximum frequency which lead to a total of $20.8$ vibration periods that were recorded.
        Since the shuttler motion is approximately periodic, the position function was smoothed by overlaying all recorded periods and applying a Savitzky-Golay filter (window length: 60 samples, polynomial order: 3) (Fig.~\ref{figSI:vibration}a).

        Afterwards, the measurement data was synchronized with the position data.  
        Because the vibration was initiated during the experiment, the exact start time, sample position, and direction of motion at that moment were known.  
        We refined the synchronization by using minima and maxima observed in the measurement to ensure that the $M_x$-versus-shuttler-position curves matched for both upward and downward motion directions.

        The AC magnetic field to be measured (\SI{20}{\hertz}, \SI{1.8}{\micro\tesla}) was generated by a two-turn coil that remained stationary during the experiment.  
        As a result, the actual magnetic field experienced by the sample depended on its position.  
        The magnetic field was therefore calculated for each sample position using the Biot–Savart law, using that the sample was located at the center axis of the coil. The spatial extent of the diamond sample was taken into account by calculating the probe in small segments.

    \section{Pulse Sequence Timing} \label{secSI:timingzdrive}
        \begin{figure}[h]
            \centering
            \includegraphics[width=0.49\textwidth]{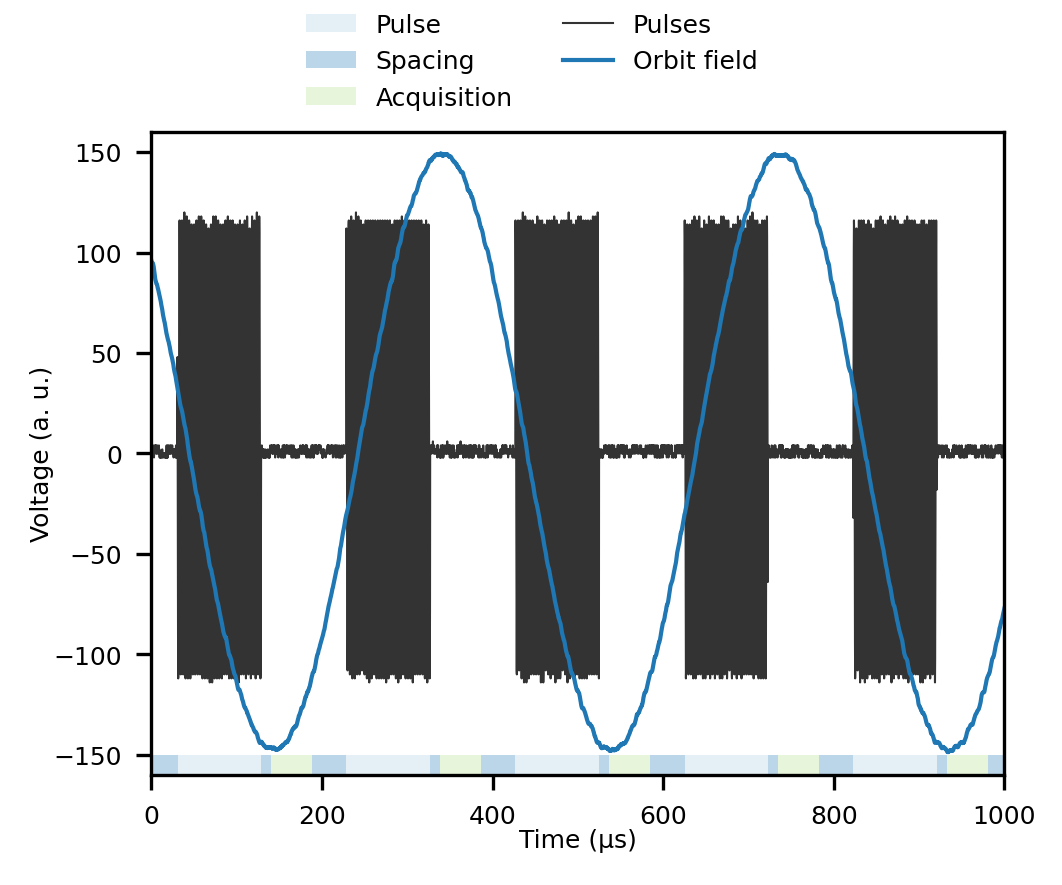}
            \caption{\textbf{Temporal profile of the pulse sequence} recorded using an oscilloscope over a total duration of \SI{1}{\milli\second}. The grey trace shows the $\xhat$-pulse, and the blue trace corresponds to the $\zhat$-drive (orbit field). Shaded regions along the time axis indicate the active pulse period (light blue), the inactive period (``spacing'', dark blue), and the acquisition window within the spacing (light green). In this example, the acquisition time was \SI{48}{\micro\second}, the spacing length was \SI{100}{\micro\second}, and the pulse duration was \SI{98}{\micro\second}.}
            \label{figSI:zdrivetiming}
        \end{figure}

        The orbit field generates the two magnetization axes and must be synchronized with the duration of two $\xhat$-pulses (along with the corresponding inter-pulse distance). Additionally, the relative phase of the orbit field with respect to the $\xhat$-pulses determines the spatial orientation of the two magnetization vector states. Phase variation induces a rotation of the eigenstates about the $\xhat$-axis. For maximum sensitivity, a large elevation angle is desirable; thus, the eigenstates should lie predominantly within the $\xhat\zhat$-plane, with the $\yhat$-component minimized. This condition can be readily achieved by synchronization of the orbit field phase with the $\xhat$-pulses.

        Fig.~\ref{figSI:zdrivetiming} shows an oscilloscope trace recorded with simultaneous acquisition of the control pulses and orbit field. In this measurement, the pulse length was \SI{98}{\micro\second}, the inter-pulse separation was \SI{100}{\micro\second}, and the acquisition window was \SI{48}{\micro\second}, starting \SI{12}{\micro\second} after the pulse end.

    \section{Simulation of Spin Dynamics} \label{secSI:simspin}
        \subsection{Simulation of Magnetization Dynamics with Rotation Matrices}
            \label{secSI:SimulationRotationMatricesBRAYDEN}
            In this section, we describe a simplified framework for modeling the driven spin dynamics using rotation matrices. If the couplings are neglected, then evolution over one cycle amounts to a rotation
            \begin{equation}
            U_{F,k} = e^{-i h_{F;k}T} \equiv e^{-i \gamma\, \mathbf{n}_k \cdot \bm{I}}
            \label{eq:one_cycle_rotation}
            \end{equation}
            where the lowercase \(h_F\) is used to distinguish the simplified Floquet Hamiltonian from the true Floquet Hamiltonian \(H_F\) which includes couplings.  Since \([h_{F;k}, \mathbf{n}_k \cdot \bm{I}] = 0\), the axis \(\mathbf{n}_k \cdot \bm{I}\) is stroboscopically conserved. Therefore, we find that the trajectories can be simulated by finding the unique real eigenvectors (eigenvalue $+1$) of the corresponding $\mathrm{SO}(3)$ rotations for each prethermal axis \(\mathbf{n}_k\) (\(k=1,2\)).
            
\begin{figure}[h]
                \centering
                \includegraphics[width=0.49\textwidth]{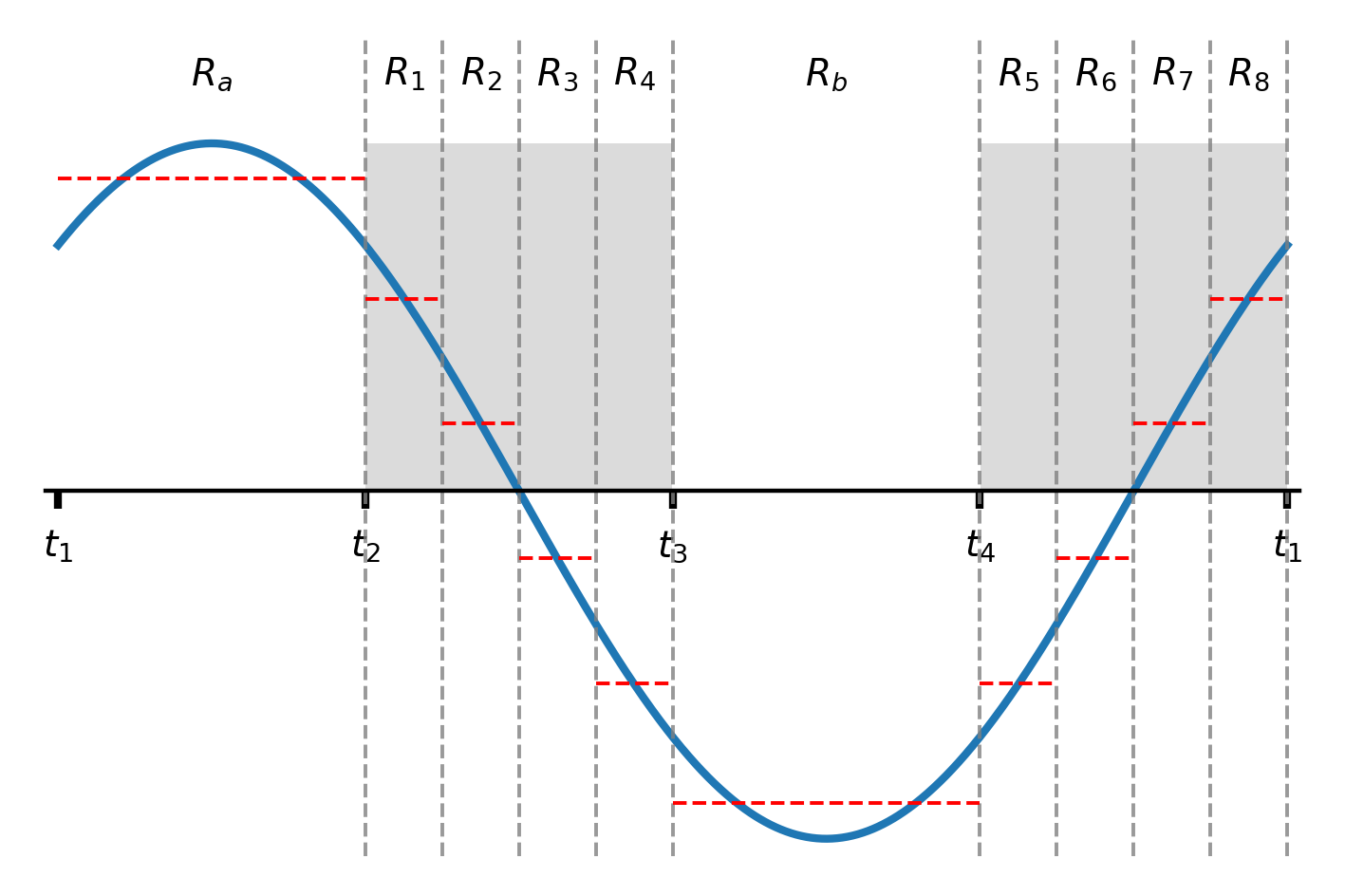}
                \caption{\textbf{Segmented-rotation approximation for one Floquet cycle.} 
                Each $\xhat$-pulse (gray area) is discretized into $N$ subintervals ($N = 4$ shown; $N = 8$ used in simulations). The orbit field (periodic $\zhat$-drive, blue) is assumed constant within each subinterval, with its value taken as the mean over that interval (red). In subinterval $j$, the Hamiltonian is approximated as $H_j = \Omega_x I_x + \omega_z(\tilde{t}_j) I_z$, where $\tilde{t}_j$ denotes the representative time point for that subinterval, producing a rotation about the axis $(\Omega_x, 0, \omega_z(\tilde{t}_j))$ by an angle $\theta_j = \Delta t\, \sqrt{\Omega_x^2 + \omega_z^2(\tilde{t}_j)}$. The windows $R_a$ and $R_b$ correspond to pure $\zhat$-rotations by angles $\phi_a$ and $\phi_b$, respectively, determined by the area under $\omega_z(t)$. The full stroboscopic rotations $R_\mathrm{I}$ and $R_{\mathrm{II}}$ are obtained as time-ordered products of these segment rotations. The prethermal magnetization aligns with the unique real eigenvectors of $R_\mathrm{I}$ and $R_{\mathrm{II}}$.}
                \label{figSI:SimulationWithRotationMatrices}
            \end{figure}

            \textbf{Free evolution (spacing).} During $\tau_a$ (\(t_1\) to \(t_2\) in Fig. \ref{figSI:SimulationWithRotationMatrices}) and $\tau_b$ (\(t_3\) to \(t_4\)) only the $\zhat$-drive acts, yielding pure $\zhat$-rotations 
            \[
            R_a = R_z(\Phi_a), \quad R_b = R_z(\Phi_b)
            \]
            with angles
            \[
            \Phi_a = \int_{t_1}^{t_2} \omega_z(t)\, dt, \quad \Phi_b = \int_{t_3}^{t_4} \omega_z(t)\, dt.
            \]

            \textbf{Finite-width pulses.} During the spin-lock (SL) pulses, evolution is given by a time-ordered exponential

            \begin{equation}
            U_{\mathrm{SL}} = \mathcal{T}\exp\left[-i \int_{\tau_{p}} \left( \Omega I_x + \omega_z(t)\, I_z \right) dt \right]
            \label{eq:unitary for pulse}
            \end{equation}
            Considering a discretized approximation to the full time-ordered operator,  
            \begin{equation}
            U_{\mathrm{SL}} \approx \prod_j \exp\left[ -i\, \Delta t\, \left( \Omega I_x + \omega_z(t_j)\, I_z \right) \right]
            \label{eq:Up1_approx}
            \end{equation}
            we partition each pulse into $N$ subintervals of equal duration $\Delta t = \tau_p/N$, where $\tau_p$ is the x-pulse duration. Within each subinterval, we approximate $\omega_z(t)$ by a constant value equal to its average over that subinterval (Fig. \ref{figSI:SimulationWithRotationMatrices} red dashed lines). The magnetization undergoes a simultaneous rotation about the $\xhat$- and $\zhat$-axes, which corresponds to a single rotation about the axis $(\Omega_x, 0, \omega_z(t_j))$ by angle
            \[
            \theta_j = \Delta t \sqrt{\Omega_x^2 + \omega_z^2(t_j)}.
            \]
            Defining the rotation vector
            \[
            \boldsymbol{\omega}_j = (\Omega_x \Delta t,\, 0,\, \omega_z(t_j)\Delta t) \equiv (\theta_{x,j},\, 0,\, \theta_{z,j}),
            \]
            the unit axis is $\mathbf{n}_j = \boldsymbol{\omega}_j / \|\boldsymbol{\omega}_j\|$ and the corresponding rotation matrix $R_j$ is obtained from Rodrigues’ formula:
            \[
            R_j = \mathbb{I} + \sin \theta_j\,[\mathbf{n}_j]_\times + (1 - \cos \theta_j)\,[\mathbf{n}_j]_\times^2,
            \]
            with $[\mathbf{n}]_\times$ the skew-symmetric cross-product matrix:
            \[
            [\mathbf{n}]_\times = 
            \begin{bmatrix}
            0 & -n_z & n_y \\
            n_z & 0 & -n_x \\
            -n_y & n_x & 0
            \end{bmatrix}.
            \]
            This construction is exact for constant $\Omega_x$ and $\omega_z$ within each subinterval and by design yields proper rotations ($R_j \in \mathrm{SO}(3)$).
            
            We then time-order the rotation matrices to form the stroboscopic rotations in the two frames (rightmost factor acts first):
            
            \textbf{Frame I (immediately after pulse 2):}
            \[
            R_\mathrm{I} = R_8 R_7 R_6 R_5 R_b R_4 R_3 R_2 R_1 R_a.
            \]
            
            \textbf{Frame II (immediately after pulse 1):}
            \[
            R_{\mathrm{II}} = R_4 R_3 R_2 R_1 R_a R_8 R_7 R_6 R_5 R_b.
            \]
            
            Here $R_1,\dots,R_4$ are the subinterval rotations within pulse~1, and $R_5,\dots,R_8$ those within pulse~2; $R_a$ and $R_b$ are the free-evolution $\zhat$-rotations (cf. Fig. \ref{figSI:SimulationWithRotationMatrices}).

            \textbf{Measurement timing.} Detection in our protocol occurs across the spacing. Since $R_a$ and $R_b$ are pure $\zhat$-rotations, they change only the azimuthal but not the elevation angle of $\mathbf{M}$. For determining the aligned direction it is therefore sufficient to treat the measurement as occurring immediately at the end of the preceding pulse (i.e., at the end of one full cycle)); including the finite detection window only adds an overall azimuthal phase.

            \textbf{Numerical accuracy and convergence.} Increasing the number of pulse subintervals $N$ improves the approximation of the time-varying $\zhat$-drive within the pulse. We found that $N \geq 8$ yields converged rotation axes and converged predicted magnetization vector positions, and we therefore used $N = 8$ in the simulations shown. This segmented-rotation scheme faithfully captures the micromotion induced by the finite pulse widths and the continuous $\zhat$-drive while retaining an efficient, interaction-free description.

        \subsection{Floquet Theory Description} \label{secSI:SimulationFloquet}
        
        To complement the rotation-matrix simulations presented in the previous section, we additionally plot the analytical predictions for the magnetization dynamics under the Floquet protocol as derived in Ref.~\cite{sahin22_trajectory}, including corrections that account for the finite pulse duration. For the concise description below, however, we work in the $\delta$-pulse limit for simplicity. Throughout, we restrict our attention to Frame~I, as defined previously and shown in Fig.~\ref{figSI:SimulationWithRotationMatrices}. In this frame, the evolution over a single Floquet cycle is

        \begin{equation}
        \begin{aligned}
        U_{F;1} ={}&
        e^{-i\pi I^x} e^{-i\epsilon I^x}
        e^{-i\tau\!\left(H_{nn} + \frac{\Phi_b}{\tau} I^z\right)} \\
        &\times
        e^{-i\pi I^x} e^{-i\epsilon I^x}
        e^{-i\tau\!\left(H_{nn} + \frac{\Phi_a}{\tau} I^z\right)} .
        \end{aligned}
        \end{equation}
        
        Here $\Phi_a$ and $\Phi_b$ are the acquired phases as defined in the previous section, $H_{\mathrm{nn}}=\sum_{i<j} d_{ij}\big(3I_i^z I_j^z-\mathbf{I}_i\cdot\mathbf{I}_j\big)$ is the secular dipolar Hamiltonian, and $\epsilon = \pi - \theta$ denotes the deviation of the spin-lock angle from a perfect $\pi$ pulse. After performing a toggling-frame expansion to eliminate the strong $\xhat$-pulses, we obtain
        
        \begin{equation}
        U_{F;1}
        = e^{-i\epsilon I^x}
          e^{-i\tau\!\left(H_{nn} - \frac{\Phi_b}{\tau} I^z\right)}
          e^{-i\epsilon I^x}
          e^{-i\tau\!\left(H_{nn} + \frac{\Phi_a}{\tau} I^z\right)} ,
        \end{equation}
        
        and, using the symmetry of the sine function, $\Phi_b = -\Phi_a$. Applying the Baker-Campbell-Hausdorff (BCH) expansion, the leading-order effective Hamiltonian (to order $O(\tau^0)$) becomes
        
        \begin{equation}
        H_{\mathrm{eff};1} = H_{nn} + w_x I^x + w_z I^z ,
        \end{equation}
        
        where $w_x = \epsilon/\tau$ and $w_z = \Phi_a/\tau$. In this limit, the elevation angle of the effective field is
        
        \begin{equation}
        \phi_{\mathrm{elevation}} = \arctan\!\left(\frac{\Phi_a}{\epsilon}\right),
        \end{equation}
        
        showing that the elevation angle is maximized near $\theta = \pi$, in excellent agreement with the experimental observations (Fig.~\ref{figSI:pulseangle}).
        
        For the second stroboscopic frame (Frame~II in the previous section), the effective Hamiltonian $H_{\mathrm{eff};2}$ is obtained from $H_{\mathrm{eff};1}$ by a $\pi$ rotation about the $\xhat$-axis:
        
        \begin{equation}
        H_{\mathrm{eff};2} = H_{nn} + w_x I^x - w_z I^z ,
        \end{equation}
        
        which directly illustrates the emergence of the two magnetization vector orientations.

        \subsection{Simulation of Suppression Factor} \label{secSI:sim_suppressionfactor}
            To benchmark the measured suppression factor against the ideal expectation, we numerically generated a noise-free reference signal (cf. Fig. \ref{fig:fig4}c). A sinusoidal RF background wave was numerically generated over a duration of \SI{1}{\second} using the sampling rate of the corresponding experiment. A second sinusoidal component was added with sign inversion of every other sample, mimicking the experimental imprint’s orbit-induced reversal and producing two signals with a $180^{\circ}$ phase shift.

            The resulting synthetic signal was processed identically to the experimental data: FFT peak amplitudes were determined before and after signal extraction, and the suppression factor calculated via Eq.~\ref{eqSI:SuppFacCalculation}. This procedure was repeated for a large set of RF background frequencies in \SI{0.5}{\hertz} steps over the range $0$–\SI{1200}{\hertz} relative to the Larmor frequency.

\section*{\T{Supplementary Audio Files}}

\textbf{Supplementary Audio 1: Raw signal with background contamination.} \label{SIaudio1}
Corresponds to the data shown in Figure \ref{figSI:soundmeasurementlong}a. This WAV file contains the time-domain signal from one orbit axis prior to the background suppression. The recording includes the target acoustic waveform superimposed with a strong, artificially generated radio-frequency (RF) background.

\vspace{1em}

\textbf{Supplementary Audio 2: Reconstructed signal after background suppression.}\label{SIaudio2}
Corresponds to the data shown in Figure \ref{figSI:soundmeasurementlong}b. This WAV file contains the acoustic waveform after processing the raw data from \hyperref[SIaudio1]{Supplementary Audio 1} with the manifold subtraction (differential signal). It demonstrates the high-fidelity recovery of the original waveform.

\end{document}